\documentclass[useAMS,usenatbib]{mn2e}

\usepackage{graphicx}			
\usepackage{amsmath}			
\usepackage{amssymb}
\usepackage{multirow}
\usepackage{array}
\usepackage{tabularx}

\newcommand{\fnl}{f_{_{\rm NL}}}

\newcommand{\mnras}{MNRAS}

\newcommand{\nat}{Nat.}
\newcommand{\aap}{A\&A }
\newcommand{\apj}{ApJ }
\newcommand{\apjs}{ApJS}
\newcommand{\apjl}{ApJ }
\newcommand{\prd}{Phys.~Rev.~D }
\newcommand{\physrep}{Physics Reports}
\newcommand{\jcap}{JCAP}
\newcommand{\na}{NewA}
\renewcommand{\arraystretch}{1.5}
\title[Non Gaussianity and Minkowski Functionals]{Non Gaussianity and Minkowski Functionals: forecasts for Planck}

\author[A. Ducout, F.\,R. Bouchet, S. Colombi, D. Pogosyan and S. Prunet]{A. Ducout$^{1}$\thanks{E-mail: ducout@iap.fr}, F.\,R. Bouchet$^{1}$, S. Colombi$^{1}$, D. Pogosyan$^{2}$ and S. Prunet$^{1}$\\
$^{1}$Institut d'Astrophysique de Paris, CNRS \& UPMC, 98bis boulevard Arago, 75014 Paris, France\\
$^{2}$Department of Physics, University of Alberta, 11322-89 Avenue, Edmonton, Alberta, T6G 2G7, Canada}

\begin{document}

\pagerange{\pageref{firstpage}--\pageref{lastpage}} \pubyear{2012}

\maketitle

\label{firstpage}

\begin{abstract}
We study Minkowski Functionals as probes of primordial non-Gaussianity in the Cosmic Microwave Background, specifically for the estimate of the primordial `local' bi-spectrum parameter $ f_{_{\rm NL}}$, with instrumental parameters which should be appropriate for the Planck experiment. We use a maximum likelihood approach, which we couple with various filtering methods and test thoroughly for convergence. We included the effect of inhomogeneous noise as well as astrophysical biases induced by point sources and by the contamination from the Galaxy. We find that, when Wiener filtered maps are used (rather than simply smoothed with  Gaussian),  the expected error on the measurement of $f_{_{\rm NL}}$ should be as small as $\Delta f_{_{\rm NL}} \simeq 10$ when combining the 3 channels at 100, 143 and 217\,GHz in the Planck extended mission setup. This result is fairly insensitive to the non homogeneous nature of the noise, at least for realistic hit-maps expected from Planck. We then estimate the bias induced on the measurement of $f_{_{\rm NL}}$ by point sources in those 3 channels. With the appropriate masking of the bright sources, this bias can be reduced to a negligible level in the 100 and 143\,GHz  channels. It remains significant in the 217\,GHz channel, but can be corrected for.  The galactic foreground biases are quite important and present a complex dependence on sky coverage: making them negligible will depend strongly on the quality of the component separation methods.
\end{abstract}


\section[Introduction]{Introduction}

The Planck satellite was launched in May 2009 \citep{2010A&A...520A...1T, 2011A&A...536A...1P} to observe the microwave sky and particularly the Cosmic Microwave Background (hereafter CMB). It will provide major pieces of information about the early evolution of the universe and the origin of structures thanks to its unprecedented combination of resolution, sensitivity, and spectral coverage.
One of the important objectives of this mission is to bring new constraints on inflation theory and primordial non Gaussianities. In the simplest inflationary models based on a single slowly rolling scalar field, primordial fluctuations should be only weakly non Gaussian \citep{2003JHEP...05..013M, 2006JCAP...06..024B}. Yet in more general models a much higher level of non-Gaussianity (hereafter NG) is
expected. We can cite models with multiple scalar fields, features in the inflaton potential, non-adiabatic fluctuations, non-standard kinetic terms, warm inflation, deviations from Bunch-Davies vacuum (review in \citealt{2004PhR...402..103B}) or completely different mechanisms, for example topological defects such as cosmic strings \citep{1984Natur.310..391K}.

Non-Gaussianity is often parametrised in a phenomenological way by the non-linear `local' coupling parameters $f_{_{\mathrm{NL}}}$, $g_{_{\mathrm{NL}}}$, ... which appear in the
perturbative development of the primordial curvature perturbation \citep{2001PhRvD..63f3002K, 2002PhRvD..66f3008O},
\begin{equation}
	\Phi(x)=\phi_{\rm L}(x)+f_{_{\mathrm{NL}}}(\phi_{\rm L}^{2}(x)-\langle\phi_{\rm L}^{2}(x)\rangle)+g_{_{\mathrm{NL}}}\,\phi_{\rm L}^{3}(x)+ ... \, ,
\end{equation}
where $\phi_{\rm L}(x)$ is the linear Gaussian part of the Bardeen curvature. We will focus here on constraining the first parameter, $f_{_{\mathrm{NL}}}$.

The latest results from WMAP \citep{2011ApJS..192...18K} with 7 years of data provide the constraint
\begin{equation}
	f_{_{\mathrm{NL}}}^{\rm } = 32 \pm 21 \label{eq:wmapfnl}
\end{equation}  
at the 68\% confidence level. With Planck, we expect the error on $f_{_{\mathrm{NL}}}^{\rm }$ to be at best of the order of $2.5$ from both temperature and polarisation \citep{2010AdAst2010E..71Y, 2009JCAP...12..022S} or of the order of $5$ for the temperature only \citep{2001PhRvD..63f3002K}. These constraints on $f_{_{\rm NL}}^{\rm }$ were obtained with bi-spectrum measurements \citep{2005ApJ...634...14K, 2007ApJ...664..680Y, 2008ApJ...678..578Y}. The CMB bi-spectrum is the harmonic transform of the three-point correlation function and it has been shown that this is theoretically an optimal estimator for $f_{_{\rm NL}}$ as it saturates the Cram\'er-Rao inequality for a weak non Gaussianity \citep{2005PhRvD..72d3003B}.

However, alternative statistics to the bi-spectrum have been also developed, that can at least serve as checks and diagnoses of the results obtained from the bi-spectrum.  One of the main reasons to study various probes of non Gaussianity is indeed that they are affected differently by different systematics and contaminants such as inhomogeneous noise and foreground residuals induced by our galaxy and point sources, as well as secondary anisotropies  \citep[e.g.,][and references therein]{2008RPPh...71f6902A} such as integrated Sachs-Wolfe effect, Sunyaev-Zeldovich effect \citep{1972CoASP...4..173S} and weak gravitational lensing,  which can all contribute to non Gaussianity of the CMB in a non trivial way. In particular, the most serious bias on $f_{_{\rm NL}}^{\rm }$ from secondary anisotropies arises from the coupling between weak gravitational lensing and integrated Sachs-Wolfe effect  \citep[see][]{1999PhRvD..59j3002G, 2008PhRvD..77j7305S, 2009PhRvD..80h3004H, 2009PhRvD..80l3007M, 2010MNRAS.401.2406M},  which adds complexity to the analyses.

Here we focus on a complete set of topological tools, the Minkowski Functionals (hereafter MFs) introduced in cosmology by \citet{1994A&A...288..697M}. MFs describe the morphological features of random fields over excursion sets, i.e. regions where the field exceeds some threshold level $\nu$. These well-known probes of primordial non-Gaussianities \citep{1997ApJ...482L...1S, 1998NewA....3...75W, 1998MNRAS.297..355S} have been widely used in two and three dimensions, for instance on WMAP CMB data \citep{2006ApJ...653...11H, 2008MNRAS.389.1439H} and on the SDSS galaxy catalogue \citep{2005ApJ...633...11P,2006ApJ...653...11H}.

For CMB studies, MFs provide a nice complement to the bi-spectrum for several reasons. Firstly, at variance with the bi-spectrum, which evolves in harmonic (or Fourier) space, MFs are defined in real space, which makes a robust implementation for MFs in practice much easier than for the bi-spectrum. Secondly, MFs are sensitive to the full hierarchy of higher-order correlations, instead of third order only, and can provide additional information on all the non-linear coupling parameters beyond the sole $f_{\rm NL}$.  Because they are measured on excursions of the density field smoothed with an isotropic window, MFs only probe angular averages of higher order statistics, leaving out the angular dependences, at variance with the bi-spectrum. This explains why MFs are less optimal than the bi-spectrum in terms of disentangling between various models of weak NG. However, the very nature of these angular averages eases considerably statistical analysis and reduces the number of parameters while performing maximum likelihood analysis. As a result, even though suboptimal, MFs represent powerful tools of investigation of NG. For instance, the first limit on the primordial non-Gaussianity in the isocurvature perturbation was measured with MFs, even though theoretical predictions where computed on the bi-spectrum \citep{2009MNRAS.398.2188H}. Note finally that the constraint on $f_{\rm NL}$ obtained from the analysis of WMAP3  with MFs reads  \begin{equation}f_{_{\mathrm{NL}}}^{\rm } = 11 \pm 40\end{equation}  at the 68\% confidence level \citep{2008MNRAS.389.1439H}, a rather competitive result, after all, compared to the bi-spectrum constraints from WMAP7 (eq.~\ref{eq:wmapfnl}). 

In this paper, we investigate constraints on $f_{_{\rm NL}}$ obtained with MFs from a practical point of view, in order to ease at best future application of MFs to Planck data. After detailing the MFs estimators and the Bayesian method we implemented, we review the effects of the main systematics that can affect/bias the results. These systematics can be of instrumental nature --inhomogeneous noise, beams--  or of astrophysical nature: the observed microwave sky contains  foreground emission from the Milky Way and from extragalactic sources. 

Our Galaxy is indeed a strong source of contamination. It is generally accounted for by masking and/or by using component separation methods and  by assuming that the final residual bias on $f_{_{\rm NL}}^{\rm }$ is negligible compared to error bars  \citep{2008MNRAS.389.1439H, 2003ApJS..148..119K}. Nevertheless it has been shown that component separation leave Galactic features in CMB maps \citep{2003ApJ...590L..65C}. Indeed the various methods of component separation \citep{2008A&A...491..597L} need to be controlled at the new level of accuracy on $f_{_{\rm NL}}$ expected for Planck. Another approach is to use foreground templates and to marginalise over them \citep{2002ApJ...566...19K, 2011ApJS..192...18K}. Here, we chose to analyse the behaviour of MFs when using masks and a na\"ive model of component separation. For that, we used foreground templates provided by the Planck Sky model (PSM) code (Delabrouille et al, 2012).

Point sources are mostly unresolved galaxies (at least in low-density regions of Galactic emission), some emitting a signal sufficiently high to be detected individually, the others forming a diffuse unresolved  background. Point sources consist first of radio-galaxies, active galactic nuclei which emit in radio frequencies with synchrotron process. They can also be dusty starburst galaxies which are observed via the thermal emission of their dust heated by the ultra-violet emission of young stars. Current instruments are not able to detect all these galaxies individually and the integrated emission of all the faint galaxies form a diffuse background, the Cosmic Infrared Background (CIB) which has been recently studied and observed \citep{1996A&A...308L...5P,2000A&A...355...17L, 2011A&A...536A..18P}. The strong point sources are accounted for by masking them but the fainter ones can still induce biases in NG studies. The point source bi-spectrum has already been measured \citep{2011ApJS..192...18K} and its effect in the estimates of $f_{_{\rm NL}}$ evaluated, for bi-spectrum measurements \citep{2008PhRvD..77l3011B, 2008PhRvD..77j7305S,2012MNRAS.tmp.2556L}.

In addition to these contributions, as mentioned above, CMB contains secondary anisotropies imprinted between the surface of last scattering and present time. While we leave the treatment of these secondary anisotropies for future work, we shall treat in detail the contamination from our galaxy and point sources. 


This paper is organised as follows. In section 2, we review the Bayesian method we use to optimise the constraint on $f_{_{\rm NL}}^{\rm }$ from MFs. In section 3, we study the effects of pixelisation and filtering. Section 4 deals with the effect of inhomogeneous noise. In sections 5 and 6, we study astrophysical sources of systematic effects: first point sources, then Galactic foregrounds. Section 7 summarises and discusses the results. For completeness, some additional technical details can be found in Appendix A, which discusses analytic predictions for MFs in the weakly non Gaussian regime, Appendix B, which details the algorithm used to measure MFs, Appendix \ref{app:numsim}, with a convergence study of our $\chi^2$,  and Appendix D, which provides some details on the Planck simulation experiments performed in this work.

The pixelisation scheme adopted in this paper is the same as in Planck processed maps, namely HEALPix\footnote{Available at http://healpix.jpl.nasa.gov }  \citep{2005ApJ...622..759G}.

Note finally that there are a number of tables to illustrate the results. Some of these tables are purposely incomplete to lighten the presentation.


\section[Method]{Method}

In this section, we set notations by first recalling basics of Minkowski functionals and their measurement inside an excursion of varying height (\S~\ref{sec:MFDEF}). Then we detail the Bayesian approach we use to estimate $f_{_{\rm NL}}^{\rm }$ (\S~\ref{sec:tng}), by comparing the measurement ${\hat y}$ of a functional or a combination of functionals to its expected value. Within a set of reasonable simplifying assumptions, in particular weak non Gaussianity, this approach reduces to a simple $\chi^2$ test (\S~\ref{sec:conchi}). The numerical estimate of this $\chi^2$ requires the accurate calculation of the covariance matrix of our estimator ${\hat y}$ from a large set of Gaussian simulations, as discussed in  \S~\ref{sec:conviss}. Additional details concerning the Gaussian simulations set up as well as the method used to test the convergence of the $\chi^2$ are given in Appendix \ref{app:numsim}. Once we have a numerically robust $\chi^2$ test, we study its sensitivity to important parameters of the problem, such as the number of bins used to explore the excursion and the range of the excursion levels, as well as the choice of the functionals and/or their combination (\S~\ref{sec:sensitivity}). 

\subsection{Minkowski Functionals}\label{sec:MFDEF}

For a two-dimensional field $\delta$ of zero mean and of variance $\sigma_{0}^2$, defined on the sphere, and smoothed with a window of typical size $\ell$ (to be defined later), we consider an excursion set of height $\nu=\delta/\sigma_{0}$, i.e. the set of points where the field exceeds the threshold $\nu$. In what follows, we shall study the topological properties of the excursion with four quantities, denoted by  $V_{k}$ ($k=0,1,2,3$). The first three ones correspond to MFs: $V_{0}$ is the fractional Area of the regions above the threshold, $V_{1}$ is the Perimeter of these regions and $V_{2}$ is the Genus, defined as the total number of connected components of the excursion above the threshold minus the total number of connected components under the threshold. The fourth one, $V_3$ that will be called the Number of clusters, also noted $N_{\rm clusters}$, is just the number of connected regions above the threshold for positive thresholds and reversely for negative thresholds \citep[these regions are also known as Betti numbers or as cold/hot spots,][]{2012ApJ...755..122C}. Analytic formulae (theoretical expectation values) for the quantities $V_k$  are known for Gaussian and weakly non Gaussian fields and are summarised in Appendix \ref{theory}. An interesting property of these functionals is that the pure spectral dependence can be factorised out: 
\begin{equation}
	V_{k}(\nu)=A_{k}v_{k}(\nu)
\label{eq:normalizedvk}
\end{equation} 
where the amplitude $A_{k}$ is determined by the shape of the power-spectrum of the field fluctuations. The renormalised functionals $v_k$ depend only on the non Gaussian corrections, i.e. on the behaviour of correlations functions beyond second order. The analysis of $v_k$ allows us to focus on non Gaussianity, which is the goal of this paper. 

In what follows, we denote by ${\hat X}$ an estimator of the quantity $X$. We calculate ${\hat v}_k={\hat V}_k/{\hat A}_k$ on a  pixelised map by proceeding as described in Appendix \ref{algorithm} to estimate ${\hat V}_k$.  The quantity ${\hat A}_k$ itself is obtained by direct measurement of $\sigma_{0}$ and $\sigma_{1}$ on each map (see Appendix \ref{theory} for definitions).

\subsection{Testing non Gaussianity with a $\chi^2$} \label{sec:tng}

To provide constraints on non-Gaussianity, we adopt a Bayesian approach similar to that of \citet{2008MNRAS.389.1439H}, comparing measurement of normalised functionals $v_k$ on the data map under
consideration to the ``theoretical predictions'' obtained from the average of measurements of $v_k$ on a large number of non-Gaussian simulations, the non Gaussian part being simply proportional to $f_{_{\rm NL}}^{\rm }$. We will focus here on the non-linear coupling parameter ‘local’ $f_{_{\mathrm{NL}}}^{\rm }$ but the method can be applied to other types of non Gaussianities. 

To perform the measurements, we consider an ensemble of $n_{\rm bins}$ values of $\nu_i$ ranging from $-\nu_{\rm max}$ and $+\nu_{\rm max}$, defining a vector ${\hat v}_k \equiv \{ {\hat v}_k(\nu_1),\cdots,{\hat v}_k(\nu_{n_{\rm bins}}) \}$ for each of the functionals.  The statistics under consideration will then be a vector of $n \geq n_{\rm functionals}\times n_{\rm bins}$ elements,  ${\hat y}={\hat v}_k$ if only one functional is used in the analysis, or ${\hat y}=\{ {\hat v}_i,{\hat v}_j,\cdots \}$ if a combination of $n_{\rm functionals}>1$ functionals is considered. 
 
The Bayes formula writes
\begin{equation} 
	P(f_{_{\rm NL}}|\, \hat{y})=\dfrac{P(\hat{y}\, | f_{_{\mathrm{NL}}}) P(f_{_{\mathrm{NL}}})}{\int P(\hat{y}\, | f_{_{\mathrm{NL}}}) P(f_{_{\mathrm{NL}}}) {\rm d}\fnl}.  
\end{equation}
In what follows, we shall take a flat prior for $\fnl$,  with $P( f_{ _{\mathrm{NL}}} )$ a constant, while the evidence $\int P(\hat{y}\, | f_{_{\mathrm{NL}}}) P(f_{_{\mathrm{NL}}}) {\rm d}\fnl$ will be just considered as a normalisation. 

\subsubsection{Construction of a $\chi^2$ test} \label{sec:conchi}

We assume that the likelihood $P(\hat{y}\, | f_{_{\mathrm{NL}}})$ is a Gaussian, which allows us to
define a simple $\chi^{2}$ test for $f_{_{\rm NL}}$. This approximation is expected to be good in the regime where the fluctuations of each component $y_i$ of the vector ${\hat y}$ remain small compared to its ensemble average $\langle y_i \rangle$, when one considers a large number of realisations of the random field.  Keeping these fluctuations small sets practical constraints on (i) the value of $\nu_{\rm max}$, which should not be too extreme to avoid sensitivity to rare events, (ii) the smoothing scale $\ell$, that should be small enough to probe a sufficient number of independent harmonic modes on the sky. Because we consider only small values of $\ell$ in most of what follows our main concern is the choice of  $\nu_{\rm max}$. We shall restrict to the range $\nu_{\rm max} < 5$ \citep[see also the discussion in][]{1990ApJ...352....1G}.

With all these simplifications, the posterior becomes
\begin{equation}
	P(f_{_{\mathrm{NL}}}|\,\hat{y}) \propto \exp\left[ -\dfrac{\chi^{2}(\hat{y},\fnl)}{2} \right]
\label{eq:postsimp}
\end{equation}
with
 \begin{eqnarray}
	\chi^{2}(\hat{y}, f_{_{\rm NL}}) & \equiv &  \left[ \hat{y}- \bar{y}( f_{_{\rm NL}} )   \right]^{T}C^{-1} \left[ \hat{y}- \bar{y}( f_{_{\rm NL}} ) \right] \label{eq:monchi22} \\
             & = & \sum_{i=1}^{n}  \sum_{j=1}^{n}C^{-1}_{ij}\left[ \hat{y}_{i}- \bar{y}_{i}( f_{_{\rm NL}} )\right] \left[\hat{y}_{j}- \bar{y}_{j}( f_{_{\rm NL}} )\right],
\label{eq:monchi2}
\end{eqnarray}
where $\bar{y}( f_{_{\rm NL}} )$ is the model under test. It  can be derived analytically in the weakly non Gaussian regime \cite[see][and appendix \ref{theory}]{2008MNRAS.389.1439H}  or by direct measurements on simulations. Since we want to take into account complex additional contributions such as inhomogeneous noise and  masks, we use here the more flexible second approach with non-Gaussian simulations provided by \citet{2009ApJS..184..264E}. So $\bar{y}( f_{_{\rm NL}} )\equiv\left\langle y( f_{_{\rm NL}} )\right\rangle $ is the mean of $y$ measured over $m_{_{\rm {NG}}}$ maps with a level non-Gaussianity $f_{_{\rm NL}}$. In practice, we shall take
\begin{equation}
	m_{\rm NG}=200.
\end{equation}

\begin{figure}
   \includegraphics[width=8.5cm]{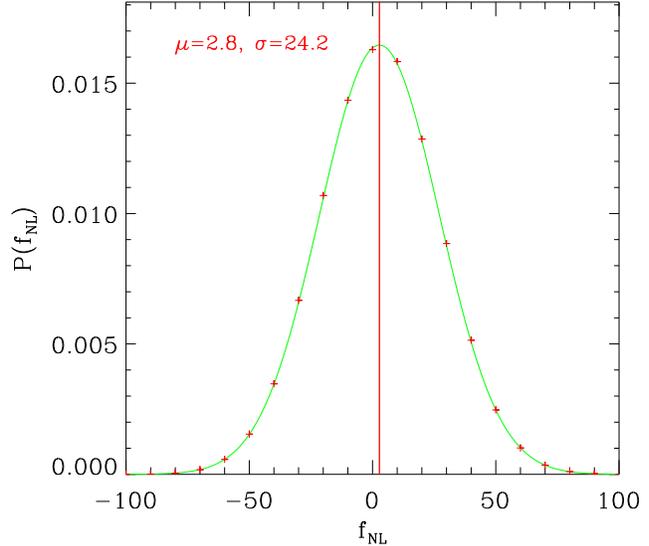}\\
  \caption{Example of measured posterior probability for $f_{_{\rm NL}}$ when using the Perimeter on one test map (red points). The green curve corresponds to a  Gaussian of average ${\hat f}_{_{\rm NL}}$ and of variance ${\widehat \Delta f}_{_{\rm NL}}$, with ${\hat f}_{_{\rm NL}}$  and ${\widehat \Delta f}_{_{\rm NL}}$ given by eqs.~(\ref{eq:fnl}) and (\ref{eq:deltafnl}). For this particular example, we used  $n_{\rm bins}=26$ and $\nu_{\rm max}=3.5$. The actual value of $f_{_{\rm NL}}$ is $f_{_{\rm NL}}=0$.}
  \label{fig:bayesian_fnl}
\end{figure}

Since non-Gaussianity is weak \citep{2011ApJS..192...18K}, one can compute the covariance matrix $C$ in a Monte-Carlo fashion relying on simulations of
{\em Gaussian} maps of the CMB (to which we add, when relevant, beam effects, noise, masks). 
In that case, the posterior probability distribution function $P(f_{_{\mathrm{NL}}}|\,\hat{y})$  is expected to be very close to a Gaussian, as illustrated by Fig.~\ref{fig:bayesian_fnl} for the Perimeter. Indeed,  in the weakly non Gaussian regime, ${\bar y}_i(f_{_{\rm NL}})$ is, at first order, a linear function of $f_{_{\rm NL}}$ (see, e.g., Appendix \ref{theory}).  

To estimate the most likely posterior value of $f_{\rm NL}$ and an uncertainty, instead of trying to find iteratively the maximum of $P(f_{_{\rm NL}})$ followed by an estimate of the local curvature using e.g. the Fisher information matrix to estimate the error, we prefer, for simplicity and robustness, to compute directly the average and the variance of the posterior distribution, $P(f_{_{\rm NL}})$. This latter is estimated using a number of equally spaced  bins, $f_i$, $i=1,\cdots,n_{_{\rm NL}}$, with, for the purpose of this paper $n_{_{\rm NL}}=21$. From this approximation of the posterior, the average and the variance are directly estimated numerically via the simple formulae
\begin{eqnarray}
	{\hat f}_{_{\rm NL}} &\equiv & \frac{\sum_i f_i P_i}{\sum_i P_i}, \label{eq:fnl}\\
	({\widehat {\Delta f}}_{_{\rm NL}})^2 & \equiv & \frac{ \sum_i (f_i-{\hat f}_{_{\rm NL}})^2 P_i}{\sum_i P_i}, \label{eq:deltafnl}
\end{eqnarray}
where $P_i$ is proportional to $P(f_{_{\rm NL}}=f_i)$. 

Finally, one might consider performing a number $m_{\rm test}$ of realisations of the data  ${\hat y}$ --also called in this paper ``test'' maps-- to have a more accurate description of the ``typical'' expected posterior probability. However such a {\em forecasting} is not obvious, as we have to define a frequentist average over Bayesian quantities, to predict the typical value expected
for ${\hat f}_{_{\rm NL}}$ and $\widehat{\Delta f}_{_{\rm NL}}$. Our choice, equivalent to Fisher
information forecast in the pure Gaussian case, consists in estimating $f_{_{\rm NL}}$ and $\Delta f_{_{\rm NL}}$ from the following averages
\begin{eqnarray}
	\langle {\hat f}_{_{\rm NL}} \rangle & = & \frac{1}{m_{\rm test}}\sum_{i=1}^{\rm m_{\rm test}} {\hat f}_{_{\rm NL},i}, \\
	\langle \widehat{\Delta f}_{_{\rm NL}}^2  \rangle & = & \frac{1}{m_{\rm test}-1} \sum_{i=1}^{\rm m_{\rm test}} \widehat{\Delta f}_{_{\rm NL},i}^2,
\label{eq:forecasterror}
\end{eqnarray} 
where ${\hat f}_{_{\rm NL},i}$ and $\widehat{\Delta f}_{_{\rm NL},i}$ are obtained from the $\chi^2$ analysis of ``data'' map number $i$ using
eqs.~(\ref{eq:fnl}) and (\ref{eq:deltafnl}).  In what follows, we shall always take (for each individual value of $f_{\rm NL}$ considered in the tests maps)
\begin{equation}
	m_{\rm test}=200.
\end{equation}

\subsubsection{Convergence issues}
\label{sec:conviss}

The covariance matrix is estimated from the average over $m$ simulations, on each of which a $y^{\rm G}$ vector of functionals is measured:
 \begin{equation}C_{ij}\equiv\left\langle \left( y_{i}^{\rm G}- \bar{y}_{i}^{\rm G}\right) \left( y_{j}^{\rm G}- \bar{y}_{j}^{\rm G}\right) \right\rangle,\end{equation}
with $\bar{y}_{j}^{\rm G}\equiv\left\langle y_{j}^{\rm G}\right\rangle $.

The calculation of the $\chi^2$ requires the inversion of the covariance matrix, an uneasy task, because $C$ can be ill-conditioned, especially if $m$ is not large enough. The number of simulations required to have a good estimate of the $\chi^2 (\hat{y}, f_{_{\rm NL}})$  function indeed depends on the number of bins, $n_{\rm bins}$, on the functionals under consideration, their number, $n_{\rm functionals}$, and on the choice of $\nu_{\rm max}$. The way we estimate $m$ is exposed in detail in Appendix \ref{app:numsim}.  The results of our analyses are summarised in table \ref{table:number_m1}, where the minimum values of $m$ required for having a better than two percent accuracy on the $\chi^2$ obtained from the perimeter are displayed for various realistic set-ups in terms of the number of bins and of the  excursion range. Similar orders of magnitude are expected for other functionals or {\em when combining a set of functionals}, as explicitly checked for the combination $V_1+V_2$, $n_{\rm bins}=26$ and $\nu_{\rm max}=3.5$, where the corresponding value found in Table~\ref{table:number_m1} remains {\em unchanged} within 5\%. This latter property comes from the fact that in practice, two different functionals are much less correlated than the same functional estimated at two successive bins of the excursion: combining functionals doubles the dimensions of matrix $C$ but does not make it significantly more degenerate. In the subsequent calculations performed in this paper, we shall take 
\begin{equation}
	m=10\,000, 
\end{equation}
a number of simulations sufficient for a good estimate of the $\chi^2$ for $n_{\rm bin} \la 50$ and $\nu_{\rm max} \ga 3.5$.

\begin{table}
\begin{center}
\begin{tabular}{l c c c}
\hline
& $\nu_{\rm max} =5$ & $\nu_{\rm max} =3.5$ & $\nu_{\rm max} =2$\\
\hline
\multirow{2}{1.7cm}{$n_{\rm bins}=11$} &   $\Delta \nu=1$ &   $\Delta \nu=0.7$ &   $\Delta \nu=0.4$\\
 & $m=7.0\times10^{3}$   & $m=8.0\times10^{3}$  & $m=8.0\times10^{3}$ \\
 \hline
 \multirow{2}{1.7cm}{$n_{\rm bins}=26$} &   $\Delta \nu=0.4$ &   $\Delta \nu=0.28$ &   $\Delta \nu=0.16$  \\
 & $m=8.0\times10^{3}$   & $m=9.0\times10^{3}$  & $m=9.0\times10^{3}$  \\
 \hline
 \multirow{2}{1.7cm}{$n_{\rm bins}=51$} &   $\Delta \nu=0.2$ &   $\Delta \nu=0.14$ &   $\Delta \nu=0.08$  \\
 & $m=9.0\times10^{3}$   & $m=10.0\times10^{3}$  & $m=10.0\times10^{3}$ \\
 \hline
 \multirow{2}{1.7cm}{$n_{\rm bins}=101$} &   $\Delta \nu=0.1$ &   $\Delta \nu=0.07$ &   $\Delta \nu=0.04$  \\
 & $m=12.0\times10^{3}$   & $m=14.0\times10^{3}$  & $m=15.0\times10^{3}$  \\
\hline
\end{tabular} 
  \caption{Number of Gaussian maps $m$ needed for a good convergence of the $\chi^2$ statistic with an accuracy better than $2$\%. The calculations are performed here for the Perimeter, ${\hat y}={\hat v}_1$, and for various values of the number of bins $n=n_{\rm bins}$ and of $\nu_{\rm max}$, but these results would not change much for other functionals or combinations of functionals, as discussed in the main text.  The details on the simulations are given in Appendix~\ref{app:numsim}. For $\nu_{\rm max} =5$ we have removed extreme thresholds and reduced $n_{\rm bins}$ to $\lbrace  9, 22, 45, 89 \rbrace$ respectively because the distribution of errors was not Gaussian for these bins which are too sensitive to rare  events.}
\label{table:number_m1} 
 \end{center}
 \end{table}

\subsection{Sensitivity of the estimator} \label{sec:sensitivity}

In this section we discuss the sensitivity of our $\chi^2$ estimator, that is the uncertainty on the estimated $f_{_{\rm NL}}$, with respect to the number of bins, $n_{\rm bins}$, the excursion range, $\nu_{\rm max}$,  and the set of functionals used, whether it is the Area, the Perimeter, the Genus, the Number of clusters, or a arbitrary combination of any of them. Our simulation setup is the same as in the previous paragraph (and detailed in Appendix \ref{app:numsim}).  We checked, in this full sky configuration with homogeneous noise, that our estimator is unbiased, $\langle {\hat f}_{_{\rm NL}} \rangle = f_{_{\rm NL}}$ (within numerical limits set by the finiteness of $m_{\rm NG}$ and $m_{\rm test}$).

 The results of our analyses are displayed in Tables \ref{table:thresholding} and \ref{table:mink_seuil}. They show that the combination 
\begin{equation}
	(\nu_{\rm max},n_{\rm bins})\simeq (3.5,26)
\end{equation}
 is fairly optimal and shall represent our choice in the subsequent analyses.  Note that it is important to notice that assuming the covariance matrix to be diagonal, as done for instance in \citet{2004ApJ...612...64E, 1990ApJ...352....1G} 
 {\em is not} a good approximation in the case of $f_{_{\rm NL}}$ and decreases by more than a factor two the constraining power of the $\chi^2$. 

The comparison between various functionals lead to the following ranking, in term of discriminative power:
\begin{equation}
{\rm Perimeter} \gtrsim {\rm Genus} > N_{\rm clusters} \gg {\rm Area}.
\end{equation}
While the Perimeter, the Genus and the Number of cluster present the same order of sensitivity, the Area is about twice less discriminative than them.  Most of the information on $f_{_{\rm NL}}$ can be extracted by a combined analysis of the perimeter and the genus, $V_1+V_2$, with a little improvement when taking into account the number of clusters, $V_1+V_2+V_3$, but the area does not carry significant pieces of additional information compared to the three others.

\begin{table}
 \begin{center}
\begin{tabular}{l c c c c}
\hline
\hspace*{0.1cm}$\langle \widehat{\Delta f}_{_{\mathrm{NL}}}^2\rangle^{1/2}$ \hspace*{0.2cm} & $n=11$ & $n=26$ & $n=51$ & $n=101$ \\
\hline
\hspace*{0.1cm}$\nu_{\rm max} =5$ &  29  & 25.5  & 25 & 25 \\
\hspace*{0.1cm}$\nu_{\rm max} =3.5$ &  27  & 25  & 25 & 25 \\
\hspace*{0.1cm}$\nu_{\rm max} =2$ &  37.5  & 37  & 37 & 36.5 \\
\hline
\end{tabular}
  \caption{Investigation of the best combination for $(\nu_{\rm max},n_{\rm bins})$. This table gives the uncertainly in $f_{_{\mathrm{NL}}}$ for different values of the number of bins $n=n_{\rm bins}$ and the threshold $\nu_{\rm max}$. The calculations are performed for the Perimeter, which is the most sensitive to $f_{_{\rm NL}}$.  The results would be nearly the same for the Genus and $N_{\rm clusters}$, while the Area is quite insensitive to $n$ and $\nu_{\rm max}$ for the range of values tested here. From this table it is fairly easy to conclude that the combination $(\nu_{\rm max},n_{\rm bins})=(3.5,26)$ is close to optimal. In particular it is not necessary in practice to go beyond $n_{\rm bins} \ga 26$. Note for completeness that the error bars computed here correspond to the specific set up of Appendix \ref{app:numsim} (see also caption of next table).}
  \label{table:thresholding}
 \end{center}
 \end{table}
\begin{table}
 \begin{center}
 \begin{tabular}{@{}  l  c   c c @{}}
\hline
 \hspace*{0.1cm} Functionals $V_{k}$ & & $\langle \widehat{\Delta f}_{_{\mathrm{NL}}}^2\rangle^{1/2}$ & \\
\hline
 \hspace*{0.1cm} $V_{0}$ & &  44 &\\
 \hspace*{0.1cm} $V_{1}$ & &  25& \\
 \hspace*{0.1cm} $V_{2}$ &  & 25& \\
 \hspace*{0.1cm} $V_{3}$ & & 28 & \\
 \hspace*{0.1cm} $V_{1}+V_{2}$ &  & 20 & \\
 \hspace*{0.1cm} $V_{1}+V_{2}+V_{3}$ &  & 19 &\\
 \hspace*{0.1cm} $V_{1}+V_{2}+V_{3}+V_{0}$ &  & 18.5 &\\
\hline
\end{tabular}
  \caption{Sensitivity of the estimator: error bars on the measurement of $f_{_{\rm NL}}$ for each functional, and for various combinations. The calculations are performed assuming $n=n_{\rm functionals} \times n_{\rm bins}$ with $n_{\rm bins}=26$ and $\nu_{\rm max} =3.5$. Note for completeness that the error bars computed here correspond to the specific set up of Appendix \ref{app:numsim} (see also Appendix~\ref{planck}): extended mission for the 3 channels and  noise filtering with a Gaussian window of full width at half maximum $\theta_{_{\rm FWHM}}^{\rm S}=10'$ and HEALPix resolution parameter $N_{\rm side}=2048$. }
  \label{table:mink_seuil}
 \end{center}
 \end{table} \setlength{\tabcolsep}{0.1cm} \renewcommand{\arraystretch}{1.5}

\section{Filtering and smoothing}

The measurement of Minkowksi Functionals requires smoothing of temperature maps in order to remove the contribution of the noise. This smoothing is performed at various scales on a pixelised map in order to extract all the statistical information available. First, one has to deal with discreteness effects brought by the pixelisation. In \S~\ref{sec:ellmaxnside}, we show that for practical measurements, it is not needed to have a large value of HEALPix resolution parameter $N_{\rm side}$ to extract all the relevant information even if this means a pixel size comparable to the size of the smoothing window: this is because we explicitly account for these pixelisations effects in the model. In \S~\ref{sec:gaussiansm}, we consider Gaussian smoothing and show that most information on $f_{_{\rm NL}}$ obtained from MFs is present at the smallest possible scales, i.e. at scales comparable to the size of the beam of the instrument. Note however that this result stands for local $f_{_{\rm NL}}$ and might vary for different types of non Gaussianity. Finally,  \S~\ref{sec:wienerfilters} deals with Wiener filtering for the field, its first and second derivatives.  We show that the results obtained with a simple combination of Wiener filters set much better constraints on $f_{_{\rm NL}}$ than na\"ive Gaussian smoothing at various scales. 

In what follows, smoothing scale will be expressed in units of the full width at half maximum (FWHM), $\theta_{_{\rm FWHM}}^{\rm S}$ (see Appendix~\ref{app:simulations} for more details).

\subsection{Pixelisation effects: choice of $N_{\rm side}$} \label{sec:ellmaxnside}

In principle, the pixel size should be small compared to the smoothing kernel size, in order to avoid systematic errors introduced by the discrete nature of the pixelisation \citep[see, e.g.][]{2000PhRvL..85.5515C,2006MNRAS.366.1201N}. We checked in practice that no bias is introduced on the measurement of $f_{_{\rm NL}}$ even when the pixel size becomes comparable to the smoothing window size, because the defects of the pixelisation are present as well in the model. However, introducing larger pixels is similar to coarse graining  and decreases the effective level of statistical richness, which in turn increases the uncertainty on $f_{_{\rm NL}}$. Furthermore, a too large pixel size would simply make the additional filtering operation inoperative: instead, what we would get in that case is simply the dominant part of filtering to be a convolution with a top-hat of the pixel shape. 

Table~\ref{table:sensitivity_nside_lmax} gives the error obtained on $f_{_{\rm NL}}$ when using the combination of all four functionals, for various values of HEALPix resolution parameter $N_{\rm side}$, different levels of noise and a Gaussian smoothing with $\theta_{_{\rm FWHM}}^{\rm S}=5'$. For Planck purpose, we also find that the improvement brought by the $N_{\rm side}=2048$ resolution is tiny, as expected for higher level of noise, and this table shows that it is not needed in practice to go beyond $N_{\rm side}=1024$, which is rather handy computationally speaking. From now on, unless specified otherwise, we shall assume
\begin{equation}
	N_{\rm side}=1024
\end{equation}
for all subsequent analyses. 

\begin{table}
 \begin{center}
 \begin{tabular}{ l   c  c  c  }
\hline
 \hspace*{0.1cm}Noise & $N_{\rm side}=512$ \hspace*{0.1cm}  & \hspace*{0.1cm}$N_{\rm side}=1024$ \hspace*{0.1cm}  & \hspace*{0.1cm} $N_{\rm side}=2048$ \\
&$\ell_{\rm max}=1000$ & $\ell_{\rm max}=2000$ & \hspace*{0.15cm} $\ell_{\rm max}=3500$ \\
\hline
\hspace*{0.1cm}0.33 $\mu$K.deg  \hspace*{0.2cm}            &   19.5        &  16.5  &   16  \\
\hspace*{0.1cm}0.5 $\mu$K.deg               &   20        &  16.5  &   16.5  \\
\hspace*{0.1cm}0.7 $\mu$K.deg               &   20     &  17     &   16.5  \\
\hspace*{0.1cm}1.1 $\mu$K.deg               &   20        &  18.5  &   18    \\
\hline
\end{tabular}
  \caption{Sensitivity $\langle\widehat{\Delta f}_{_{\rm NL}}^2\rangle^{1/2}$ versus $N_{\rm side}$ for various levels of noise (see Appendix~\ref{planck}), when using the combination $V_{0}+V_{1}+V_{2}+V_{3}$ of all four functionals and Gaussian smoothing with $\theta_{_{\rm FWHM}}^{\rm S}=5'$. Other parameters used to perform the simulations are $n_{\rm bins}=26$, $ \nu_{\rm max}=3.5$, $m=10\,000$ Gaussian maps for the covariance matrix, $m_{\rm NG}=200$ reference maps with different levels of $f_{_{\rm NL}}$ for the model and 200 test maps with $f_{_{\rm NL}}=0$. Note that, in the weakly non Gaussian regime considered here, the forecast $\langle\widehat{\Delta f}_{_{\rm NL}}^2\rangle^{1/2}$  does not depend significantly on the actual value of $f_{_{\rm NL}}$.  For each value of $N_{\rm side}$, calculations in harmonic space are band limited to $\ell \leq \ell_{\rm max} \simeq 2 N_{\rm side}$. This cut-off at $\ell_{\rm max}$ does not affect significantly the results.}
  \label{table:sensitivity_nside_lmax}
 \end{center}
 \end{table}
 
\subsection{Gaussian Smoothing} \label{sec:gaussiansm}

Smoothing with a Gaussian kernel depends only on angular scale, $\theta_{_{\rm FWHM}}^{\rm S}$. {\it A priori}, each type non-Gaussianity is characterised by a specific scale range where the sensitivity of the estimator of $f_{_{\rm NL}}$ is the best and we have to stay aware that we are limited here to only one particular case of non Gaussianity, although quite typical. 

We tested different smoothing scales, including even very small scales ($\theta_{_{\rm FWHM}}^{\rm S}=5'$), smaller than size of the beam ($\theta_{_{\rm FWHM}}^{\rm b}=7.2'$ in the case of the combined channels), and just above the size of the pixel, which might look awkward, but in fact does not introduce any significant bias on the measurement of $f_{_{\rm NL}}$, as already argued in \S~\ref{sec:ellmaxnside} when discussing about pixelisation effects. 

One issue about smoothing at large scales is that it reduces the number of independent modes available on the sky. This in turns reduces the quality of the measurement of the MFs in the tails (large values of $|\nu|$) and can make the likelihood function non Gaussian, as discussed in the beginning of \S~\ref{sec:conchi}, particularly if $\nu_{\rm max}$ is too large.  For instance, with $\nu_{\rm max}=4$, the Gaussian assumption for the likelihood is legitimate for $\theta_{_{\rm FWHM}}^{\rm S}=5'$ but not for $\theta_{_{\rm FWHM}}^{\rm S}=40'$. Here we checked that our $\chi^2$ analysis was still valid and well converged at smoothing scales as large as $\theta_{_{\rm FWHM}}^{\rm S}=40'$ for our default choice for the parameters, $\nu_{\rm max}=3.5$, $n_{\rm bins}=26$ and $m=10\,000$.

Table~\ref{table:mink_scale} summarises the results of our analyses for the case when the combination of all functionals is used at various scales or various combinations of scales. Note that when combining 4 scales, we have a large number of entries in the data vector,  $n=n_{\rm bins}\times n_{\rm functionals}\times n_{\rm scales}\simeq 400$, but we checked that this did not affect the convergence of the calculation of the covariance matrix with $m=10\,000$. The results of Table~\ref{table:mink_scale} show that there is no significant statistical information for $\theta_{_{\rm FWHM}}^{\rm S}\ga 20'$. Most of the signal is captured by the combination $\theta_{_{\rm FWHM}}^{\rm S}=5'$ with $\theta_{_{\rm FWHM}}^{\rm S}=10'$.

Note that the scales we consider here are much smaller than those considered in Table 3 of \citet{2006ApJ...653...11H} that correspond to $\theta_{_{\rm FWHM}}^{\rm S}=\lbrace 11.75', 23.5', 47' \rbrace$, which explains the better constraints we obtain on $f_{_{\rm NL}}$. 
\begin{table}
 \begin{center}
 \begin{tabular}{@{} l   c  c c @{}}
\hline
 \hspace*{0.1cm} $\theta_{_{\rm FWHM}}^{\rm S}$ && $\langle \widehat{\Delta f}_{_{\rm NL}}^2\rangle^{1/2}$& \\
\hline
\hspace*{0.1cm} 5'        && 16.5 & \\
\hspace*{0.1cm} 10'       &&  20 & \\
\hspace*{0.1cm} 20'     && 26.5 & \\
\hspace*{0.1cm} 40'      && 34  &\\
\hspace*{0.1cm} 5' + 10'       && 14 & \\
\hspace*{0.1cm} 5' + 10' + 20'      && 13.5 & \\
\hspace*{0.1cm} 5' + 10' + 20' +40'     && 13.5 & \\
\hline
\end{tabular}
  \caption{Sensitivity of the estimator of $f_{\rm NL}$ versus Gaussian smoothing scale $\theta_{_{\rm FWHM}}^{\rm S}$ when the combination of all functionals, $V_{0}+V_{1}+V_{2}+V_{3}$, is used. The forecasted
quantity   $\langle \widehat{\Delta f}_{_{\rm NL}}^2\rangle^{1/2}$ is given for the combined 100, 143 and 217\,GHz channels in the extended Planck mission (see appendix \ref{planck}). }
  \label{table:mink_scale}
 \end{center}
 \end{table}

\subsection{Wiener filters} \label{sec:wienerfilters}

Now we consider Wiener filtering, which is an optimal way of recovering the signal of a data map $D=S+N$ in case both the underlying signal $S$ and the noise $N$ are Gaussian. The Wiener filter writes, in harmonic space,
\begin{equation}
	W_{\rm M}=\dfrac{C_{\ell}}{C_{\ell}+N_{\ell}}
\label{eq:wiener1}
\end{equation}
where $C_{\ell}$ is the forecasted power-spectrum of the signal 	we want to analyse and $N_{\ell}$ is the known power spectrum of the noise. Here, signal will refer to the dominant, Gaussian, cosmological part of the CMB. Obviously, this is sub-optimal, since we are after the non Gaussian part of the CMB, while all the rest should be considered as noise. However, using a Wiener filter designed for extracting the non Gaussian signal would require a stronger prior, making the measurement of $f_{\rm NL}$ potentially more accurate but only for a restricted class of non Gaussianities.

Similarly, one can define in harmonic space a ``derivative''  Wiener operator,
\begin{equation}
	W_{\rm D1}=\sqrt{\ell (\ell +1) } \dfrac{C_{\ell}}{C_{\ell}+N_{\ell}},
	\label{eq:wiener2}
\end{equation}
and a ``second derivative'' Wiener filter optimal for recovering the Laplacian of the signal, $\Delta S$,
\begin{equation}
	W_{\rm D2}  =  \ell (\ell +1) \dfrac{C_{\ell}}{C_{\ell}+N_{\ell}}.
	\label{eq:wiener3}
\end{equation}

\begin{figure}
\begin{center}
\includegraphics[width=8cm]{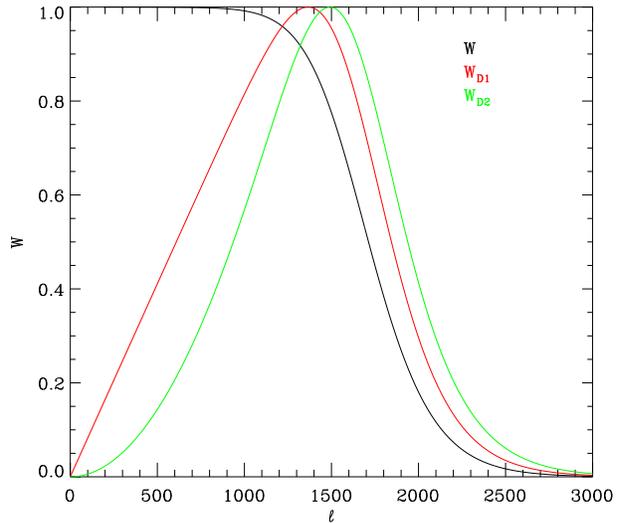}
  \caption{The three Wiener filters, $W_{\rm M}$, $W_{\rm D1}$ and $W_{\rm D2}$ given by eqs.~(\ref{eq:wiener1}), (\ref{eq:wiener2}) and (\ref{eq:wiener3}).}
  \label{fig:wiener_filter_curves}
\end{center}
\end{figure}

In practice, we use a smoothed version of the quantity ${C_{\ell}}/({C_{\ell}+N_{\ell}})$, a ``Wiener-like'' function used for component separation method (CMB removal) in \citet{2011A&A...536A...6P}. Before applying the Wiener filters we also correct the map for the Gaussian beam corresponding to the channel configuration. The three resulting filters, $W_{\rm M}$, $W_{\rm D1}$ and $W_{\rm D2}$ are represented on Fig.~\ref{fig:wiener_filter_curves}.

Table~\ref{table:wiener} summarises the results of our analyses using them. It shows, not surprisingly, that $W_{\rm M}$ alone does much better than Gaussian smoothing (Table~\ref{table:mink_scale}) and a significant improvement is obtained when combining $W_{\rm M}$ with $W_{\rm D1}$ resulting in an overall reduction of the error on $f_{\rm NL}$ of 30\% compared to the best results obtained with Gaussian smoothing. On the other hand, $W_{\rm D2}$ does not bring anything interesting, but this is somewhat expectable: for a stationary and isotropic random field, there is no correlation between the field and its first derivatives, but there is a strong correlation between the field and its second derivatives.

\begin{table*}
 \centering
 \begin{minipage}{140mm}
 \begin{tabular}{ @{}  l  c  c  c   c   c   c @{}  }
\hline
  \hspace*{0.2cm}Functional  & $W_{\rm M}$ \hspace*{0.2cm} & \hspace*{0.2cm}$W_{\rm D1}$\hspace*{0.2cm}  & \hspace*{0.2cm}$W_{\rm D2}$ \hspace*{0.2cm}& \hspace*{0.2cm}$W_{\rm M}+W_{\rm D1}$ \hspace*{0.2cm} & \hspace*{0.2cm}$W_{\rm M}+W_{\rm D2}$\hspace*{0.2cm} & \hspace*{0.2cm}$W_{\rm M}+W_{\rm D1}+W_{\rm D2}$   \\
\hline
\hspace*{0.2cm}$V_{0}$         &  51 &   & & & & \\
\hspace*{0.2cm}$V_{1}$         &  14 &  & & & & \\
\hspace*{0.2cm}$V_{2}$         &  21 &  & & & & \\
\hspace*{0.2cm}$V_{3}$         &  20 &  & & & & \\
\hspace*{0.2cm}$V_{1}+V_{2}$    & 13 &   &  & & &\\
\hline
\hspace*{0.2cm}$V_{0}+V_{1}+V_{2}+V_{3}$  \hspace*{0.2cm}  & 12.5 & 32  & 27 & 9 & 12.5 & 9\\
\hline
\end{tabular}
  \caption{Sensitivity $\langle\widehat{\Delta f}_{_{\rm NL}}^2\rangle^{1/2}$ when Wiener filters are applied to the data maps prior to MFs measurements. It is calculated in the framework of the extended mission for the 3 channels.}
  \label{table:wiener}
 \end{minipage}
 \end{table*}

\section{Inhomogeneous Noise}
\label{sec:anisotropicnoise}

\begin{figure}
\begin{center}
 \includegraphics[angle=90,width=8.5cm]{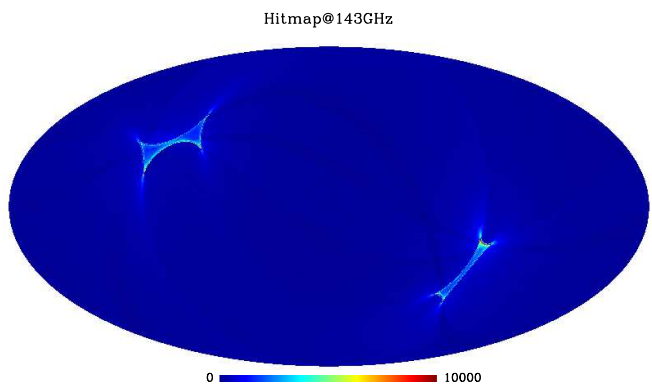}
  \caption{A typical hit map in Galactic coordinates, obtained for $N_{\rm side}=2048$. Pixel values quantify the number of observations of the pixel. Areas near the ecliptic poles are observed several times more frequently than regions of the sky near the ecliptic plane. The lowest values are about 25 while the highest are about $20\,000$.}
  \label{fig:hitmap}
\end{center}
\end{figure}

\begin{figure}
\begin{center}
 \includegraphics[angle=90,width=8.5cm]{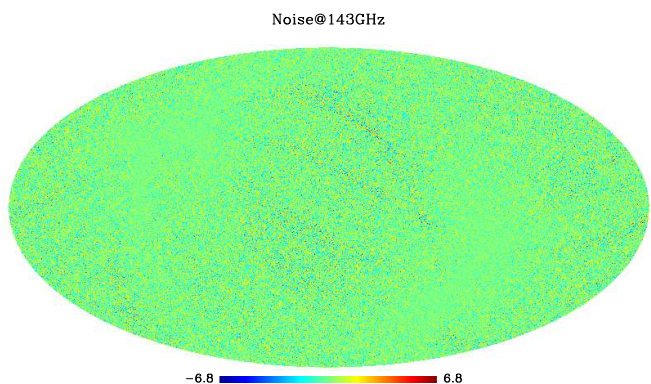}
  \caption{A typical map of inhomogeneous noise, using equation (\ref{eq:noisemeth}) on the hitmap of Fig.~\ref{fig:hitmap}. The map is shown in units of its standard deviation.}
  \label{fig:anis_noise_map}
\end{center}
\end{figure}

Due to the scanning strategy and the orientations of the different horns and bolometers \citep{2005A&A...430..363D}, the distribution of the noise in the raw sky maps is correlated, anisotropic and inhomogeneous. In what follows, we treat effects of inhomogeneity and anisotropy of the noise, by relying on {\em hitmaps} \citep[generated with the software {\tt MADAM} of][]{2005MNRAS.360..390K}, in which each pixel value represents the number of times the pixel has been observed by the satellite. Our modelling of the noise in each pixel of the map $i$ then reads
\begin{equation} 
\mathrm{noise}\,(i)=\sigma_{\rm isotropic\: noise} \times {\cal N}(0,1)  \times \sqrt{\dfrac{\langle{\rm hitmap}\rangle}{\mathrm{hitmap}\, (i)}}. 
\label{eq:noisemeth}
\end{equation}
Note thus that, since correlations are neglected,  anisotropy of the noise is modelled only partly, through the anisotropy of the hit map.   Even more realistic analyses would take into account of the effects of correlations in the noise, which will be addressed in future work.  Figures \ref{fig:hitmap} and \ref{fig:anis_noise_map} show a hit map and the corresponding noise map for the 143\,GHz channel in a simulation of Planck signal \citep[which used the characteristics of the instrument described in][]{2011A&A...536A...6P}. 

To analyse the impact of various levels of realism in modelling of the noise on the determination of $f_{_{\rm NL}}$, we consider two configurations: one where inhomogeneous noise is included in all the steps of the $\chi^2$ analysis, and one where only each of the $m_{\rm test}=200$ test maps has its own realisation of inhomogeneous noise -- relying on the same hit map\footnote{The one obtained in the 143\,GHz channel, to be specific.} but with different random seeds. We test here the null hypothesis $f_{_{\rm NL}}=0$ but we checked that the conclusions do not change significantly for other values of $f_{_{\rm NL}}$. All the simulations are performed as detailed in Appendix~\ref{planck} for full sky surveys with no foregrounds, $N_{\rm side}=2048$, and a Gaussian smoothing with $\theta_{_{\mathrm{FWHM}}}^{\rm S}=5'$ to filter out the noise, which corresponds to a rather (almost the most) pessimistic case in terms of inhomogeneous noise.  

\begin{figure*}
 \begin{center}
 \begin{minipage}{150mm}
\begin{tabular}{c c}
\includegraphics[bb=0cm 1.6cm 25cm 17.5cm,clip,scale=0.4]{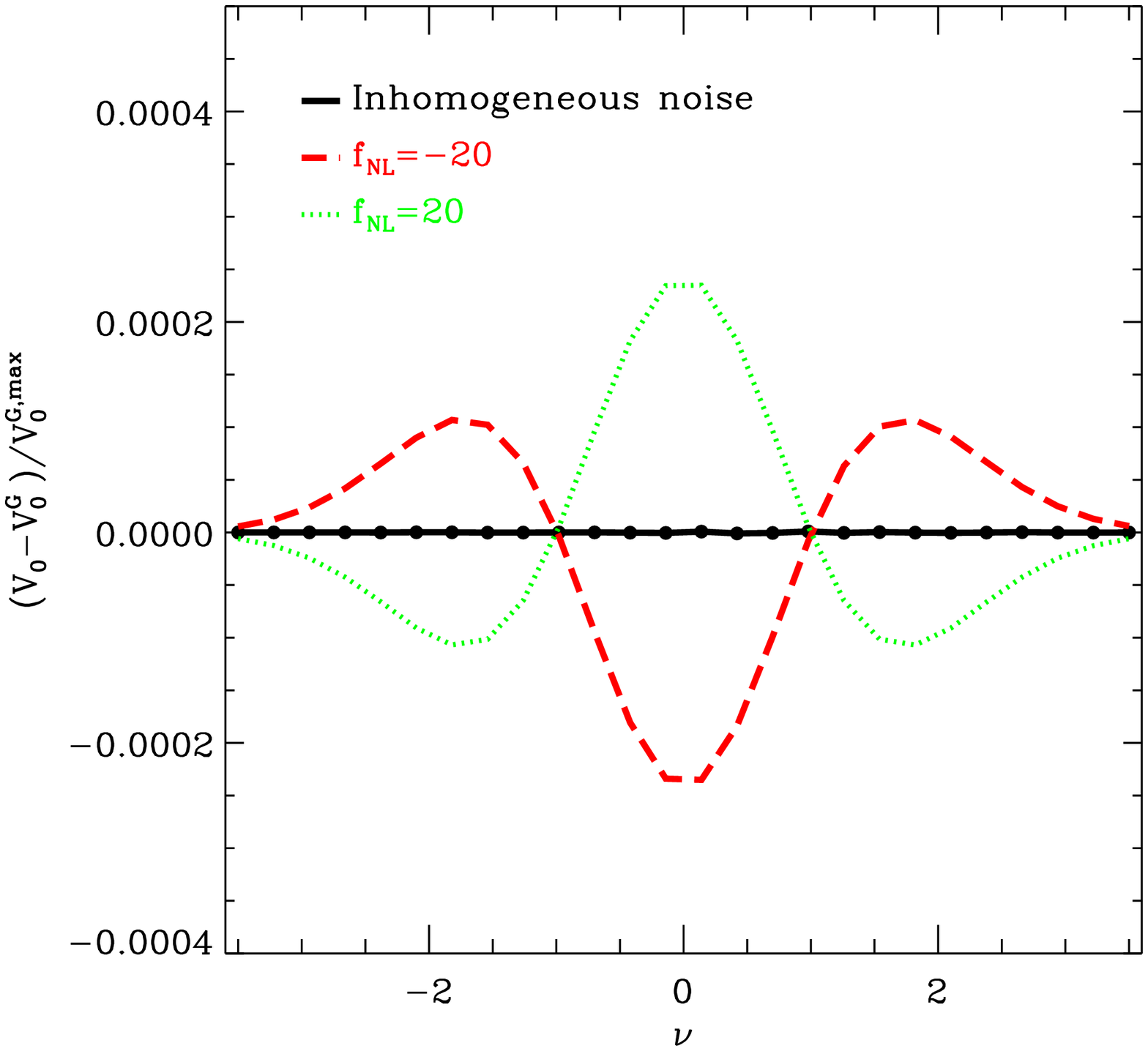} &
\includegraphics[bb=0cm 1.6cm 25cm 17.5cm,clip,scale=0.4]{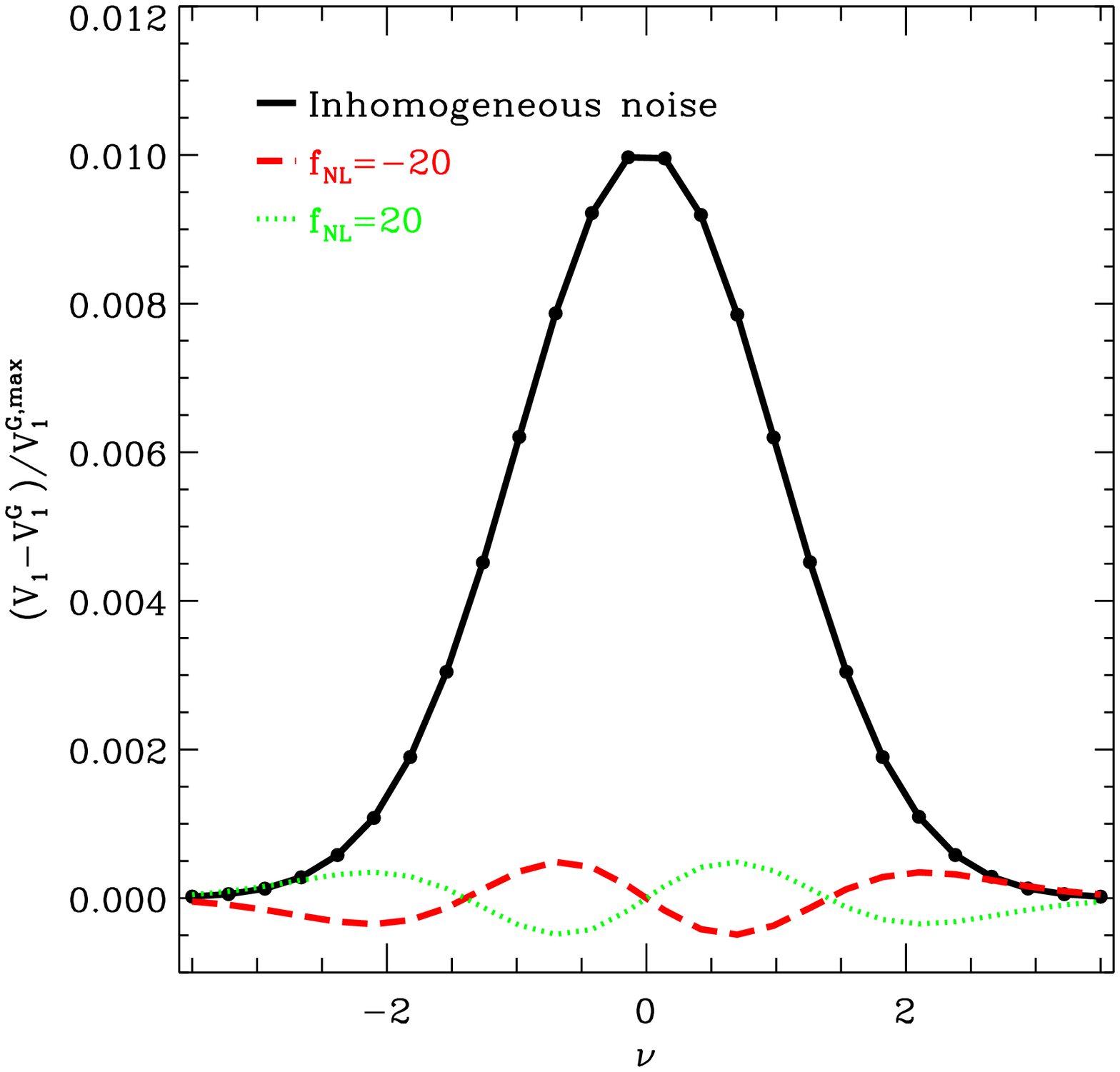}  \\
\includegraphics[bb=0cm 0cm 25cm 17.5cm,clip,scale=0.4]{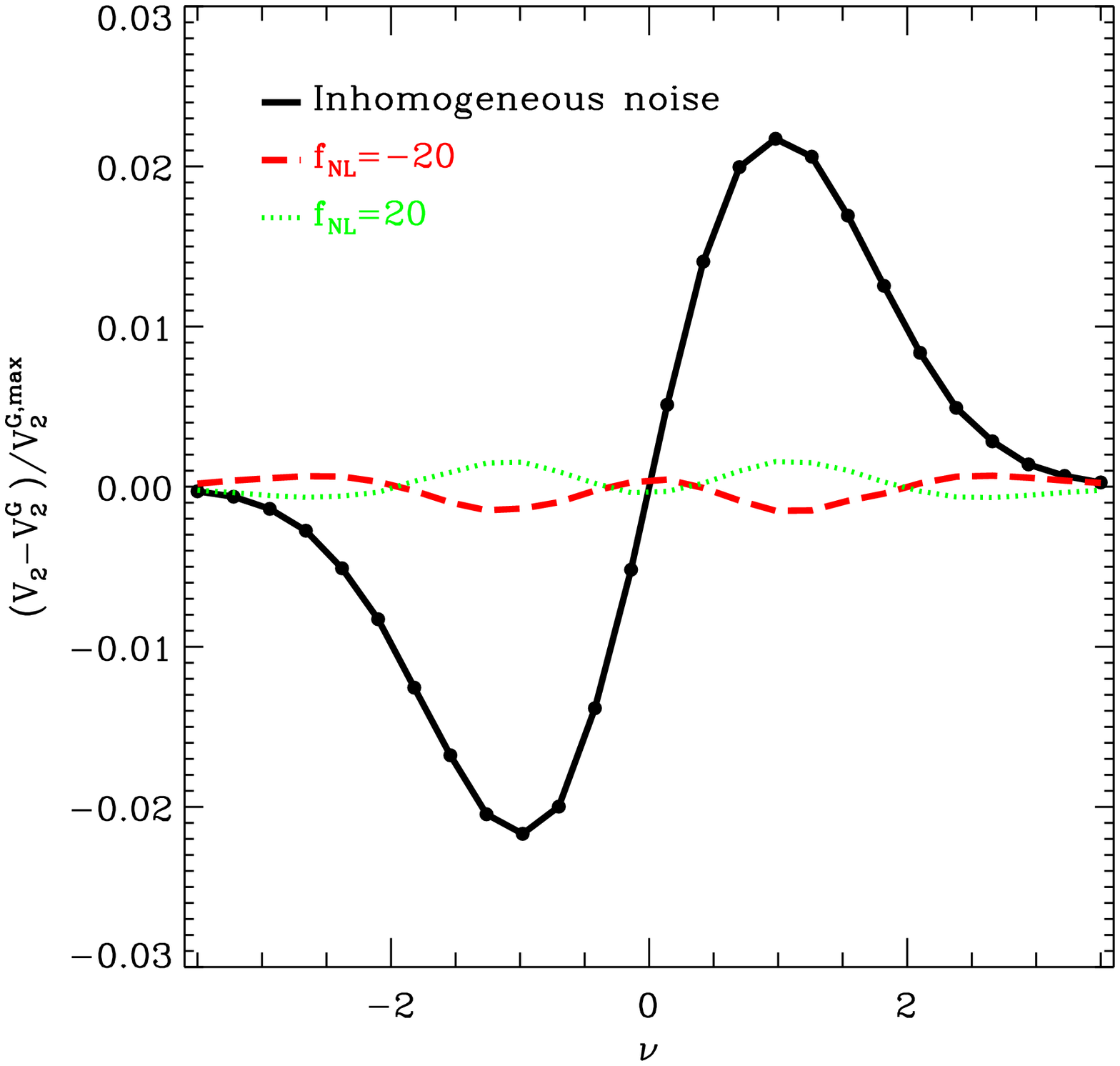}  &
\includegraphics[bb=0cm 0cm 25cm 17.5cm,clip,scale=0.4]{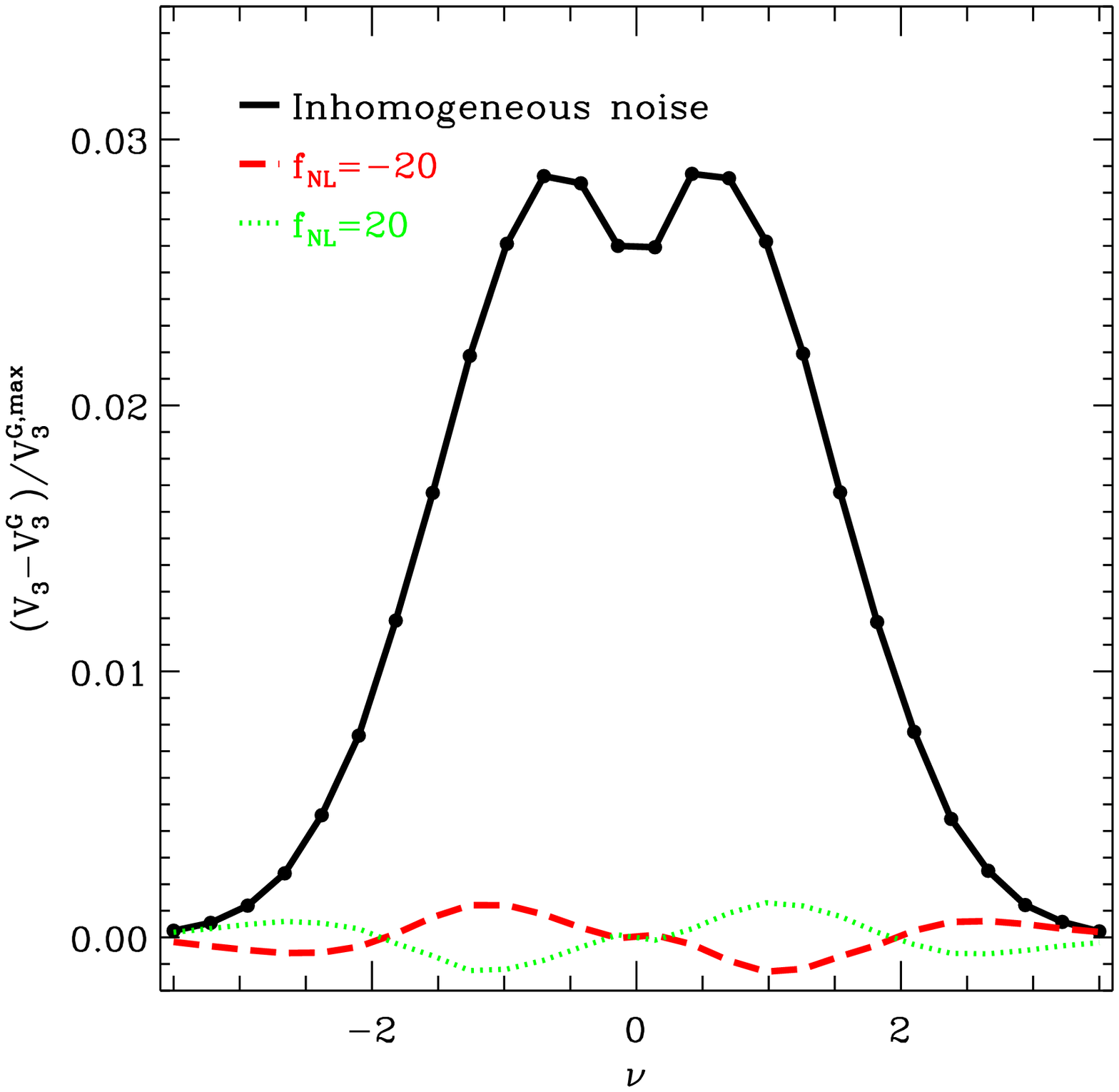}  \\
\end{tabular}
\caption{Effect of inhomogeneous noise on the ``raw'' functionals $V_k$ for the nominal mission in the 143\,GHz channel and a Gaussian smoothing with $\theta_{_{\mathrm{FWHM}}}^{S}=5'$. The relative difference between the $V_{k}$ measured in different types of maps and the Gaussian limit $V_{k}^{\rm G}$ is displayed as a function of $\nu$. Each panel corresponds to an individual functional. The curves represented on each panel are calculated by the average over 200 realisations, with $f_{_{\rm NL}}=0$ and inhomogeneous noise for the black thick curve and  with primordial non Gaussianity ($f_{_{\rm NL}} \neq 0$) and homogeneous noise for the green and red curves, while $V_k^{\rm G}$ was computed alike with homogeneous noise.}
\label{fig:courbes_hitmap_nonorm}
\end{minipage}
\end{center}
\end{figure*}

\begin{figure*}
 \begin{center}
 \begin{minipage}{150mm}
\begin{tabular}{c c}
\includegraphics[bb=0cm 1.6cm 25cm 17.5cm,clip,scale=0.4]{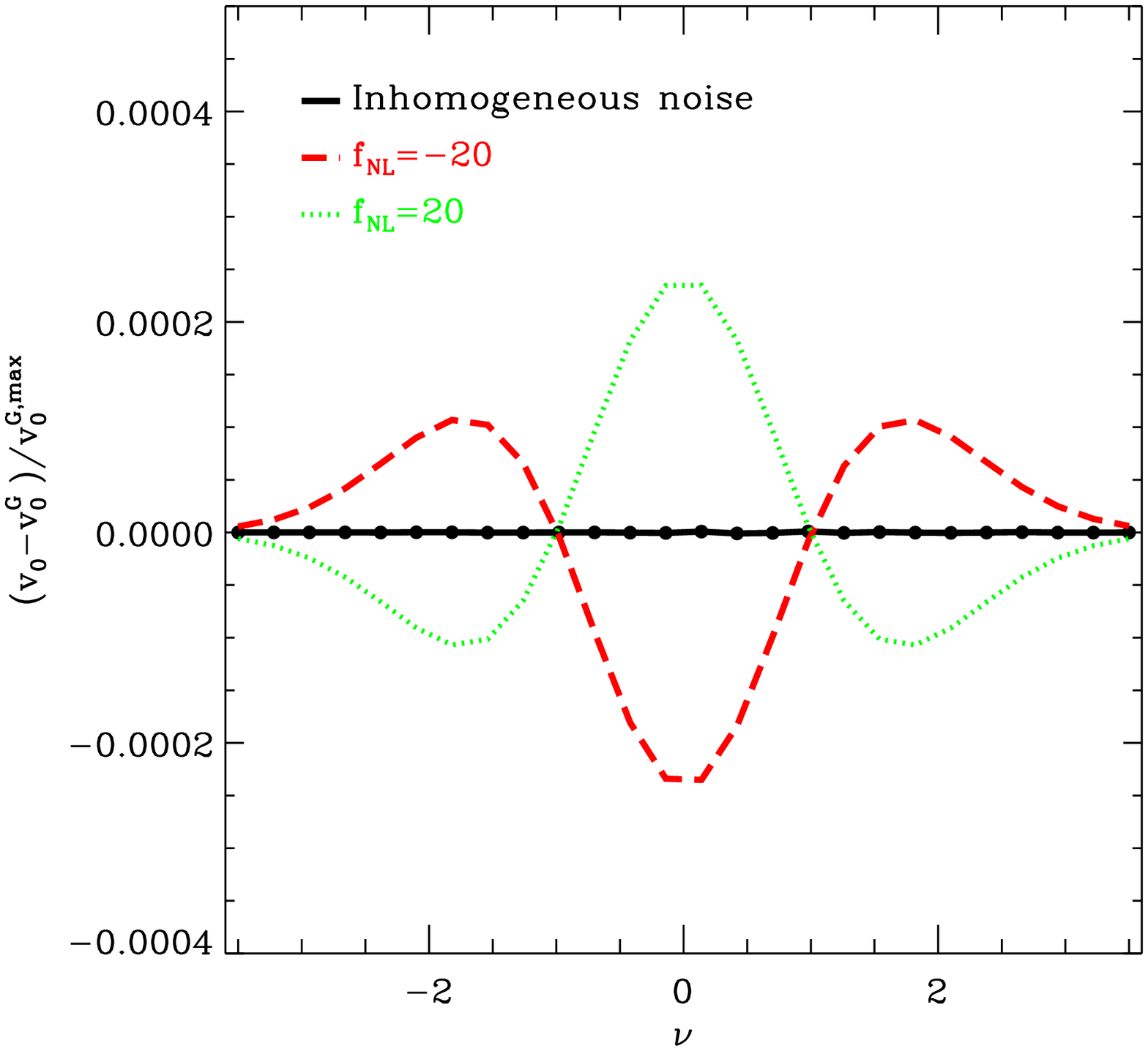} &
\includegraphics[bb=0cm 1.6cm 25cm 17.5cm,clip,scale=0.4]{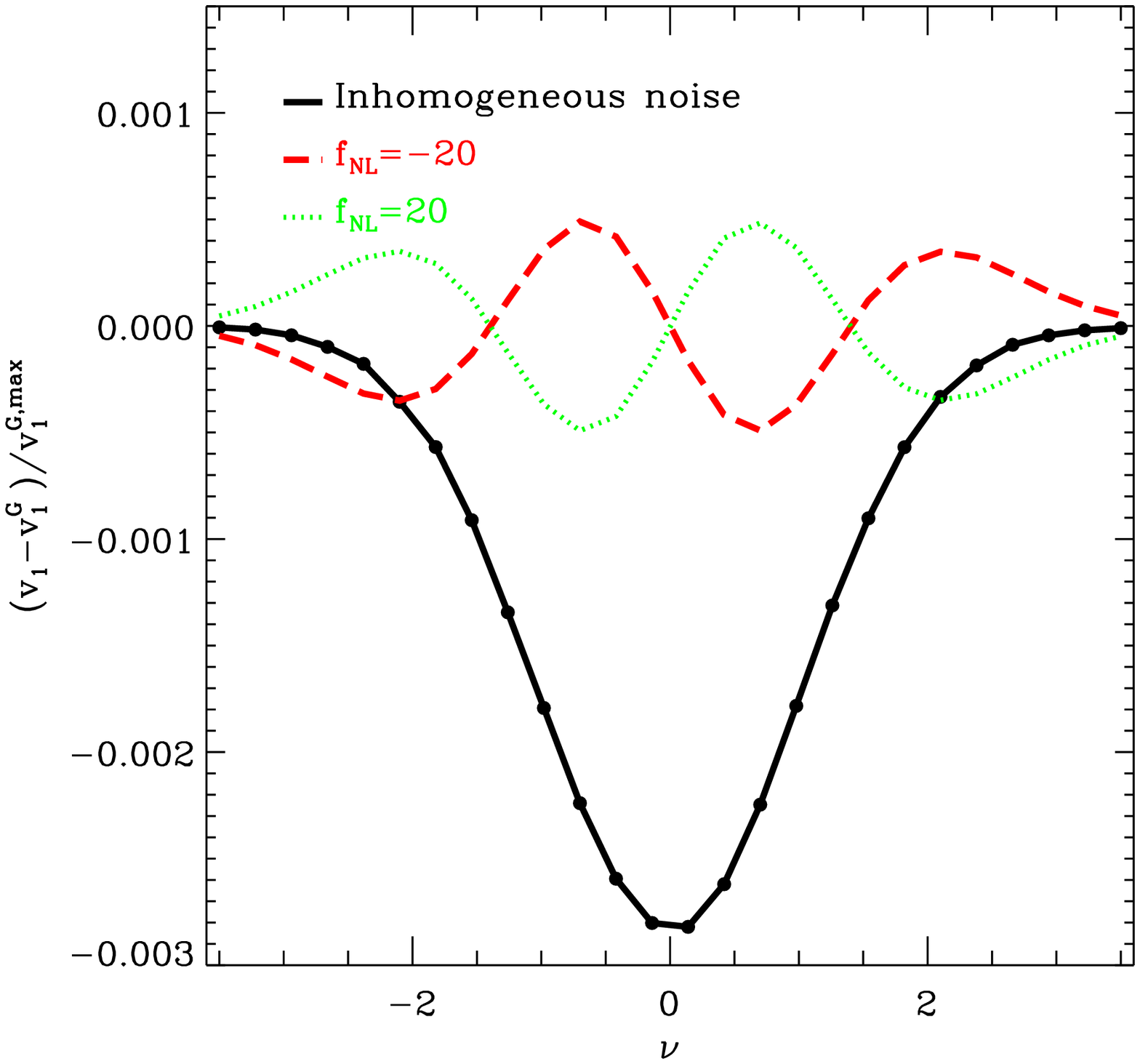}  \\
\includegraphics[bb=0cm 0cm 25cm 17.5cm,clip,scale=0.4]{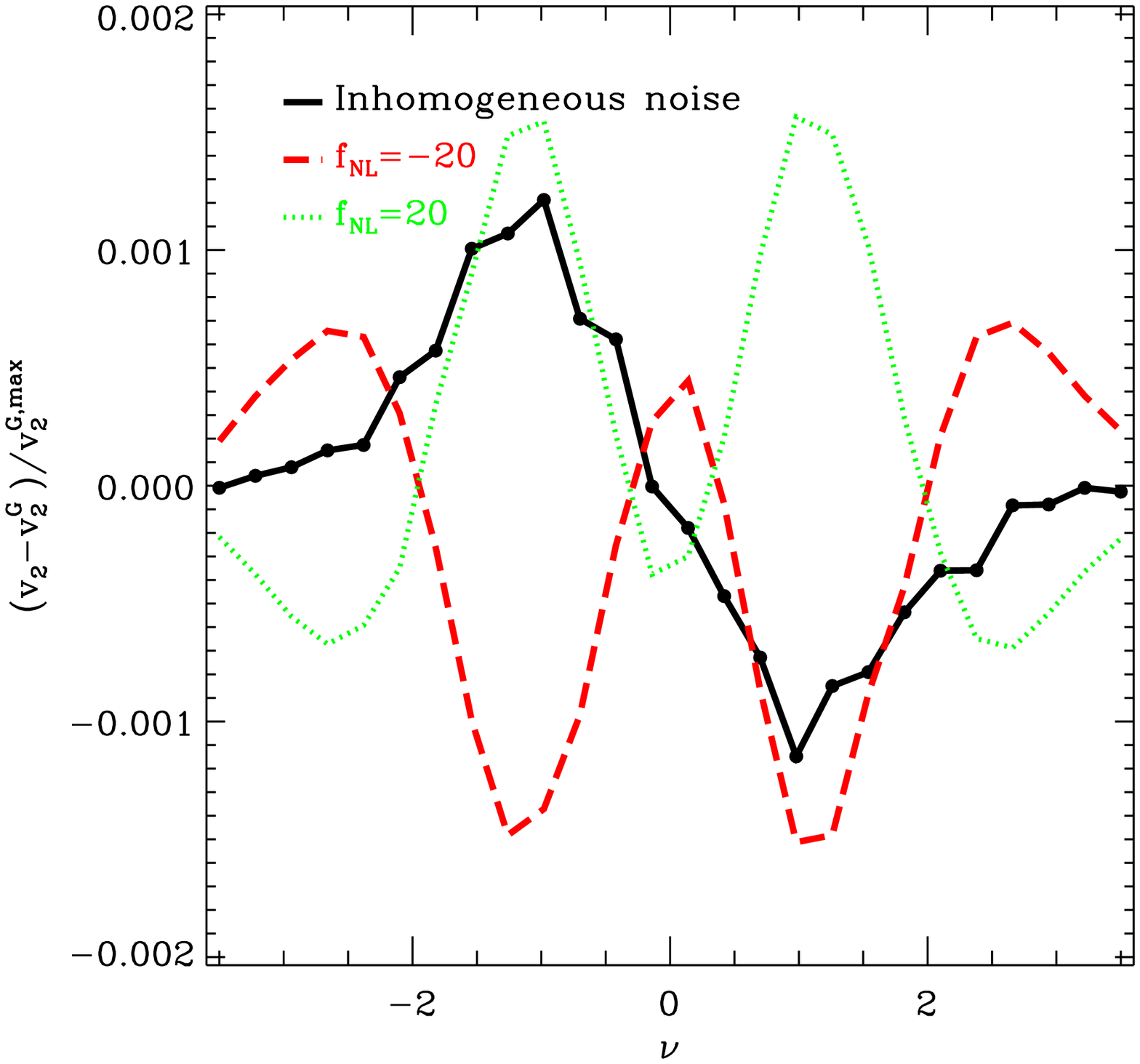}  &
\includegraphics[bb=0cm 0cm 25cm 17.5cm,clip,scale=0.4]{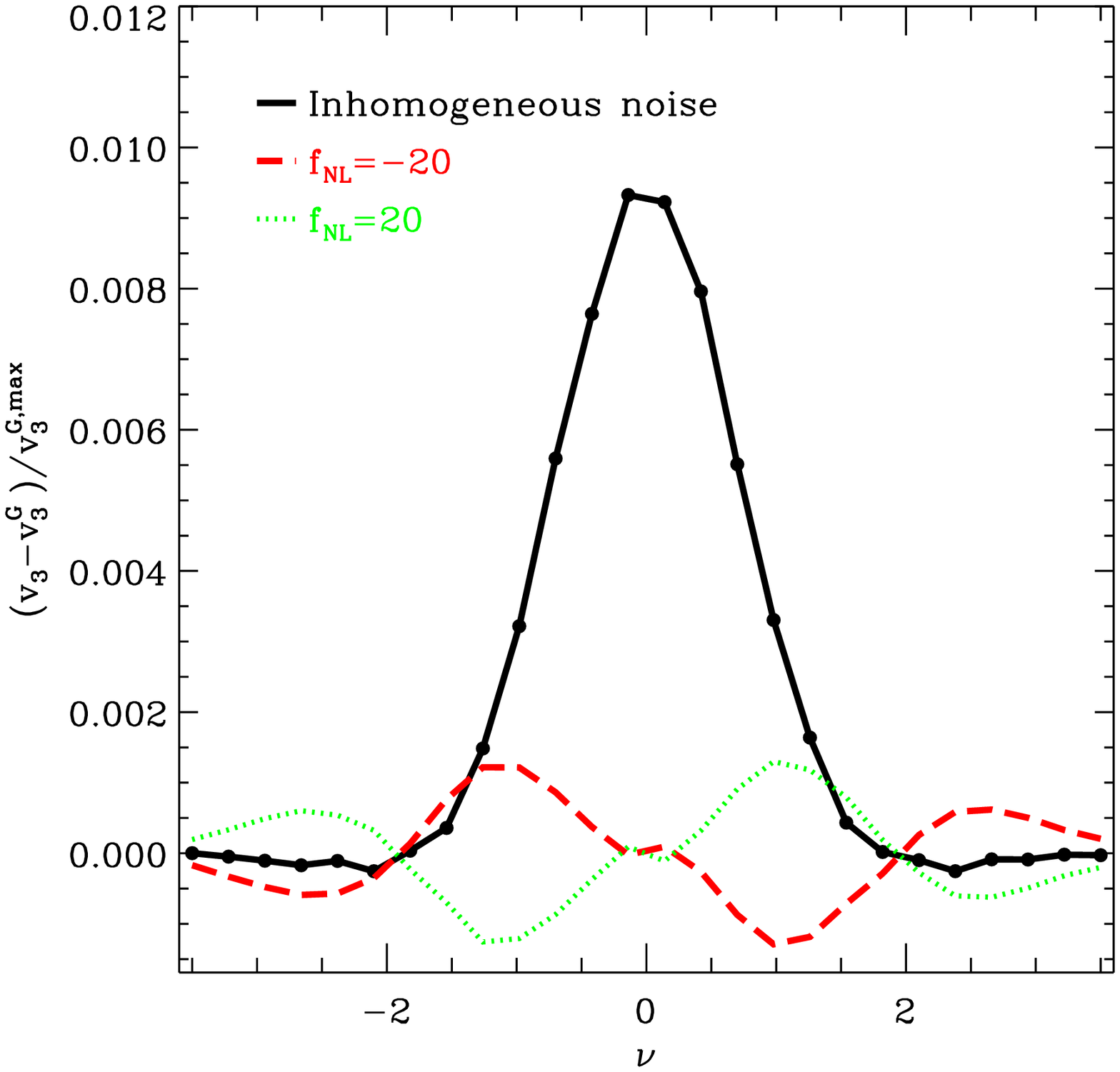}  \\
\end{tabular}
\caption{Same as in Fig.~\ref{fig:courbes_hitmap_nonorm}, but for the normalised functionals $v_k$, showing that these latter are much less sensitive to the effects of inhomogeneous noise than the ``raw'' functionals, $V_k$.}
\label{fig:courbes_hitmap_norm}
\end{minipage}
\end{center}
\end{figure*}

The results of our analyses are summarised in table~\ref{table:mink_anis1} and table~\ref{table:mink_anis2} for four levels of noise which are likely to bracket the actual sensitivity of Planck. To understand the results displayed in the tables,  Figures~\ref{fig:courbes_hitmap_nonorm} and \ref{fig:courbes_hitmap_norm} compare the effect of neglecting inhomogeneous noise to the presence of a ``true'' $f_{_{\rm NL}}$ on the functionals for the nominal mission in the 143\,GHz channel. 

The Area functional, $V_0$, seems fairly insensitive to the effect of the inhomogeneity of the noise, which in turns makes the determination of $f_{_{\rm NL}}$ from the Area quite robust to that respect. This is not very surprising: the area is proportional to the cumulated one point distribution function (pdf). The presence of inhomogeneous noise locally induces a convolution of this pdf with a Gaussian of varying width depending on the value of the number of hits in the map. The effect of this convolution is negligible when the r.m.s. $\sigma_{\rm noise}$  of the local noise  is small compared to the r.m.s. $\sigma_{0}$ of the field under consideration.\footnote{Note that this argument would be valid as well for a non Gaussian noise.} Here this is the case: the  additional Gaussian smoothing with $\theta_{_{\mathrm{FWHM}}}^{\rm S}=5'$ reduces the typical local rms of the smoothed noise map to $\sigma_{\rm noise} \la 4 \times 10^{-6}$ whatever the channel considered, to be compared to $\sigma_{0}\sim 4\times 10^{-5}$. 

The examination of Fig.~\ref{fig:courbes_hitmap_nonorm} shows that other functionals are rather sensitive to the effect of inhomogeneous noise, which is also expected. Indeed, we can guess that the presence of inhomogeneous noise augments the contrasts between cold and hot spots compared to the homogeneous case,  hence building up a signal in $V_{1}$, $V_{2}$ and $V_{3}$. However, this signal is also contained in the parameter $A_k$ in eq.~(\ref{eq:normalizedvk}) which tends to compensate for the effect on $V_k$. Hence, the normalised functionals, $v_k$, appear to be less affected by inhomogeneous noise than the ``raw'' functionals, $V_k$, as illustrated by Fig.~\ref{fig:courbes_hitmap_norm}. There is still some rather significant residual signal, at least for the small smoothing scale considered here. However, the parity of the black curves in the different panels of Fig.~\ref{fig:courbes_hitmap_norm} is opposite to that of the curves corresponding to a true primordial $f_{_{\rm NL}}$ (green and red curves): we do not expect in that case the presence of inhomogeneous noise to introduce any bias on the measurement of $f_{\rm NL}$. These simple statements are confirmed by the examination of Table~\ref{table:mink_anis1}. On the other hand, the presence of inhomogeneous noise makes the uncertainty on $f_{_{\rm NL}}$ slightly larger, increasingly with the average level of noise, as shown in Table~\ref{table:mink_anis2}. Fortunately, for the three combined channels in extended Planck mission, the effects of the inhomogeneity of the noise become nearly negligible and can be fairly ignored, as we shall do from now on.

\setlength{\tabcolsep}{0.08cm}

\begin{table*}
 \begin{center}
\begin{minipage}{156mm} 
 
 \begin{tabular}{  l  c  c  c  c  c  c  c  c }
\hline
\hspace*{0.1cm}Noise ($\mu$K.deg)  & \multicolumn{2}{c}{1.1}  & \multicolumn{2}{c}{0.7}  & \multicolumn{2}{c}{0.5}     &\multicolumn{2}{c}{0.33} \\
\hline
\hspace*{0.1cm}Functional      &\hspace*{0.3cm} $ \langle \hat{f}_{_{\rm NL}} \rangle$ & \hspace*{0.3cm}$\langle \widehat{\Delta f}_{_{\rm NL}}^2 \rangle^{1/2}$ &\hspace*{0.3cm} $\langle \hat{f}_{_{\rm NL}} \rangle$ & \hspace*{0.3cm}$\langle \widehat{\Delta f}_{_{\rm NL}}^2 \rangle^{1/2}$ & \hspace*{0.3cm}$ \langle \hat{f}_{_{\rm NL}} \rangle$   &  \hspace*{0.3cm}$ \langle {\widehat{\Delta f}}_{_{\rm NL}}^2 \rangle^{1/2}$ & \hspace*{0.3cm}$ \langle \hat{f}_{_{\rm NL}} \rangle$ & \hspace*{0.3cm}$ \langle {\widehat{\Delta f}}_{_{\rm NL}}^2 \rangle^{1/2}$\\
\hline
\hspace*{0.1cm}$V_{0}$         & &  & 1 & 38.6  &    &   &  & \\
\hspace*{0.1cm}$V_{1}$         &  &  & -4 & 23.6  &    &   &  & \\
\hspace*{0.1cm}$V_{2}$         &  &  & -0.4 & 22  &    &   &  & \\
\hspace*{0.1cm}$V_{3}$         &  &  & -0.2 & 24.7  &    &   &  & \\
\hspace*{0.1cm}$V_{0}+V_{1}+V_{2}+V_{3}$ \hspace*{0.4cm}  & -4.4 & 18.8 & -3.8 & 17.9  & -2.3  & 17  & -0.8 & 16.7 \\
\hline
\end{tabular}
  \caption{Effect of neglecting the presence of inhomogeneous noise when estimating ${f}_{_{\rm NL}}$ and the error bar on it in the null hypothesis, $f_{_{\rm NL}}=0$. Each of the first two columns corresponds to a given level of noise expected in some specific Planck channel, respectively the 217\,GHz and the 143\,GHz channels, while the third and the fourth ones correspond to the combination of the  100, 143 and 217\,GHz channels  in the nominal and the extended mission case, respectively (see  Appendix \ref{planck}). The numbers in this table assume a Gaussian smoothing of the data maps with $\theta_{_{\mathrm{FWHM}}}^{\rm S}=5'$.}
  \label{table:mink_anis1}
  \end{minipage}
 \end{center}
 \end{table*}

\setlength{\tabcolsep}{0.1cm}

\begin{table*}
 \begin{center}
 \begin{minipage}{174mm}
 \begin{tabular}{@{}  l  c  c  c  c  c  c  c  c @{}}
\hline
\hspace*{0.2cm}Noise ($\mu$K.deg) & \multicolumn{2}{c}{\hspace*{1cm} 1.1 \hspace*{1.5cm} }  & \multicolumn{2}{c}{\hspace*{1cm} 0.7 \hspace*{1.5cm} }  & \multicolumn{2}{c}{\hspace*{1cm} 0.5 \hspace*{1.5cm} }     &\multicolumn{2}{c}{ 0.33  } \\
\hline
 \hspace*{0.2cm}Functionals: $V_{0}+V_{1}+V_{2}+V_{3}$ \hspace*{0.3cm}& $ \langle \hat{f}_{_{\rm NL}} \rangle$ & \hspace*{0.2cm}$\langle \widehat{\Delta f}_{_{\rm NL}}^2 \rangle^{1/2}$ &\hspace*{0.1cm} $\langle \hat{f}_{_{\rm NL}} \rangle$ & \hspace*{0.2cm}$\langle \widehat{\Delta f}_{_{\rm NL}}^2 \rangle^{1/2}$ & \hspace*{0.1cm}$ \langle \hat{f}_{_{\rm NL}} \rangle$   & \hspace*{0.2cm}$ \langle {\widehat{\Delta f}}_{_{\rm NL}}^2 \rangle^{1/2}$  & \hspace*{0.1cm}$ \langle \hat{f}_{_{\rm NL}} \rangle$ & \hspace*{0.2cm}$ \langle {\widehat{\Delta f}}_{_{\rm NL}}^2 \rangle^{1/2}$\\
\hline
\hspace*{0.2cm} Configuration 1\footnote{For the reference maps and the test maps, we introduce isotropic noise.} & 0.19 & 17.8  & -0.05 & 16.5  &  $0.044$ & 16.2 &  $0.07$ & $16$ \\
\hspace*{0.2cm} Configuration 2\footnote{For the reference maps, we introduce isotropic noise and for test maps we introduce inhomogeneous noise.}   & -4.4 & 18.8 & -3.8 & 17.9  & -2.3  & 17  & -0.8 & 16.7 \\
\hspace*{0.2cm} Configuration 3\footnote{For the reference maps and the test maps, we introduce inhomogeneous noise.}   & 0.002 & 25.5 & -0.44 & 23 & -0.15  & 19.9 & -0.15 & 17 \\
\hline
\end{tabular}
  \caption{Effect of inhomogeneous noise when estimating ${\hat f}_{_{\rm NL}}$ and the error bar $\Delta f_{_{\rm NL}}$.  Three settings are considered, as detailed below. The table is otherwise similar to Table~\ref{table:mink_anis1}.}
  \label{table:mink_anis2}
 \end{minipage}
\end{center} 
  \end{table*}

\renewcommand{\arraystretch}{-0.5}
\setlength{\tabcolsep}{-1cm}

\renewcommand{\arraystretch}{-0.5}
\setlength{\tabcolsep}{-1cm}

\setlength{\tabcolsep}{0.1cm}
\renewcommand{\arraystretch}{1.5}

\section[Foregrounds I]{Foregrounds I: Point sources}

``Point sources'' refer to the (large) number of radio and infra-red galaxies that are detectable at the CMB frequencies. These galaxies are in general not all resolved by CMB all-sky experiments like WMAP or Planck, even if the brightest objects are detected individually. The faint ones contribute to an inhomogeneous sky background. In this section, we test for the first time the effect of these sources on the estimate of $f_{_{\rm NL}}$ with Minkowski Functionals. In previous studies relying on MFs analyses \citep[e.g.,] []{2008MNRAS.389.1439H, 2010MNRAS.408.1658N}, the effect of point sources was indeed supposed to be completely subtracted off by an appropriate masking or simply negligible compared to error bars. 

This section is divided into two parts: \S~\ref{sec:PSMmethod} describes in details the simulation pipeline we used to perform our analyses while \S~\ref{sec:PSMresults} discusses the results of our investigations. 

\subsection{Method}
\label{sec:PSMmethod}

To perform our simulations of the data, we need to compute accurately the contribution of point sources. To do so, we use the {\it Planck Sky Model} (PSM) code described in \citet{PSM}\footnote{{\tt http://www.apc.univ-paris7.fr/}$\sim${\tt delabrou/PSM/psm.html}}. As reviewed in \S~\ref{sec:PSM}, the point source contribution depends significantly on frequency. As a result, our simulations and analyses will consider separately the three cosmological channels, 100, 143 and 217\,GHz, in the extended mission configuration for the noise level (see Appendix \ref{planck}).  In particular each channel will have different masking treatment for the brightest point-sources. Since masks represent a crucial part of the treatment of point sources, we discuss about them in \S~\ref{sec:PSMmasks}. Other technical details about our simulations are provided in \S~\ref{sec:PSMprocess}.

\subsubsection{Point sources simulations}
\label{sec:PSM}

The {\it Planck Sky Model} (PSM) code \citep{PSM} is specifically designed to simulate all relevant sky emissions at Planck frequencies, including secondaries and foreground emission, as they were known before the launch of Planck. In this paper we used only the part of the PSM that deals with point sources, the rest of the simulation pipeline being detailed in \S~\ref{sec:PSMprocess}. 

Firstly, we use the PSM code to add radio sources, namely Active Galactic Nuclei (AGN), to our simulations. These AGNs are observed via their synchrotron emission.  The PSM relies on numerous surveys of radio sources at frequencies ranging from 0.85\,GHz to 4.85\,GHz to model this emission. In regions not observed by surveys or with shallower observations, sources are copied from other regions, until a coverage down to at least 20 {\it m}Jy at 5\,GHz is achieved over the full sky. Then flux densities are extrapolated at all frequencies by using a power law approximation for the spectra, of the form $S_{\nu}\propto \nu^{-\alpha}$. For the spectral index $\alpha$ estimates, sources are classified into steep or flat spectrum class and $\alpha$ is drawn from a Gaussian distribution with mean and variance corresponding to its class \citep{2006A&A...445..465R} and matching WMAP data \citep{2003ApJS..148...97B}, in several frequency ranges. Besides, WMAP sources are accounted for in the simulations. Source counts at 5 and 20\,GHz are found to be consistent with the model of \citet{1998MNRAS.297..117T} and an updated version of the model of \citet{2005A&A...431..893D}. These sources are known to contribute essentially at low frequencies, from 30 to 90\,GHz but they have been detected at higher frequencies up to 217\,GHz \citep{2011A&A...536A...1P, 2011A&A...536A...7P}. 

We note that radio sources are nearly Poisson distributed on the sky, so they essentially contribute to the CMB as a {\it shot noise}.

Secondly, we add infra-red (IR) sources to the simulations. Indeed, at high frequencies (above 150\,GHz), a thermal emission arising from dust heated by the UV emission of young stars also contributes. In addition to normal stars surrounded by a disk, numerous starburst galaxies which form stars at extreme rates contribute to this thermal emission.  In the PSM, sources are taken from the IRAS Point Source Catalogue (PSC) \citep{1988iras....1.....B} and the Faint Source Catalogue (FSC) \citep{1992ifss.book.....M}. The flux densities are extrapolated to Planck frequencies by adopting a model with modified black body spectra and the gaps in sky coverage are filled up using the same procedure as for the radio sources. 

Finally, we add to the simulations the Cosmic Infra-red Background (CIB) which is possibly the dominating component. Distant starburst galaxies are not all detected individually and the cumulated emission from the fainter ones form a diffuse background of anisotropies. The PSM adopts the count model of \citet{2006ApJ...650...42L}, which is consistent with SCUBA and MAMBO surveys. Sources are clustered following model 2 in \citet{2004MNRAS.352..493N} and the spatial distribution follows the method of \citet{2005ApJ...621....1G}, then flux densities are extrapolated to all frequencies. The CIB power spectra of the simulation agree sufficiently well with those measured by Planck \citep{2011A&A...536A..18P} and ACT \citep{2011ApJ...739...52D} for our forecast studies. 

The main difference between the distribution of radio and IR sources on the sky is that IR sources, being either observed as individual entities or as a background, are clustered in their host dark-matter halos. So their power spectrum is not flat as for the radio sources.

\subsubsection{Masks}
\label{sec:PSMmasks}

Brightest point sources can be detected individually and can be masked properly when their flux density is beyond a chosen detection threshold when compared to the level $\sigma=\sigma_{\rm noise}$ of the underlying noise in the CMB data. Here, we create three sets of point source masks corresponding to three flux density cuts (referred simply as flux cuts). The choice of these different flux cuts has been mainly derived from the Early Release Compact Source Catalogue (ERCSC) published by the Planck Collaboration \citep{2011A&A...536A...7P}. The first set of masks refers to sources beyond a $10\sigma$ level in the 3 bands 100, 143 and 217\,GHz, corresponding to respective flux density thresholds at $0.5$, $0.33$, $0.28$ Jy as chosen in the ERCSC. The second set of masks concerns sources beyond the $5\sigma$ level, which corresponds to the threshold choice in the cleanest parts of the sky of the ERCSC. The third set of masks corresponds to the $3\sigma$ level, which is not mentioned in the ERCSC, but we use it because we believe it represents a more appropriate set of masks for cosmological purposes. Indeed, in the ERCSC, the goal was to set flux density thresholds to have a sufficiently good signal to noise ratio for reliable analysis of the point sources properties. Our goal here is to {\it remove} the contribution from the point sources, which requires a much less stringent criterion on the quality of their detection.
Furthermore,  the ERCSC signal to noise level does not match that of the nominal mission and by no mean that of the extended mission. 

\begin{figure}
\begin{center}
\includegraphics[angle=90,width=8.5cm]{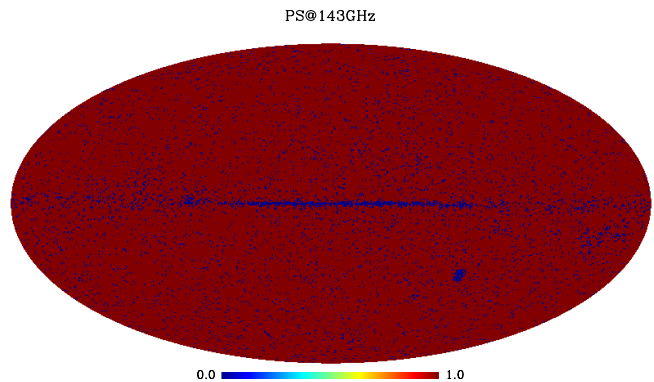}
  \caption{Masks of point sources at 3$\sigma$ (ERCSC reference), designed for the 143\,GHz map, drawn from the PSM \citep{PSM}.}
  \label{fig:map_mask_ps}
\end{center}
\end{figure}

Each mask associated to an individual point source is a disk of radius 3 times the FWHM of the beam of the instrument in the channel under consideration. When adding up the contributions of all the sources, a certain fraction $1-f_{\rm sky}$ of the sky is masked, as illustrated by Fig.~\ref{fig:map_mask_ps}. The value of the sky fraction which is then used, $f_{\rm sky}$, ranges from $f_{\rm sky}=0.90$ for the 100\,GHz channel up to $f_{\rm sky}=0.99$ for the 217\,GHz channel. These differences come from two factors on which the construction of masks depend: the beam width and the number of point sources detected beyond the threshold of interest. These two parameters decrease when passing from 100 to 217\,GHz.

 \subsubsection{Simulation pipeline}
\label{sec:PSMprocess}

To test the effects of point sources in the estimation of $f_{_{\rm NL}}$ or more specifically the approximation of neglecting their presence, we add their contribution only to the ``data'' (test) maps, ${\hat y}$ in eq.~(\ref{eq:monchi2}). We simulate $m_{\rm test}=200$ of these tests maps with $f_{_{\rm NL}}^{\rm prim}=\lbrace-10, 0, 10, 50\rbrace$, where $f_{\rm NL}^{\rm prim}$ stands for the ``primordial'' $f_{_{\rm NL}}$ (to be contrasted later with other contributions to the effective ${\hat f}_{_{\rm NL}}$ arising from biases induced by unaccounted point sources).  The $m=10\,000$ Gaussian maps used to compute the covariance matrix $C$ as well as the $m_{\rm NG}=200$ non Gaussian maps used to calculate the model prediction ${\bar y}(f_{_{\rm NL}})$ in eq.~(\ref{eq:monchi2}) neglect this contribution, but have exactly the same treatment otherwise, including sky coverage and instrumental noise as detailed below. This way, our analysis will be able to confirm if appropriate masking is enough to render the effects of point sources negligible on the measurement of $f_{_{\rm NL}}$. 

The details of our simulation pipeline now follow. CMB maps are created first  with the beam corresponding to each channel frequency $f$ (see Appendix \ref{planck}), are supplemented with point sources (only for the test maps) convolved with the same beam and with the noise corresponding to each channel $f$ for the extended mission. Next, point source masks are applied to the maps. These masks depend on the channel $f$ -- so the beam width $\theta_{\rm FWHM}^{\rm b}(f)$  is a parameter-- and on the chosen flux cut $\beta$.  The punched holes are filled by diffusive in-painting.\footnote{Choosing a lexical order (defined by HEALPix), the values inside masked pixels are computed using the average over the values inside neighbouring pixels, when available, whether it is from an unmasked pixel or a pixel inside the mask that was calculated with the algorithm in a previous step. To achieve convergence, the process is reiterated a number $n_{\rm it}$ of times. We take $n_{\rm it}=30$, which is sufficient in practice for the mask size we have in our simulations.} Then, the maps are smoothed with a Gaussian window of size $\theta_{_{\rm FWHM}}^{\rm S}=5'$. Finally, a galactic mask is applied to the maps, here corresponding to a valid fraction of the sky $f_{\rm sky}=0.80$. The procedure to construct the galactic mask will be described in \S~\ref{sec:lagalaxie}. The complete pipeline is summarised as follows:
\begin{eqnarray}
{\rm map} &= & {\rm CMB}\:(f_{_{\rm NL}})*{\rm beam}[\theta_{\rm FWHM}^{\rm b}(f)] \nonumber \\
                &+ & {\rm foreground \: of \: sources}*{\rm beam}[\theta_{\rm FWHM}^{\rm b}(f)] \nonumber \\
                &+ & {\rm noise \,}(f) + {\rm\, point \: sources \: mask }\, (\beta, f) \: {\rm inpainted} \nonumber\\
                &+ & {\rm smoothing}\: (\theta_{_{\rm FWHM}}^{\rm S}=5'\,) \nonumber \\
                &+ & {\rm galactic} \: {\rm mask} \, (f_{\rm sky}).
\label{eq:psmeth}
\end{eqnarray}
We checked that the results derived in this section are qualitatively the same for other values of $\theta_{_{\rm FWHM}}^{\rm S}$ and for the Wiener filters studied in \S~\ref{sec:wienerfilters}. Of course, a quantitative calculation of the biases induced by point sources will require a new analyse each time a new filter is considered.

\subsection{Results}
\label{sec:PSMresults}

Table~\ref{table:ps0}  shows the estimate on $f_{_{\rm NL}}$ obtained for different channels as a function of source masking level and primordial non Gaussianity, $f_{_{\rm NL}}^{\rm prim}$.  Again, a frequentist average of the posterior averages is performed over 200 test maps realisations, and is noted $\langle \hat{f}_{_{\rm NL}} \rangle$. Note that while point sources can introduce a significant {\rm bias} on the estimate of $f_{_{\rm NL}}^{\rm prim}$, they do not change significantly the error bars, $\Delta f_{_{\rm NL}}$, that depend linearly on square root of sky coverage given the overall level of noise (Table~\ref{table:mask_gal_sensitivity}). Therefore, error bars on the measured $f_{_{\rm NL}}$ are not mentioned further in this section, for simplicity. We now discuss the results obtained in Table~\ref{table:ps0}, starting with the 100 and 143\,GHz channels, dominated by radio sources (\S~\ref{sec:PSMpart1}) and finishing with the 217\,GHz channel, where one has to account for the additional IR source contribution (\S~\ref{sec:PSMpart2}).

\setlength{\tabcolsep}{0.1cm}

\begin{table*}
 \begin{center}
 \begin{minipage}{160mm}
 \begin{tabular}{   @{} l| c | c |  c @{}}
\hline

 \hspace*{0.1cm}Flux cut (detection    &         $f=100$\,GHz           &         $f=143$\,GHz             &       $f=217$\,GHz    \\
 \hspace*{0.1cm}level in the ERCSC)  & $\theta_{_{\rm FWHM}}^{\rm b}=10'$, noise=$0.7\, \mu$K.deg & $\theta_{_{\rm FWHM}}^{\rm b}=7.2'$, noise=$0.5\, \mu$K.deg & $\theta_{_{\rm FWHM}}^{\rm b}=5'$, noise=$0.7\, \mu$K.deg \\
\hline
\hspace*{0.1cm}$f_{_{\rm NL}}^{\rm prim}=0$ & $\langle \hat{f}_{_{\rm NL}} \rangle$ & $\langle \hat{f}_{_{\rm NL}} \rangle$ &  $\langle \hat{f}_{_{\rm NL}} \rangle$  \\
\hline
\hspace*{0.1cm}10$\sigma$   &  $14$ &   $10$ &  $5$   \\
\hspace*{0.1cm}5$\sigma$   &  $8$ &   $4$ &   $0.4$     \\
\hspace*{0.1cm}3$\sigma$   &  $3$ &  $1.4$ &  $0.6$   \\
\hline
\hspace*{0.1cm}$f_{_{\rm NL}}^{\rm prim}=-10$ & $\langle \hat{f}_{_{\rm NL}} \rangle$ &  $\langle \hat{f}_{_{\rm NL}} \rangle$ &   $\langle \hat{f}_{_{\rm NL}} \rangle$    \\
\hline
\hspace*{0.1cm}10$\sigma$   &  $2$ &   $-0.2$ &  $-10$   \\
\hspace*{0.1cm}5$\sigma$   & $-1.5$ &   $-6$ &   $-15$    \\
\hspace*{0.1cm}3$\sigma$   &  $-6.5$ &   $-9$ &   $-16$    \\
\hline
\hspace*{0.1cm}$f_{_{\rm NL}}^{\rm prim}=10$ &  $\langle \hat{f}_{_{\rm NL}} \rangle$ &   $\langle \hat{f}_{_{\rm NL}} \rangle$ &   $\langle \hat{f}_{_{\rm NL}} \rangle$   \\
\hline
\hspace*{0.1cm}10$\sigma$   &  $25$ &   $21$ &  $21$   \\
\hspace*{0.1cm}5$\sigma$   &  $19$ &   $14$ &   $16$   \\
\hspace*{0.1cm}3$\sigma$   &  $13$ &  $12$ &   $15.5$     \\
\hline
\hspace*{0.1cm}$f_{_{\rm NL}}^{\rm prim}=50$ &  $\langle \hat{f}_{_{\rm NL}} \rangle$ &   $\langle \hat{f}_{_{\rm NL}} \rangle$ &   $\langle \hat{f}_{_{\rm NL}} \rangle$    \\
\hline
\hspace*{0.1cm}10$\sigma$   &  64 &  61 &   72   \\
\hspace*{0.1cm}5$\sigma$   &  57 &  54    &   69  \\
\hspace*{0.1cm}3$\sigma$   &  52  &  51.7  &  69      \\
\hline
\end{tabular}
  \caption{Estimates of $f_{_{\rm NL}}$ in the presence of point sources. To create the test maps, we used the procedure described in eq.~(\ref{eq:psmeth}) with a noise at the level of the extended mission in each band. The analyses are performed for combined MFs, $V_{0}+V_{1}+V_{2}+V_{3}$.}
  \label{table:ps0}
  \end{minipage}
 \end{center}
 \end{table*}

\subsubsection{100 and 143\,GHz: effect of radio sources}
\label{sec:PSMpart1}

\begin{figure*}
 \begin{center}
\vspace{0.2cm}
\begin{tabular}{|c | c |}
\hline
\includegraphics[width=8cm]{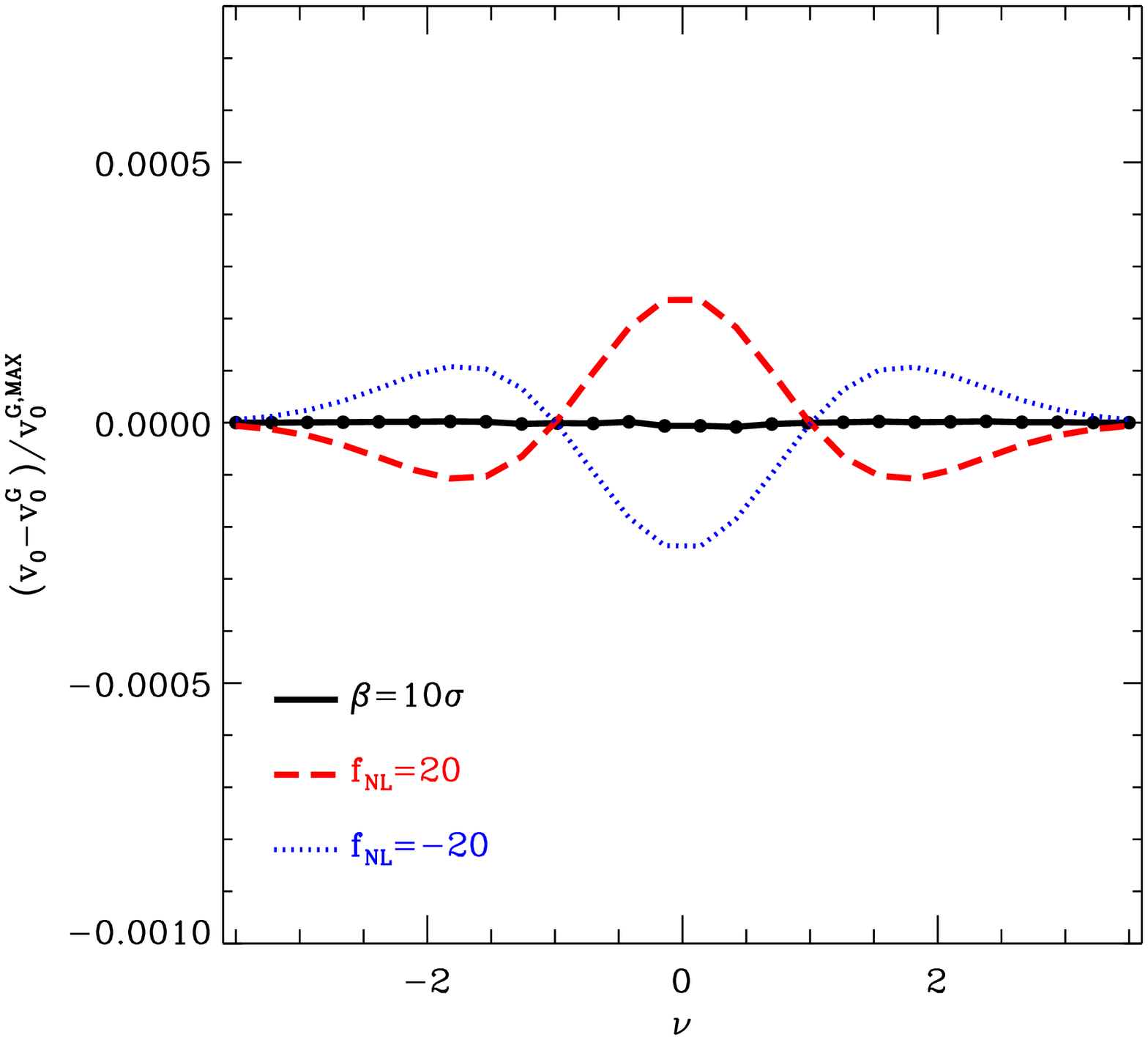} & \includegraphics[width=8cm]{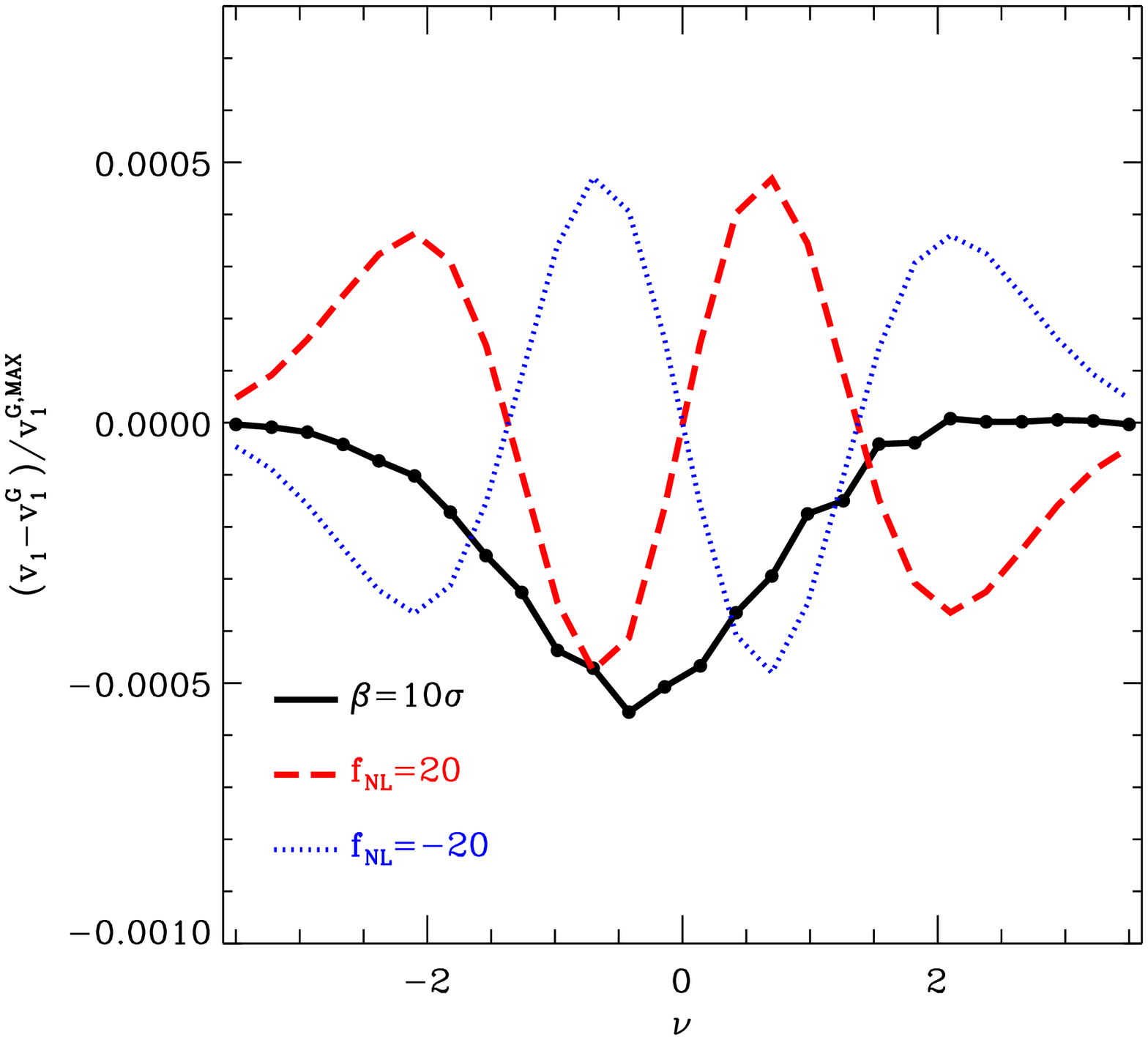}  \\
\hline
\includegraphics[width=8cm]{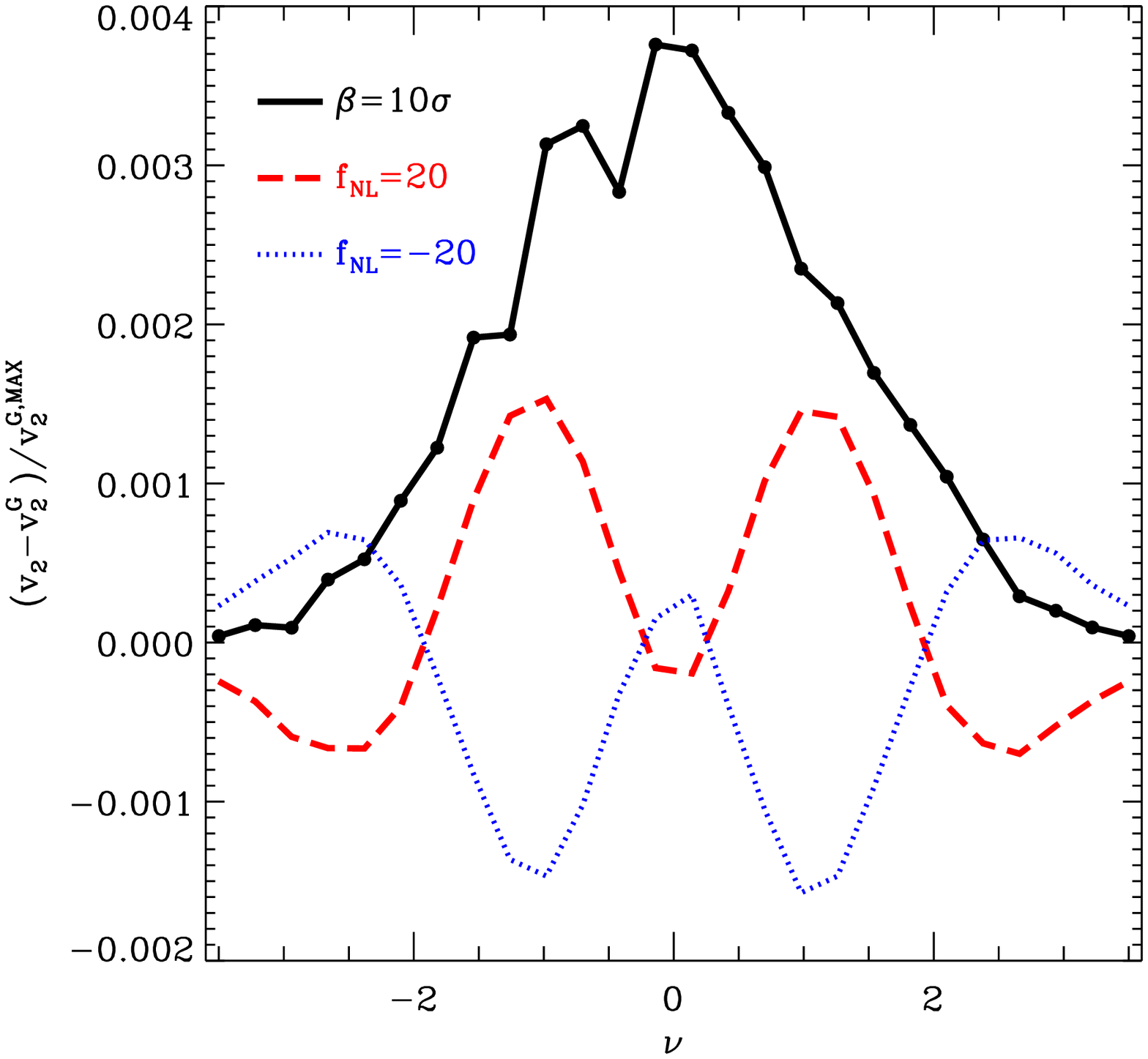} & \includegraphics[width=8cm]{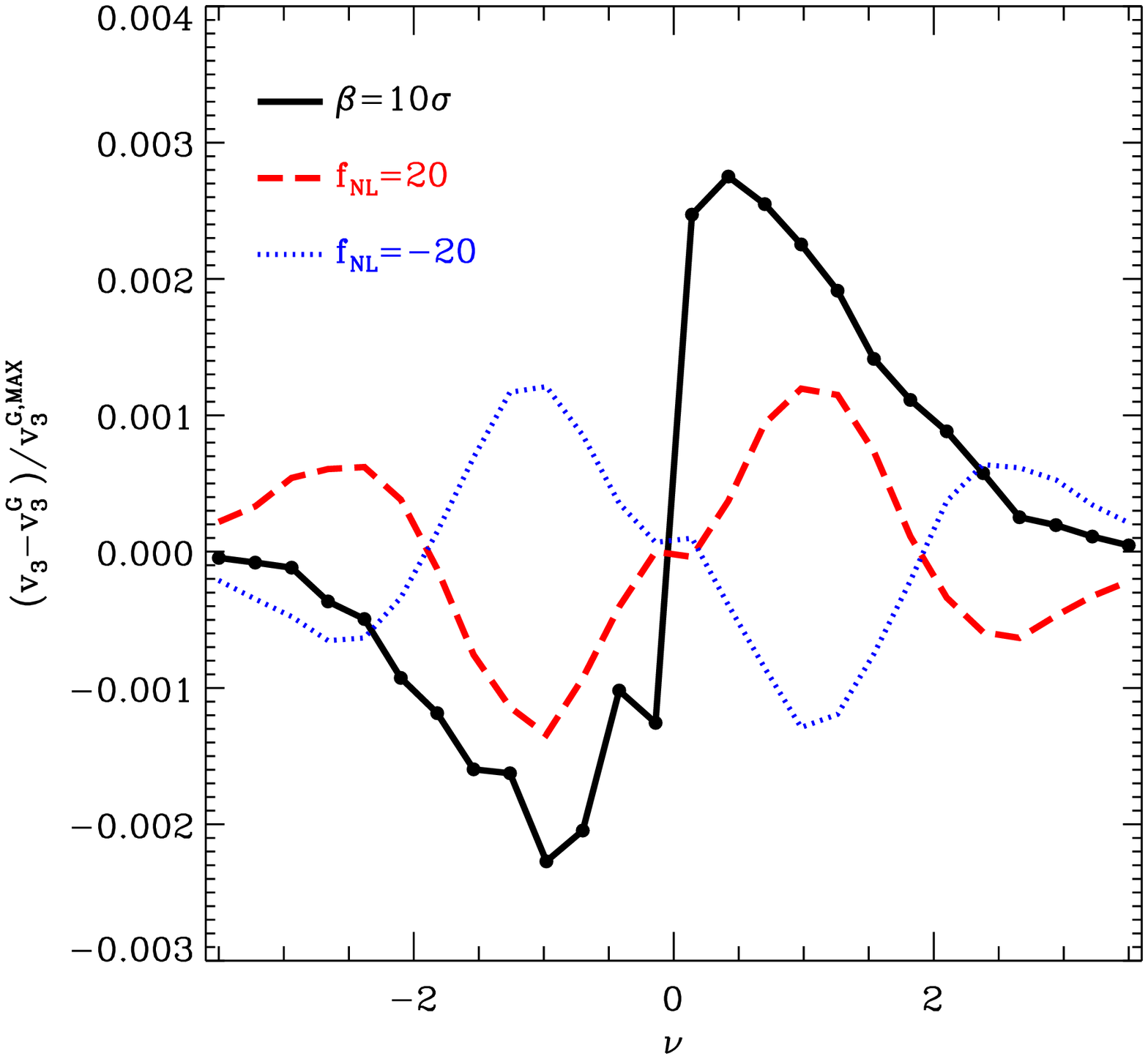}  \\
\hline
\end{tabular}
\caption{Effect of the presence of point sources on the measurement of Minkowski Functionals in the 143\,GHz channel. Here, only the brightest point sources, with an ERCSC signal to noise ratio larger than  $\beta=10\sigma$, are masked out. The configuration is the same as in Fig.~\ref{fig:courbes_hitmap_norm}: normalised functionals $v_k$ are plotted as functions of the threshold $\nu$, after subtracting the Gaussian limit prediction, $v_k^{\rm G}$. The black thick curve corresponds to $f_{_{\rm NL}}=0$ but with point sources, while the two other one correspond to the expected curves in presence of primordial $f_{_{\rm NL}}$ as shown inside each panel.}
\label{fig:courbes_ps143}
\end{center}
\end{figure*}

In the first two bands, 100 and 143\,GHz, faint point sources are composed mainly of radio sources, as can be seen in the ERCSC. Radio sources are not clustered: their power spectrum is known to be flat so they act as a positive, uncorrelated noise. To understand the effect of such a noise, we study in details the configuration of a minimal mask in the 143\,GHz channel as illustrated for each functional by Fig.~\ref{fig:courbes_ps143}. 

As a positive uncorrelated noise, radio sources do not affect significantly the Area MF. Indeed, the positive nature of such a noise is subtracted out when computing the density contrast $\delta$ of the map and, as discussed in \S~\ref{sec:anisotropicnoise}, the convolution effect of a zero average distribution on the Area is negligible as long as its variance is small compared to the variance of the signal. 

The presence of point sources brings an excess of positive clusters; it also slightly shifts the curve $V_k(\nu)$ to smaller values of $\nu$ as an effect of subtracting the average from the temperature map when computing the density contrast, as can be easily noticeable on the Perimeter panel of Fig.~\ref{fig:courbes_ps143}. These effects induce an overall positive bias for the Genus and the Perimeter, while $N_{\rm cluster}$ presents an excess for positive thresholds but a deficit (of under-dense regions) for negative thresholds. These biases remain after renormalisation, i.e. when passing from $V_k$ to $v_k$, except for the perimeter, where division by the factor $A_k$ inverts the bias. Indeed, the presence of point sources increases the ratio $\sigma_1/\sigma_0$, hence the measured value of $A_k$, $k > 0$ (see Appendix~\ref{theory}). 

Table~\ref{table:ps_143GHz_separate_mink} shows the corresponding bias on the measurement of $f_{\rm NL}$ in the null hypothesis $f_{\rm NL}^{\rm prim}=0$. The results of this table can partly be  inferred intuitively from the examination of Fig.~\ref{fig:courbes_ps143}: small bias for the Area, positive bias for the Genus and $N_{\rm clusters}$, negative bias for the Perimeter, resulting in an overall positive bias for the combination of all functionals. The examination of other  masking levels confirms the results of this analysis: the effect of point sources in the 100 and 143\,GHz is a positive bias on the measured $f_{\rm NL}$ which, in addition, does not depend on the primordial level of non Gaussianity,
\begin{equation}
	\hat{f}_{_{\rm NL}}= f_{_{\rm NL}}^{\rm prim} + f_{_{\rm NL}}^{\rm bias},
	\label{eq:bias_ps100143}
\end{equation} 
and decreases, as expected, when more point sources are excluded by the masks, as illustrated by Table~\ref{table:ps_bias_100_143GHz}. In particular, the bias induced by point sources becomes nearly negligible compared to expected error bars ($\Delta f_{_{\rm NL}} \ga 10$, Table~\ref{table:mask_gal_sensitivity}) when masks are set up at the $3\sigma$ level.

\setlength{\tabcolsep}{0.3cm}

\begin{table}
 \begin{center}
 \begin{tabular}{ c   c  c  c  c  c  }
\hline
 $f_{_{\rm NL}}^{\rm prim}=0$ & $V_{0}$  & $V_{1}$ & $V_{2}$ & $V_{3}$ & All \\
\hline
$\langle \hat{f}_{_{\rm NL}} \rangle $     &  0.8  & -8  & 14 &  2.3 & 10 \\
\hline
\end{tabular}
  \caption{Bias on the measurement of $f_{_{\rm NL}}$ introduced by point sources at 143\,GHz in case of weak masking at the $10 \sigma$ detection level in the ERCSC. Each column corresponds to using a specific functional in the $\chi^2$ analysis, or, for the last one, the combination of all functionals. Here, the null hypothesis $f_{_{\rm NL}}^{\rm prim}=0$ is tested, but in practice, the effective bias does not depends on the value of $f_{_{\rm NL}}^{\rm prim}$ (eq.~\ref{eq:bias_ps100143}).}
  \label{table:ps_143GHz_separate_mink}
 \end{center}
 \end{table}
 
\begin{table}
 \begin{center}
 \begin{tabular}{ c   c  c  }
\hline
 Flux cut & $ f_{_{\rm NL}}^{\rm bias} $ @100\,GHz & $ f_{_{\rm NL}}^{\rm bias} $ @143\,GHz \\
\hline 
10$\sigma$   & $14$ &  $ 10$ \\
5$\sigma$   &  $8$ &  $ 4$ \\
3$\sigma$   &  $3$ &  $ 1.5$  \\
\hline
\end{tabular}
  \caption{Bias on the measurement of $f_{_{\rm NL}}$ introduced by point sources at 100 and 143\,GHz as a function of masking level, expressed here in terms of ERCSC signal to noise threshold, assuming that the combination of all functionals is used to measure $f_{_{\rm NL}}$. This bias, modelled by eq.~(\ref{eq:bias_ps100143}), does not depend on the actual level of primordial non Gaussianity, $f_{_{\rm NL}}^{\rm prim}$, and can therefore be easily corrected for. Here, as detailed in \S~\ref{sec:PSMprocess}, the bias is obtained in the following configuration for the extended mission: $N_{\rm side}=1024$, $\ell_{\rm max}=2000$, Gaussian smoothing with $\theta_{_{\rm FWHM}}^{S}=5'$, $n_{\rm bins}=26$ and $\nu_{\rm max}=3.5$. With a different set up, a new estimate of the bias would be needed, but this is an easy exercise.}
  \label{table:ps_bias_100_143GHz}
 \end{center}
 \end{table}

\subsubsection{217\,GHz: effect of radio and IR sources} \label{sec:PSMpart2}

In the 217\,GHz band, in addition to radio sources, an IR background contributes to the faint point sources, which results in a new bias on the measurement of $f_{_{\rm NL}}$, as Table~\ref{table:ps0} shows.

The contribution from unclustered radio sources is a decreasing function of  frequency and masking. It should act, as in the 100 and 143\,GHz channels, as a positive bias on the measurement of $f_{_{\rm NL}}$ that does not depend on the value of $f_{_{\rm NL}}^{\rm prim}$ (see eq.~\ref{eq:bias_ps100143}). 

\begin{figure*}
 \begin{center}
\vspace{0.2cm}
\begin{tabular}{|c | c |}
\hline
\includegraphics[width=8cm]{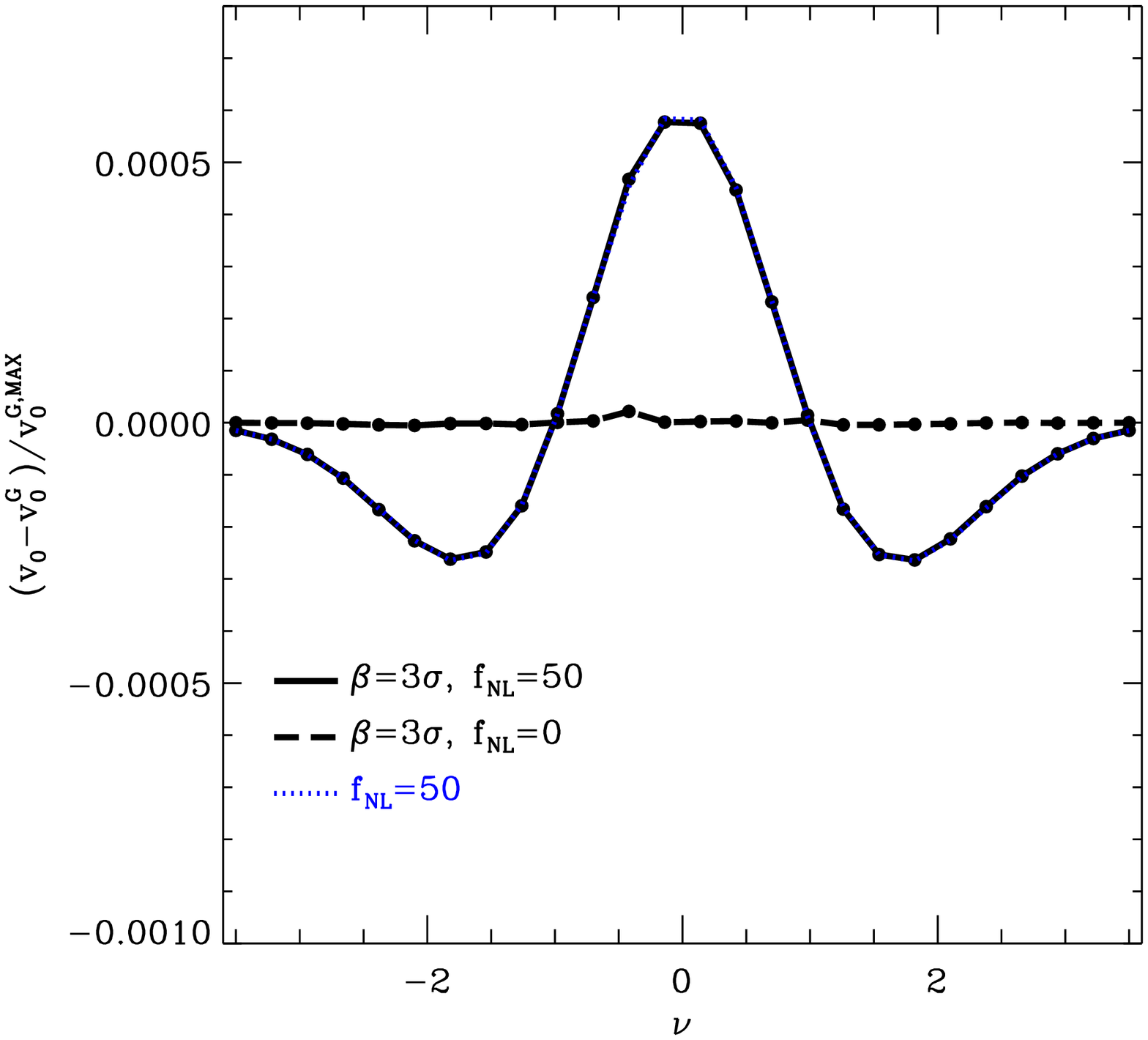} & \includegraphics[width=8cm]{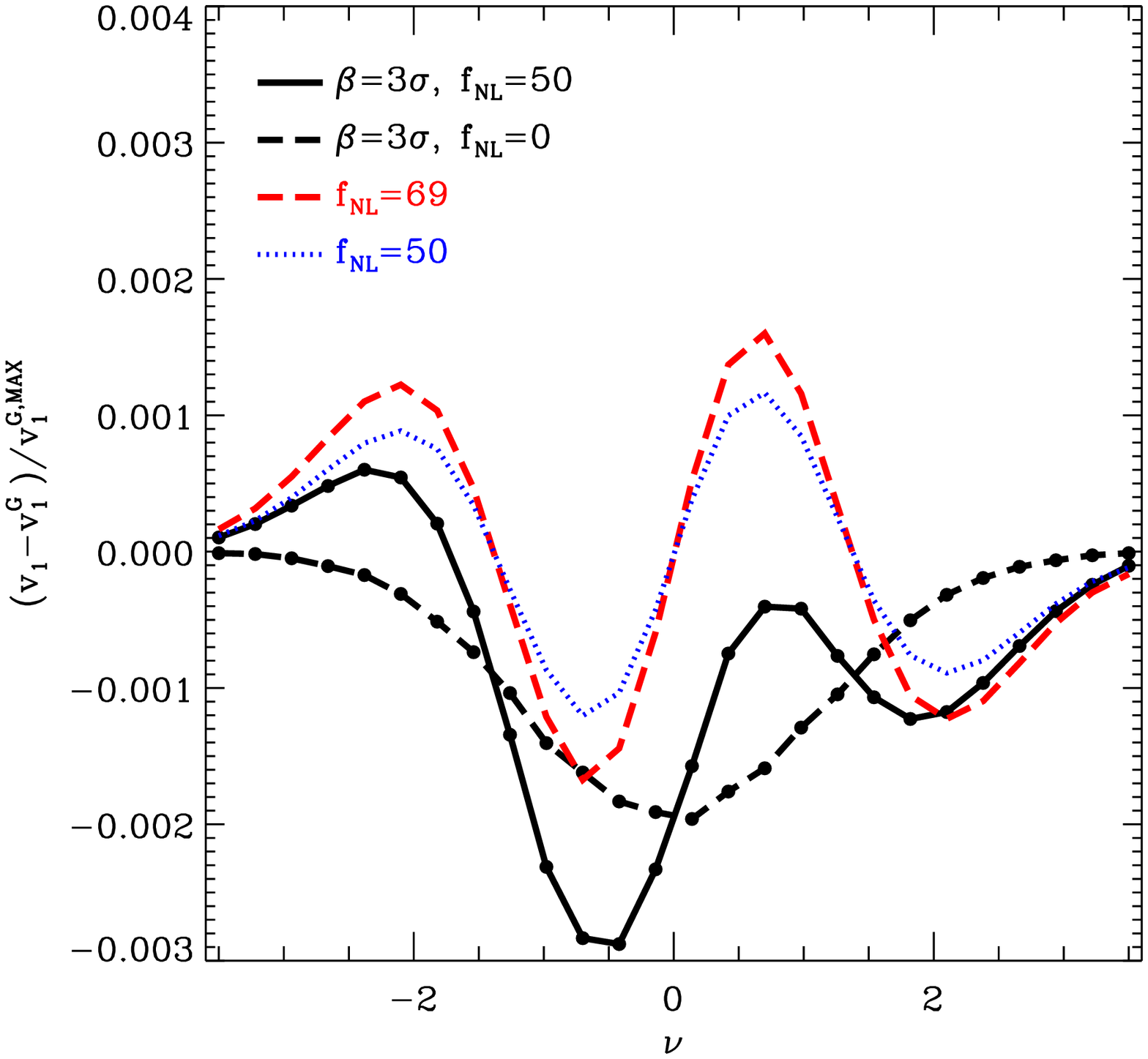}  \\
\hline
\includegraphics[width=8cm]{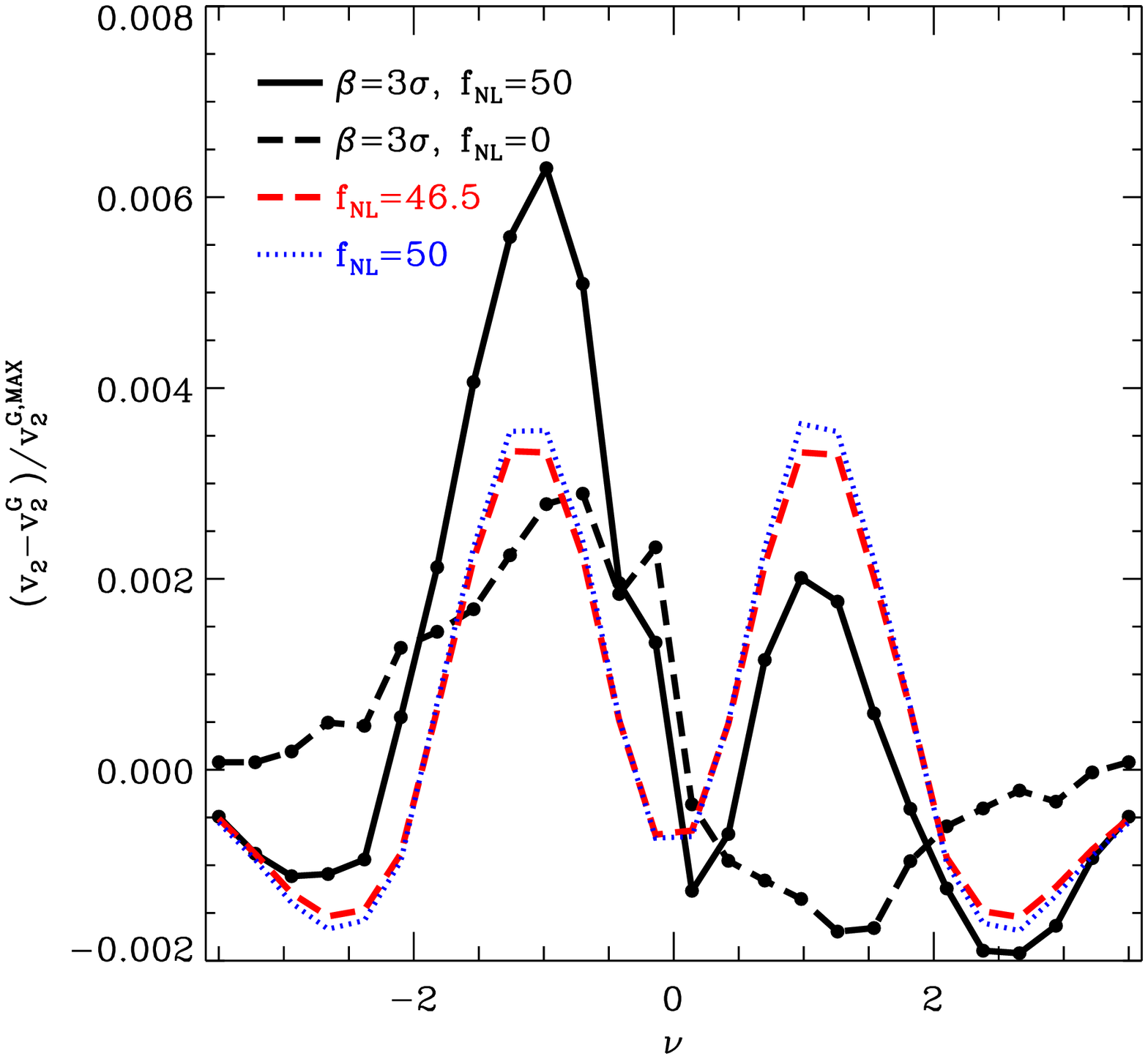} & \includegraphics[width=8cm]{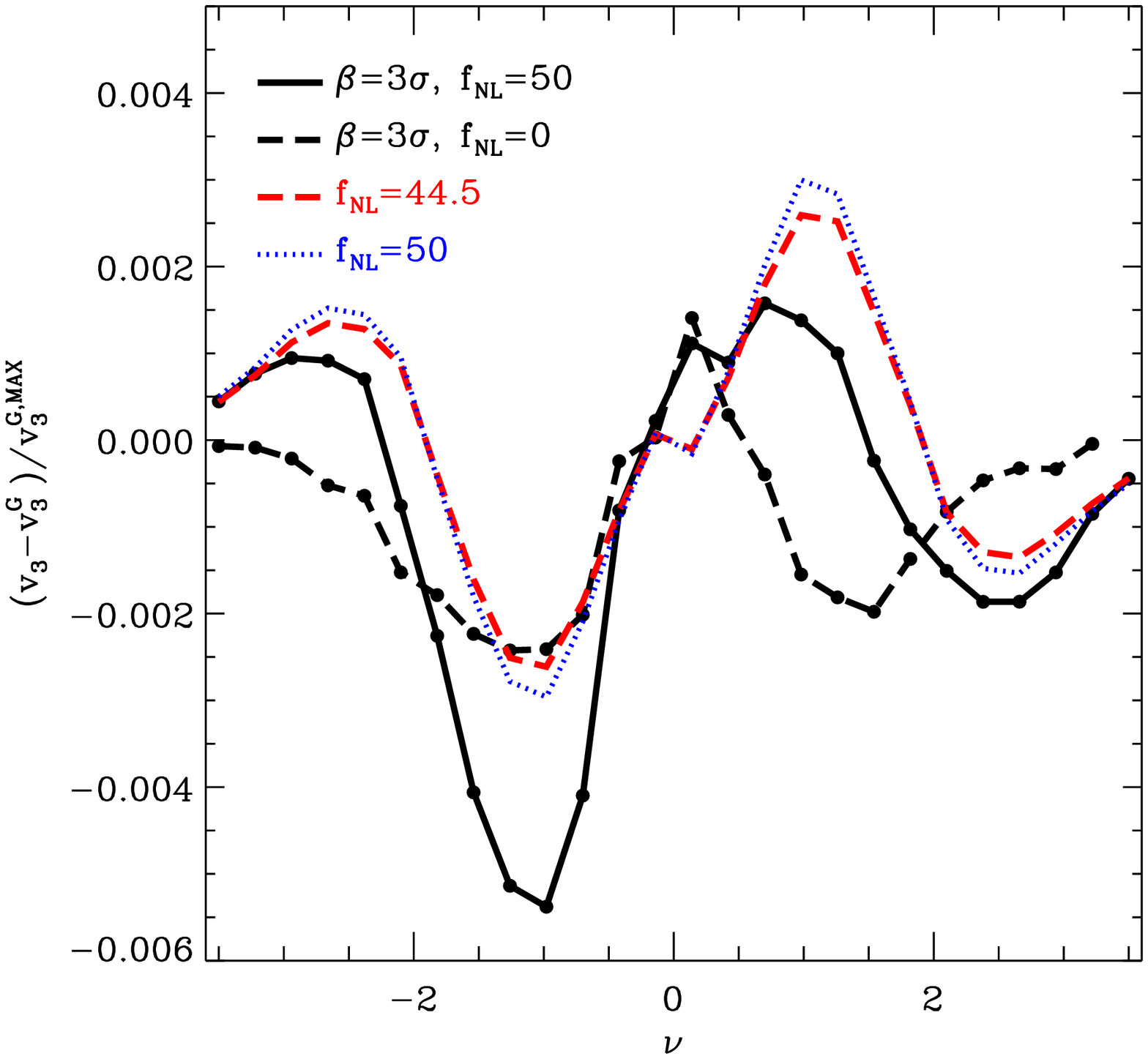}  \\
\hline
\end{tabular}
\caption{Effect of the presence of point sources on the measurement of Minkowski Functionals, similarly as in Fig.~\ref{fig:courbes_ps143}, but for the 217\,GHz channel. The black thick solid and dashed curves correspond respectively to primordial $f_{_{\rm NL}}=50$ and $f_{_{\rm NL}}=0$, plus point sources, while the two other ones 
correspond to the case of primordial $f_{_{\rm NL}}$ only, as shown inside each panel, with $f_{_{\rm NL}}=50$ for the blue curves, and with the value of $f_{_{\rm NL}}$ found when estimating this quantity in the foreground-biased maps (Table~\ref{table:ps_217GHz_separate_mink_fnl50}), in red. Note that on upper left panel, there is no red curve, and the black curve superposes exactly to the blue one. To emphasise the effect of the clustered IR background, a $3\sigma$ level masking was performed to subtract as much as possible the contribution from radio-sources.}
\label{fig:courbes_ps217}
\end{center}
\end{figure*}

On the other hand, IR sources are clustered and form mainly a diffuse, unresolved background, which cannot be dealt with masks. They induce a bias on the measurement of $f_{_{\rm NL}}$ which appears to depend on the value of $f_{_{\rm NL}}^{\rm prim}$ as can be inferred from Table~\ref{table:ps0}. To analyse this bias more in depth, we concentrate on a configuration where the contribution of radio sources is masked out as much as possible and nearly negligible, with the 3$\sigma$ flux cut setting for the masks. Figure~\ref{fig:courbes_ps217} displays the functionals obtained in two cases, $f_{_{\rm NL}}^{\rm prim}=0$ and $f_{_{\rm NL}}^{\rm prim}=50$. It is interesting to notice that the curve obtained for $f_{_{\rm NL}}^{\rm prim}=50$ is nearly exactly the sum of the curve for $f_{_{\rm NL}}^{\rm prim}=50$ with no point source contribution and  the curve for  $f_{_{\rm NL}}^{\rm prim}=0$ with point sources. Unfortunately, this linear property does not translate in a simple way in terms of bias on the measured $f_{_{\rm NL}}$, when performing the $\chi^2$ analysis, as illustrated by Tables~\ref{table:ps_217GHz_separate_mink_fnl0} and \ref{table:ps_217GHz_separate_mink_fnl50}. 
What we find, instead, is the bias due to IR sources to be roughly proportional to $f_{_{\rm NL}}^{\rm prim}$, when one considers the combination of all functionals to perform the measurements. 

Our final approximation for the total bias in the 217\,GHz channel is therefore
\begin{equation}
	 \hat{f}_{_{\rm NL}} \: \mathrm{ @217GHz} = f_{_{\rm NL}}^{\rm prim} + f_{_{\rm NL}}^{\rm bias, radio} +f_{_{\rm NL}}^{\rm bias, IR}
	\label{eq:bias1_ps217}
\end{equation} 
with $f_{_{\rm NL}}^{\rm bias, radio}$ depending only of the masking level, as in \S~\ref{sec:PSMpart1}. Here, this bias grows from negligible for the $3\sigma$ masks to $f_{_{\rm NL}}^{\rm bias, radio} \simeq 5$ for the $10 \sigma$ masks (right column part of Table~\ref{table:ps0} with $f_{_{\rm NL}}^{\rm prim}=0$). The other term, $f_{_{\rm NL}}^{\rm bias, IR}$ does not depend on masking but is approximately proportional to $f_{_{\rm NL}}^{\rm prim}$ for moderate values of $|f_{_{\rm NL}}^{\rm prim}|$:
\begin{equation}
	f_{_{\rm NL}}^{\rm bias, IR} \simeq \dfrac{f_{_{\rm NL}}^{\rm prim}}{2}, \quad |f_{_{\rm NL}}^{\rm prim}| \la 50.
\label{eq:bias2_ps217}
\end{equation} 
Note that with Planck extended mission signal to noise, one expects to be able to characterise more accurately the IR background. It might then be possible to account for it in a better way, by e.g. including it in the model itself instead of ignoring it due to lack of precise knowledge.

\begin{table}
 \begin{center}
 \begin{tabular}{ c   c  c  c  c  c }
\hline
 $f_{_{\rm NL}}^{\rm prim}=0$                     & $V_{0}$  & $V_{1}$ & $V_{2}$ & $V_{3}$ & All\\
\hline
$\langle \hat{f}_{_{\rm NL}} \rangle$     &   1.5     &    -7.5       & 3         & 2          & 0.6 \\
\hline
\end{tabular}
  \caption{Point sources bias at 217\,GHz for $f_{_{\rm NL}}=0$: for a flux cut corresponding to the $3\sigma$ detection level in the ERCSC, i.e. an important masking of radio sources, we see the effect of the clustered IR background, in the case of a null primordial non Gaussianity, $f_{_{\rm NL}}^{\rm prim}=0$.}
  \label{table:ps_217GHz_separate_mink_fnl0}
 \end{center}
 \end{table}
 
 \begin{table}
 \begin{center}
 \begin{tabular}{ c   c  c  c  c  c }
\hline
 $f_{_{\rm NL}}^{\rm prim}=50$                  & $V_{0}$  & $V_{1}$ & $V_{2}$ & $V_{3}$ & All\\
\hline
$\langle \hat{f}_{_{\rm NL}} \rangle$     &   50     &  69          & 46.5       & 44.5      & 69 \\
\hline
\end{tabular}
  \caption{Point sources bias at 217\,GHz for $f_{_{\rm NL}}=50$: same as in Table~\ref{table:ps_217GHz_separate_mink_fnl0} but for a significant level of primordial non Gaussianity.}
  \label{table:ps_217GHz_separate_mink_fnl50}
 \end{center}
 \end{table}

\section[Foregrounds II]{Foregrounds II: Galactic Residuals and Galactic Mask}
\label{sec:lagalaxie}

Galactic signals are a major issue for cosmological studies of the CMB and have to be accurately assessed. It is usually treated in two ways: masking alone or in conjunction with component separation. Here, we test the effects of the Galactic residuals left from these methods on the estimation of $f_{_{\rm NL}}$ with Minkowski Functionals by using simulations of the Foreground emission, masks constructed from these simulations for different sky coverages, and a na\"ive model of component separation quality. 

The philosophy adopted here is similar to that in the previous section: we do not include in the model templates of Galactic foregrounds. Instead, we analyse the biases on $f_{_{\rm NL}}$ introduced by neglecting the presence of galactic emission. Such biases are expected to be small if one restricts to regions far from the galactic plane or/and if proper component separation has been performed prior to the measurements. 

This section is divided into four parts. The first one, \S~\ref{sec:methgal}, details our simulation settings. The second one,\S~\ref{sec:skysensi}, looks at the statistical uncertainty expected on the measured $f_{\rm NL}$ from Minkowski functionals as a function of sky coverage.  The third one, \S~\ref{sec:galforsec}, examines, as functions of sky coverage, the biases brought by Galactic foregrounds, if these latter would not be removed at all, and the expected improvements on such biases brought by component separation. 

\subsection{Method} \label{sec:methgal}

To perform our simulations, we need a  way to generate realistic maps of the Galactic emission, which is reviewed in \S~\ref{sec:simugal}. To do so, we use again the PSM \citep{PSM}.  \S~\ref{sec:galmask} explains how masks are generated, in order to exclude regions where Galactic emission is too strong. \S~\ref{sec:alpha} details the very simple method we decided to use to assess the affect of component separation quality. Other technical details of our simulations are provided in \S~\ref{sec:simpipe}.

\subsubsection{Simulations of the foreground emission: the different components} \label{sec:simugal}

We use again the PSM to construct the Galactic emission. The code uses template maps interpolated at the desired frequencies. The detailed modelling of each component is described in \citet{PSM}. Here are the list of the components we simulated in our map of the Foreground emission.

The diffuse Galactic emission arises from of the interstellar medium (ISM) in the Milky Way. The ISM is composed of different phases, from cold molecular clouds to hot ionised regions, of magnetic fields and cosmic rays. The intensity of the corresponding emission depends on Galactic latitude. It is stronger near the center of the Galaxy and decreases at lower/higher latitudes following approximately a cosecant law $1/\sin (|b|)$.
More precisely, we usually classify the different ISM components according to their physical emission processes:
\begin{enumerate}
\item synchrotron radiation is emitted by relativistic cosmic rays spiralling in the Galactic magnetic field. Its intensity depends on the cosmic ray density and on the magnetic field strength perpendicular to the line of sight;
\item free-free emission originates from the ionised medium in the ISM, as a result of the interaction of free electrons with positively charged nucle\"i. It comes principally from star forming regions in the Galactic plane;
\item there is also a thermal emission coming from dust grains heated by stars, which is the dominant contribution above 70\,GHz;
\item another type of emission have been found at microwave frequencies which is probably due to small spinning dust particles (``Anomalous emission'');
 \item At these frequencies, there are also molecular lines emerging from molecular clouds, particularly those of  $^{12}$CO at 100 and 217\,GHz.
 \end{enumerate}

For the last contribution, i.e. CO lines, templates and models of the emission are mostly unknown at present time and we choose not to model it and to simulate only the 143\,GHz part of the Galactic foreground. However, concerning noise level, our analyses correspond to a combination of the 100, 143 and 217\,GHz channels in the framework of the Planck extended mission. 
The specific CO contribution, even if sub-dominant compared to the other physical processes, will be studied when Planck templates of CO will be available.

\begin{figure}
 \begin{center}
\vspace{0.2cm}
\begin{tabular}{c}
\includegraphics[angle=90,width=8cm]{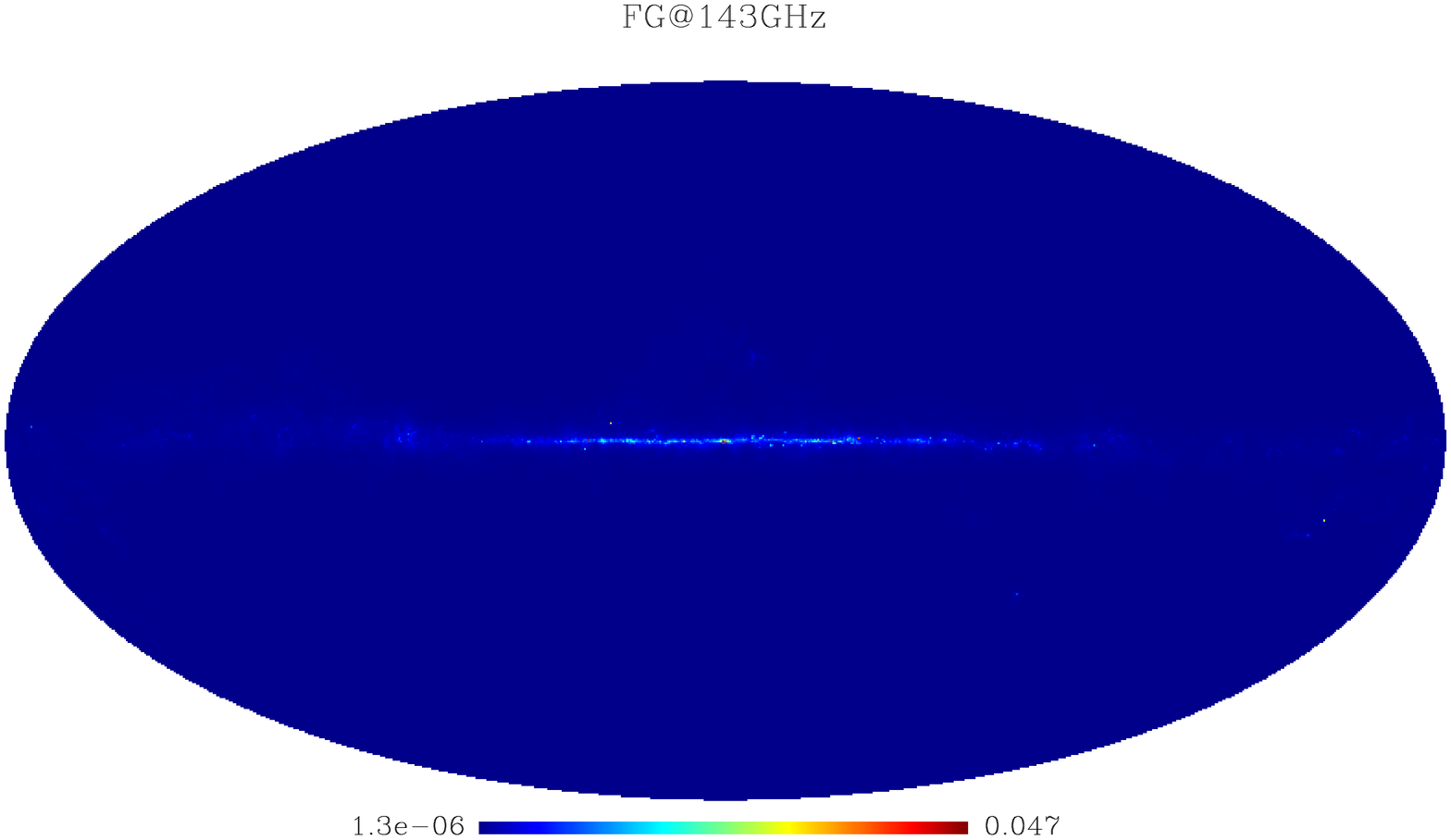} \\
Foreground map (linear colour scale) \\
 \includegraphics[angle=90,width=8cm]{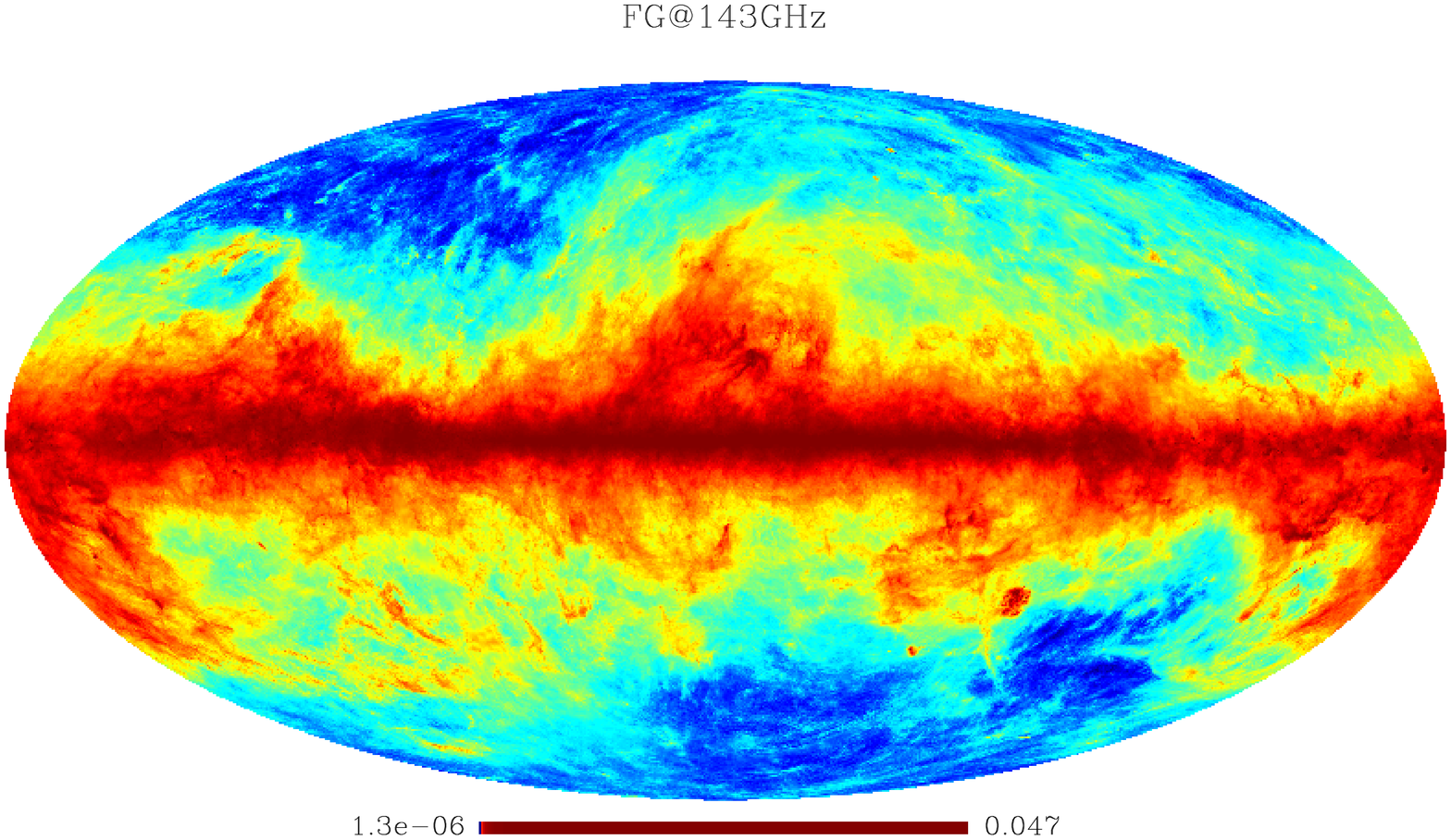}  \\
 Foreground map  (histogram-equalised colour scale in HEALPix)  \\
\end{tabular}
\caption{Map of Galactic emission at 143\,GHz, units are in Kelvin. In the top panel, the colour scale is linear. In the bottom one, the colour scale is histogram–equalised to increase the dynamic range and make visible both the regions of low and high emission intensity. We have included synchrotron, free-free, thermal emission and emission from spinning dust particles.}
\label{fig:map_galactic}
\end{center}
\end{figure}

The resulting foreground map is represented on Fig.~\ref{fig:map_galactic}, with two colour scales enhancing different aspects of this emission. 

\subsubsection{Galactic Mask}
\label{sec:galmask}

\begin{figure}
\begin{center}
  \includegraphics[angle=90,width=8.5cm]{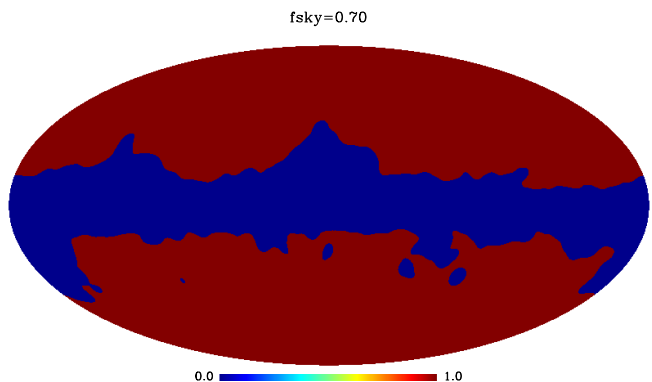}
  \caption{Example of a galactic mask with $f_{\rm sky}=0.70$, drawn from the PSM. It relates directly to the Galactic emission shown in Fig.~\ref{fig:map_galactic}.  }
  \label{fig:map_mask_gal}
\end{center}
\end{figure}

To create Galactic masks, we consider the fraction $f_{\rm sky}$  of the sky we aim to keep, which sets up an intensity threshold for the Galactic foreground  above which the corresponding region of the sky is masked out. Once the mask function $M$ is set-up,  which is equal to one for valid pixels and zero for excluded pixels, convolution of this function is performed with a Gaussian kernel of size $\theta_{_{\rm FWHM}}^{\rm S}=~5^{\circ}$, to obtain a smoothed version $M_{\rm smoothed}$. New masks with smooth boundaries are extracted from this map, by selecting pixels with $M_{\rm smoothed} \geq M_{\rm thresh}$ and excluding the others, where the value of $M_{\rm thresh}$ is tuned to get the correct value of $f_{\rm sky}$.  An example of masks constructed that way is given on Figure~\ref{fig:map_mask_gal}.

\subsubsection{Component separation quality}
\label{sec:alpha}

\begin{figure}
 \begin{center}
\vspace{0.2cm}
\begin{tabular}{c}
\includegraphics[angle=90,width=8cm]{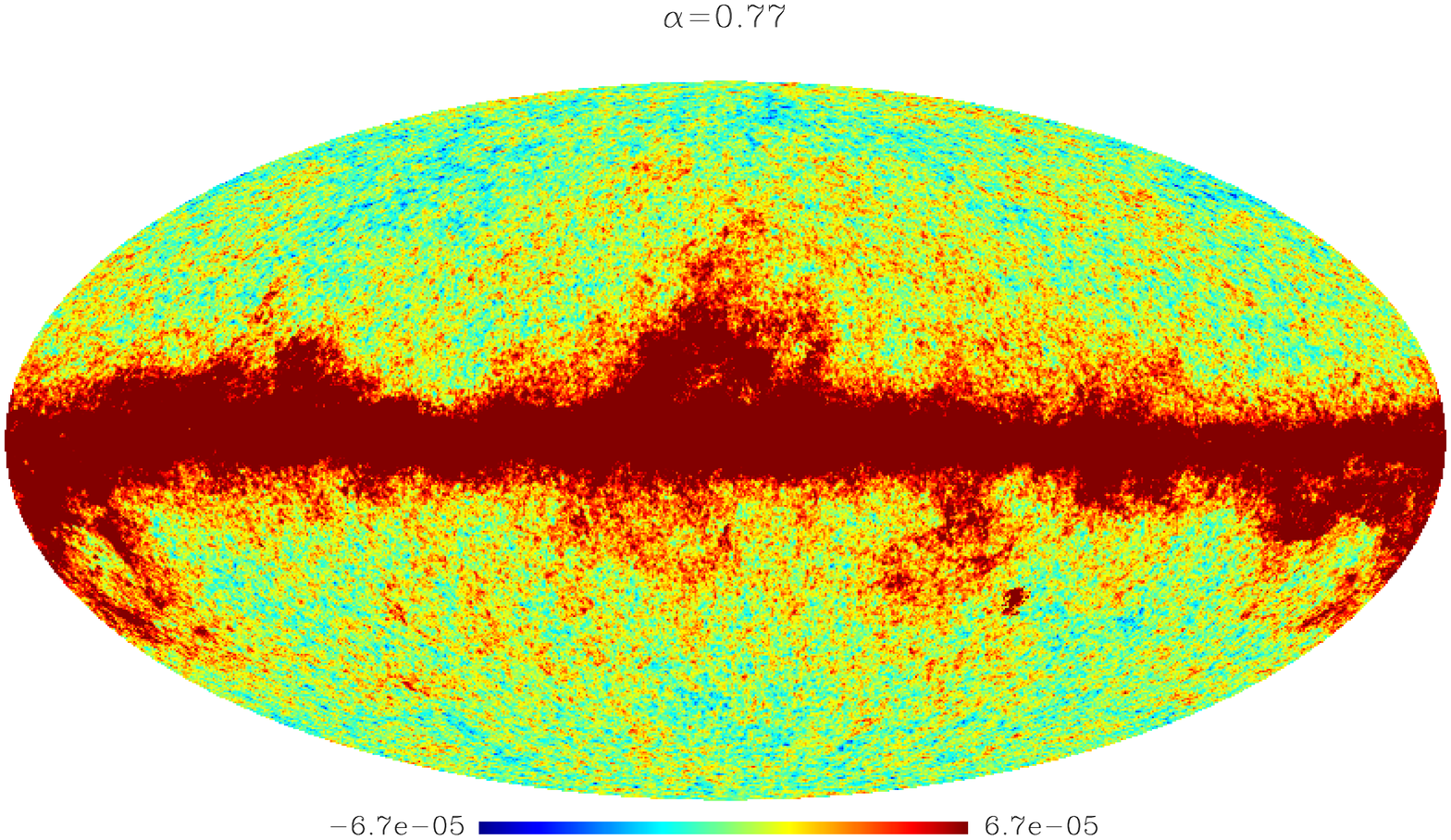}  \\
\includegraphics[angle=90,width=8cm]{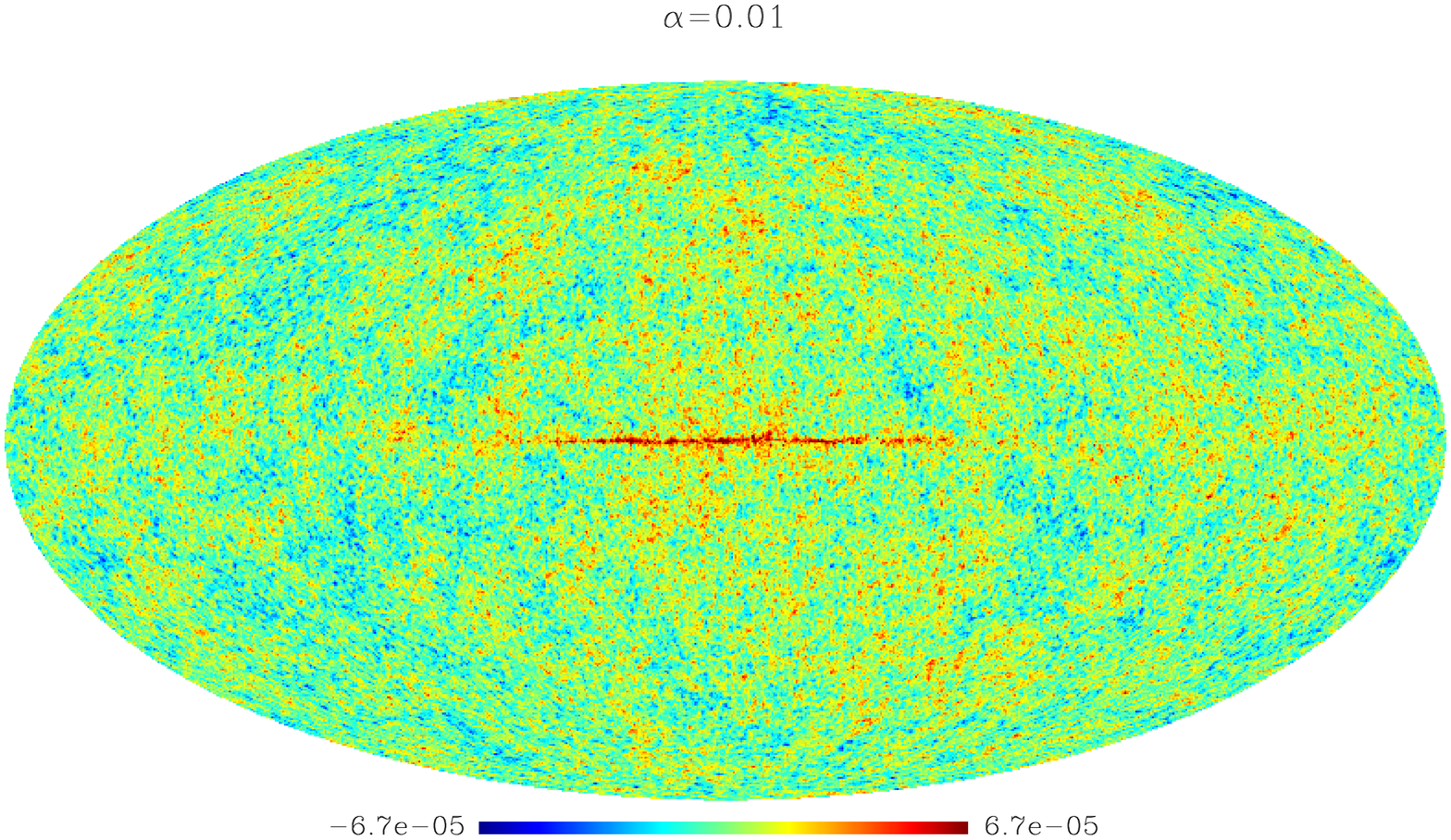} \\
 \end{tabular}
\caption{Simulations of a microwave observation at 143\,GHz, including CMB and Galactic emission in two cases: one with no component separation $\alpha=0.77$ and one with a ``good'' component separation $\alpha=0.01$. Units are in Kelvin.}
\label{fig:map_component_separation}
\end{center}
\end{figure}

In order to asses simply the impact of the residuals of component separation, i.e. of its quality, we model component separation results by simply adding to the CMB the map of Galactic foregrounds multiplied by a scaling factor $\dfrac{\sigma_{\mathrm{CMB@HL}} }{\sigma_{\mathrm{FG@HL}}}\alpha$, where $\sigma_{\mathrm{CMB@HL}}$ and $\sigma_{\mathrm{FG@HL}}$ are respectively the standard deviations of the CMB and foreground maps at high latitudes, i.e. measured in pixels outside the mask with $f_{\rm sky}=0.80$. So for a ``quality factor” $\alpha=1$, the normalised contributions of Galactic foreground and CMB at high latitudes will be the same.  The value $\alpha=0.77$ corresponds to the initial level of foreground emission, so it is equivalent to no component separation (top panel of Fig.~\ref{fig:map_component_separation}). Realistic values of $\alpha$, when examining maps obtained from actual component separation methods, rather appear to range typically between $\alpha=0.01$ (bottom panel of Fig.~\ref{fig:map_component_separation}) and $\alpha=0.05$. 

Obviously our modelling of Galactic residuals after component separation is very rough, but it should be sufficient to assess what should be the value of $f_{\rm sky}$ for making the biases on $f_{_{\rm NL}}$ induced by these residuals negligible. A better modelling of the residuals would require detailed examination of the results obtained from actual component separation. Furthermore, a new analysis  would be required each time a new component separation method is considered: this is far beyond the scope of this paper. 

\subsubsection{Simulations pipeline}
\label{sec:simpipe}
The simulation strategy used here is the same as in \S~\ref{sec:PSMprocess}: the Galactic foregrounds are added only to the test maps, or, in other words, the ``data'' maps.  The $m_{\rm NG}=200$ maps used to compute the model prediction and the $m=10\,000$
used to estimate the covariance matrix in the $\chi^2$ function do not have the foregrounds, but the treatment is the same otherwise. We test 4 values of $f_{_{\rm NL}}^{\rm prim}=\lbrace -10, 0, 10, 50 \rbrace$, for each of which $m_{\rm test}=200$ test maps are generated, to which we add Galactic foregrounds in the $f=143$\,GHz channel, with a level $\alpha$ as described in \S~\ref{sec:alpha}. Then Gaussian beaming is performed with the beam of the instrument in this channel,  $\theta_{_{\rm FWHM}}^{\rm b}=7.2'$, that corresponds as well as to the beam for the 3 combined channels (that we assume to be used for this analysis) and the noise at the level of the extended mission (for the 3 combined channels) is added (see Appendix~\ref{planck}). Then filtering is performed using either a Gaussian window of size  $\theta_{_{\mathrm{FWHM}}}^{\rm S}$ or Wiener filters discussed in \S~\ref{sec:wienerfilters}. Finally, the Galactic mask calculated as in \S~\ref{sec:galmask}  is added, for a given $f_{\rm sky}$. Note that at variance with the point-source analysis, the Galactic mask is not inpainted, due to its rather large size: the pixels inside the mask are just ignored by the MFs code. The complete pipeline is summarised as follows:
\begin{eqnarray}
{\rm map} &=& {\rm CMB } (f_{_{\rm NL}}^{\rm prim})*\mathrm{beam}[\theta_{_{\rm FWHM}}^{\rm b}(f)] \nonumber \\
 &+ & \alpha  \times \dfrac{\sigma_{\mathrm{CMB@HL}}}{\sigma_{\mathrm{FG@HL}}} \times {\rm Gal.~FG}(f)*{\rm beam}[\theta_{_{\rm FWHM}}^{\rm b}(f)]  \nonumber \\
 & + & {\rm noise}(f) \nonumber \\
 & + & \mathrm{smoothing}[\theta_{_{\rm FWHM}}^{\rm S}] {\rm \ or \ Wiener \ filtering} \nonumber\\
  & + & \mathrm{galactic} \: \mathrm{mask} \, (f_{\rm sky}).
 \label{eq:galmeth}
\end{eqnarray}

\subsection{Sensitivity versus sky coverage}
\label{sec:skysensi}
First, we test the sensitivity of our $\chi^2$ estimator to sky coverage, using maps of the CMB without Galactic foreground i.e. $\alpha=0$ and just looking at the resulting error bars, ${ \Delta f}_{_{\rm NL}}$. Table~\ref{table:mask_gal_sensitivity} shows $\langle {\widehat{ \Delta f}}_{_{\rm NL}}^2 \rangle^{1/2}$ as a function of $f_{\rm sky}$, obtained from  the combination of all four functionals  for various Gaussian smoothing scales $\theta_{_{\rm FWHM}}^{\rm S}$ and different Wiener filterings. As expected, $\langle \widehat{\Delta f}_{_{\rm NL}}^2 \rangle^{1/2}$ is approximately a linear function of $\sqrt{f_{\rm sky}}$.  Note that these results do not change significantly in the presence of Galactic foregrounds as long as $f_{\rm sky}\lesssim 0.80$. 

\setlength{\tabcolsep}{0.1cm}

\begin{table}
 \begin{center}
 \begin{tabular}{ l |  c  c   c  c  c }
\hline
\hspace*{0.1cm} $\theta_{_{\rm FWHM}}^{\rm S}$ & & & $\langle \widehat{\Delta f}_{_{\rm NL}}^2 \rangle^{1/2}$ & & \\
\hspace*{0.1cm} or Wiener & $f_{\rm sky}$=0.4 & $f_{\rm sky}$=0.6 & $f_{\rm sky}$=0.7 & $f_{\rm sky}$=0.8 & $f_{\rm sky}$=1 \\
\hline
\hspace*{0.1cm}5' &  28    & 23  &21 & 19 & 16.5 \\
\hspace*{0.1cm}10'&     32  & 27 &  26 & 24 & 20 \\
\hspace*{0.1cm}5'+10' & 25  & 19& 18 & 17 & 14\\
\hspace*{0.1cm}$W_{\rm M}$ & 21.5  & 17.5 & 16 & 14.5 & 12.5\\
\hspace*{0.1cm}$W_{\rm M}+W_{\rm D1}$ & 15.5 & 12 & 11 & 9.5 & 9 \\
\hline
\end{tabular}
  \caption{Sensitivity of the estimator versus sky coverage: $\langle \widehat{\Delta f}_{_{\rm NL}}^2 \rangle^{1/2}$ is the error bar estimated from the combination of all functionals, $V_{0}+V_{1}+V_{2}+V_{3}$. It is approximately a linear function of $\sqrt{f_{\rm sky}}$. The estimates of the error on $f_{\rm NL}$  are performed for the combination of the 3 channels in the extended mission configuration.}
  \label{table:mask_gal_sensitivity}
 \end{center}
 \end{table}

\subsection{Effect of Galactic foreground} \label{sec:galforsec}

In the following, we study the actual effects of the Galactic foregrounds, which present a rather complex behaviour as a function of sky coverage, $f_{\rm sky}$. Our analyses restrict to Gaussian smoothing and put aside Wiener filtering. From the qualitative point of view, the latter indeed leads to results similar to what is obtained with the smallest Gaussian smoothing scale. This section is divided into two parts. Section \ref{sec:twobehaviors} assumes no component separation and analyses in details the biases brought on the measurement of $f_{_{\rm NL}}$ by the Galactic foregrounds, while \S~\ref{sec:compalpha} examines the biases as functions of component separation quality factor, $\alpha$. 

\subsubsection{Two behaviours, two components}  \label{sec:twobehaviors} 

We now examine what kind of biases the Galactic foregrounds induce on $f_{_{\rm NL}}$.  
Table~\ref{table:gal1} shows our forecast for $\langle {\hat f}_{_{\rm NL}} \rangle$ as a function of $f_{\rm sky}$  and $f_{_{\rm NL}}^{\rm prim}$, at the smallest Gaussian smoothing scale, $\theta_{_{\rm FWHM}}^{\rm S}=5'$. Remind that Galactic foregrounds are completely present, without any attempt to remove them with component separation.

 \begin{table*}
 \begin{center}
 \begin{minipage}{115mm}
 \begin{tabular}{@{}  l  c c c c @{} }
\hline
\hspace*{0.1cm}$\alpha=0.77$ \hspace*{0.4cm} & \hspace*{0.4cm}$f_{_{\rm NL}}^{\rm prim}=0$\hspace*{0.4cm} &  $f_{_{\rm NL}}^{\rm prim}=-10$ \hspace*{0.4cm} & $f_{_{\rm NL}}^{\rm prim}=10$ \hspace*{0.4cm} & $f_{_{\rm NL}}^{\rm prim}=50$   \\
\hline
\hspace*{0.1cm}$f_{\rm sky}=0.90$&$\langle \hat{f}_{_{\rm NL}} \rangle $&$\langle\hat{f}_{_{\rm NL}}\rangle$&$ \langle\hat{f}_{_{\rm NL}}\rangle $ & $\langle \hat{f}_{_{\rm NL}}\rangle $    \\
\hline
\hspace*{0.1cm}$V_{1}+V_{2}+V_{3}$ \hspace*{0.4cm}           & -22 &   -23    &  -21    &  -15    \\
\hspace*{0.1cm}$V_{0}+V_{1}+V_{2}+V_{3}$ \hspace*{0.4cm} & -26  &  -27   &  -25     & -21   \\
\hline
\hspace*{0.1cm}$f_{\rm sky}=0.80$      & $\langle \hat{f}_{_{\rm NL}} \rangle$  & $\langle \hat{f}_{_{\rm NL}} \rangle$  & $\langle \hat{f}_{_{\rm NL}} \rangle$   & $\langle \hat{f}_{_{\rm NL}} \rangle$    \\
\hline
\hspace*{0.1cm}$V_{0}$                  & -61     &         &         &         \\
\hspace*{0.1cm}$V_{1}$                  & -42    &          &        &            \\
\hspace*{0.1cm}$V_{2}$                  & -12     &          &        &               \\
\hspace*{0.1cm}$V_{3}$                  &  -58    &           &        &             \\
\hspace*{0.1cm}$V_{1}+V_{2}+V_{3}$ \hspace*{0.4cm}            & 0.8   &   -12      &  10.6     &  53     \\
\hspace*{0.1cm}$V_{0}+V_{1}+V_{2}+V_{3}$ \hspace*{0.4cm} & -5  &  -16.2    &  6.5       & 50   \\
\hline
\hspace*{0.1cm}$f_{\rm sky}=0.70$      & $\langle \hat{f}_{_{\rm NL}} \rangle$  & $\langle \hat{f}_{_{\rm NL}} \rangle$ &   $ \langle\hat{f}_{_{\rm NL}}\rangle$   &  $\langle \hat{f}_{_{\rm NL}}\rangle$    \\
\hline
\hspace*{0.1cm}$V_{0}$                  & -15   &     &         &           \\
\hspace*{0.1cm}$V_{1}$                  &  -5    &      &          &         \\
\hspace*{0.1cm}$V_{2}$                  &  8      &      &          &           \\
\hspace*{0.1cm}$V_{3}$                  &  -22  &    &           &            \\
\hspace*{0.1cm}$V_{1}+V_{2}+V_{3}$ \hspace*{0.4cm}           &  17  &   6     & 27      &  63    \\
\hspace*{0.1cm}$V_{0}+V_{1}+V_{2}+V_{3}$ \hspace*{0.4cm} & 16  &   5     &  26     & 63     \\
\hline
\hspace*{0.1cm}$f_{\rm sky}=0.60$      &  $\langle \hat{f}_{_{\rm NL}} \rangle$ & $\langle \hat{f}_{_{\rm NL}} \rangle$  &  $\langle \hat{f}_{_{\rm NL}} \rangle$   &  $\langle \hat{f}_{_{\rm NL}} \rangle$   \\
\hline
\hspace*{0.1cm}$V_{1}+V_{2}+V_{3}$ \hspace*{0.4cm}           & 13    &   4    &  23    &  60     \\
\hspace*{0.1cm}$V_{0}+V_{1}+V_{2}+V_{3}$ \hspace*{0.4cm} & 12    &  3     &  22    & 60     \\
\hline
\hspace*{0.1cm}$f_{\rm sky}=0.40$      & $\langle \hat{f}_{_{\rm NL}} \rangle$ &   $ \langle\hat{f}_{_{\rm NL}} \rangle$  &  $\langle \hat{f}_{_{\rm NL}} \rangle$   & $ \langle\hat{f}_{_{\rm NL}}\rangle$    \\
\hline
\hspace*{0.1cm}$V_{1}+V_{2}+V_{3}$ \hspace*{0.4cm}           & 9  &   -0.1    &  18.5  &  56  \\
\hspace*{0.1cm}$V_{0}+V_{1}+V_{2}+V_{3}$ \hspace*{0.4cm} & 9  &  -0.16   &  18.5  & 56    \\
\hline
\end{tabular}
  \caption{Effect of Galactic foregrounds on the measurement of $f_{\rm NL}$ as a function of sky coverage $f_{\rm sky}$ and level of primordial non Gaussianity $f_{_{\rm NL}}^{\rm prim}$. The results displayed on this table assume Gaussian smoothing with $\theta_{_{\rm FWHM}}^{\rm S}=5'$ in the simulations used to perform the calculations (\S~\ref{sec:simpipe}).}
  \label{table:gal1}
  \end{minipage}
 \end{center}
 \end{table*}

Numerous interesting results can be extracted from Table~\ref{table:gal1}. 

\begin{enumerate}
\renewcommand{\theenumi}{(\alph{enumi})}
\item First, close to the galactic plane, so for  $f_{\rm sky}$ close to unity, a very important signal, that we denote by $f_{_{\rm NL}}^{\rm Gal-plane}$, overrides the primordial one. Indeed, when $f_{\rm sky}=0.90$, $\langle \hat{f}_{_{\rm NL}} \rangle$ does not depend much on the value of $f_{_{\rm NL}}^{\rm prim}$ and can be approximated as follows:
\begin{equation}
 \langle \hat{f}_{_{\rm NL}} \rangle = f_{_{\rm NL}}^{\rm Gal-plane}= f_{_{\rm NL}}^{\rm Gal-plane, const}+\dfrac{f_{_{\rm NL}}^{\rm prim}}{10}, \quad f_{\rm sky}=0.9
\label{eq:blabli}
 \end{equation}
with $f_{_{\rm NL}}^{\rm Gal-plane, const}\simeq -22$ for the $V_{1}+V_{2}+V_{3}$ combination and $f_{_{\rm NL}}^{\rm Gal-plane, const}\simeq -26$ for the $V_{0}+V_{1}+V_{2}+V_{3}$ combination. 

\item At higher latitudes, a second type of signal less powerful than $f_{_{\rm NL}}^{\rm Gal-plane}$ appears. This new contribution, denoted by $f_{_{\rm NL}}^{\rm Gal-high}(f_{\rm sky})$, brings a positive bias on the measurement of $f_{_{\rm NL}}$ and does not hide the primordial signal.  For  $f_{\rm sky}=0.8$, both $f_{_{\rm NL}}^{\rm Gal-plane}$ --which brings a negative bias--  and $f_{_{\rm NL}}^{\rm Gal-high}$ --which brings a positive bias-- contribute:
 \begin{equation}
 \langle \hat{f}_{_{\rm NL}} \rangle = f_{_{\rm NL}}^{\rm Gal-plane}+f_{_{\rm NL}}^{\rm Gal-high}(0.8)+f_{_{\rm NL}}^{\rm prim}, \quad f_{\rm sky}=0.8
 \end{equation}
with $f_{_{\rm NL}}^{\rm Gal-high}(f_{\rm sky}=0.8)\simeq 20$ and $f_{_{\rm NL}}^{\rm Gal-plane}$ given by eq.~(\ref{eq:blabli}).  
Those two signals compensate each other and it is remarkable to see that {\em if only three functionals are used, excluding the Area, the bias on $f_{_{\rm NL}}$ is almost negligible} (apart from the $f_{\rm NL}^{\rm prim}/10$ contribution)! 
 
\item Finally, for smaller $f_{\rm sky}$, the signal from the Galactic plane is totally hidden and only $f_{_{\rm NL}}^{\rm Gal-high}(f_{\rm sky})$ contributes as a positive bias:
\begin{equation}
 \langle \hat{f}_{_{\rm NL}} \rangle = f_{_{\rm NL}}^{\rm Gal-high}(f_{\rm sky})+f_{_{\rm NL}}^{\rm prim}, \quad f_{\rm sky} \la 0.7.
 \end{equation}
The quantity $f_{_{\rm NL}}^{\rm Gal-high}(f_{\rm sky})$ is shown in Table~\ref{table:gal_bias_fsky} for various sky coverages. As expected,   $f_{_{\rm NL}}^{\rm Gal-high}(f_{\rm sky})$ is a decreasing function of $f_{\rm sky}$. 
\end{enumerate}

\begin{table}
 \begin{center}
 \begin{tabular}{ l c c c c  }
\hline
$f_{\rm sky}$   & 0.80 & $0.70$& $0.60$& $0.40$\\
\hline
$f_{_{\rm NL}}^{\rm Gal-high}$ for $V_{1}+V_{2}+V_{3}$             &  22       &  17                    & 13                     & 9                 \\
 $f_{_{\rm NL}}^{\rm Gal-high}$ for $V_{0}+V_{1}+V_{2}+V_{3}$  & $21$       &  $ 16$              & 12                     &  9                      \\
\hline
\end{tabular}
  \caption{Galactic bias at high latitudes, $f_{_{\rm NL}}^{\rm Gal-high}$, as a function of sky coverage, $f_{\rm sky}$, as explained in the main text. The results displayed on this table assume Gaussian smoothing with $\theta_{_{\rm FWHM}}^{\rm S}=5'$ in the simulations used to perform the calculations (\S~\ref{sec:simpipe}).}
  \label{table:gal_bias_fsky}
 \end{center}
 \end{table}

 \subsubsection{Smoothing}
 
To understand more deeply the effects of the two types of Galactic foregrounds we found above, we examine what happens when the smoothing scale is varied. Table~\ref{table:gal_bias_smoothing} gives $\langle \hat{f}_{_{\rm NL}} \rangle$ as a function of $\theta_{_{\rm FWHM}}^{\rm S}$, for various sky coverages.  Obviously, one can infer from this table that the positive bias increases with smoothing scale, suggesting that the Galactic signal at high latitude dominates at large scales, while the Galactic plane signal is present with its negative bias only at small scales. This result, in addition to the analyses performed in previous paragraphs, show that Minkowski Functionals remain helpful in understanding and isolating the different biases induced by Galactic foregrounds, similarly as for the bi-spectrum. This demonstrates again the discriminative power of MFs. 
\begin{table}
 \begin{center}
 \begin{tabular}{ l | c c c c  }
\hline
& & $\langle \hat{f}_{_{\rm NL}} \rangle$ & & \\
$\theta_{_{\rm FWHM}}^{\rm S}$    & $f_{\rm sky} =0.80$ & $f_{\rm sky} =0.70$& $f_{\rm sky} =0.60$ & $f_{\rm sky} =0.40$ \\
\hline
\hspace*{0.1cm}$5'$     &  -5     & 16  &  12  &  9  \\
\hspace*{0.1cm}$10'$   &    12    &    25   &   17 & \\
\hspace*{0.1cm}$20'$     &    32  &    39    &  22  & \\
\hspace*{0.1cm}$40'$     &    37   &   36    &  25  & \\
\hline
\end{tabular}
  \caption{Galactic foreground bias as a function of smoothing scale $\theta_{_{\rm FWHM}}^{\rm S}$  for various sky coverages, $f_{\rm sky}$, in the null hypothesis $f_{_{\rm NL}}^{\rm prim}=0$. The measured quantity  $\langle \hat{f}_{_{\rm NL}} \rangle$ is given for the combination $V_{0}+V_{1}+V_{2}+V_{3}$,}
  \label{table:gal_bias_smoothing}
 \end{center}
 \end{table}

\subsubsection{Component separation}
\label{sec:compalpha}

The analyses in previous paragraph were performed in the most pessimistic case, when no component separation is performed, i.e. $\alpha=0.77$ in eq.~(\ref{eq:galmeth}). Now we consider the case when the Galactic foregrounds have been largely removed with a component separation method and that only a fraction remains, $\alpha < 0.77$. Table~\ref{table:gal_alpha} shows the bias expected on the measured $f_{_{\rm NL}}$ due to Galactic foreground residues as a function of $\alpha$ and for various sky coverages. Here, we focus on small scales, with Gaussian smoothing at $\theta_{_{\mathrm{FWHM}}}^{\rm S}=5'$ and on the measurement of $f_{_{\rm NL}}$ with the combination 
$V_{0}+V_{1}+V_{2}+V_{3}$.

\begin{table}
\begin{center}
 \begin{tabular}{ l   c  c  c  c }
\hline
\hspace*{0.1cm}$\langle \hat{f}_{_{\rm NL}} \rangle$ & $f_{\rm sky}=0.8$   & $f_{\rm sky}=0.7$  & $f_{\rm sky}=0.6$ & $f_{\rm sky}=0.4$   \\
\hline
\hspace*{0.1cm}$\alpha$=0.77   & -5  &   16    &  12     &  9   \\
\hspace*{0.1cm}$\alpha$=0.35   &  14   &   10    &  8       &   3.3  \\
\hspace*{0.1cm}$\alpha$=0.1     &  6.5  &    3.6  &  2       &   1.2  \\
\hspace*{0.1cm}$\alpha$=0.05   &  3     &   2      &  0.9    &  0.3 \\
\hspace*{0.1cm}$\alpha$=0.01   &  0.9  &   0.4   &  0.2    &  0.03      \\
\hline
\end{tabular}
 \end{center}
\caption{Galactic foregrounds bias as a function of component separation quality factor $\alpha$ and sky coverage. The forecasted value $\langle \hat{f}_{_{\rm NL}} \rangle$ is given in the null hypothesis $f_{_{\rm NL}}^{\rm prim}=0$. Note that this bias would also stand approximately for a value to add to $f_{_{\rm NL}}^{\rm prim}$ in the case $f_{_{\rm NL}}^{\rm prim}\neq 0$.} 
 \label{table:gal_alpha}
\end{table}

Table~\ref{table:gal_alpha} shows that with a good but realistic component separation 
($0.01 < \alpha < 0.05$), the bias due to the Galactic foregrounds becomes negligible compared to error bars (Table~\ref{table:mask_gal_sensitivity}) even for rather large sky coverage, $f_{\rm sky}=0.8$.

\section[Conclusion]{Conclusion}

In this article we have studied in detail the ability of Minkowski Functionals\footnote{and the number of clusters, that we call a Minkowski Functional here to simplify the presentation.} of excursions of the temperature fields to estimate primordial non Gaussianity, $f_{_{\rm NL}}$, in a Planck like experiment. To do that we used a standard Monte-Carlo approach to define a $\chi^2$ statistics assuming the weakly non Gaussian regime, where the errors are dominated by the Gaussian part of the signal. We first assessed the numerical limits in the $\chi^2$ approach. Then we studied the effects of inhomogeneous noise, point sources and galactic foregrounds on the determination of $f_{_{\rm NL}}$. The main results of our investigations, performed for the 100, 143 and 217\,GHz channels,  are the following:
\begin{enumerate}

\item It is best to measure normalised functionals, $v_k=V_k/A_k$, to reduce the effects of inhomogeneous noise. 

\item The functionals are all sensitive to non Gaussianity, but the following hierarchy can be set: Perimeter  $\gtrsim$ Genus $> N_{\rm clusters}  \gg $ Area. It is worth combining several Minkowski functionals to obtain better constraints on $f_{_{\rm NL}}$, although the Area does not improve much the results. 

\item To extract most of the information of interest while keeping the $\chi^2$ approach valid, it is convenient to perform the analyses in the ``$3.5$ sigma” excursion range with a number of bins $n_{\rm bins}=26$ for each of the functionals.

\item To have proper convergence of the $\chi^2$, it is required in practice to perform about $m=10\,000$ Gaussian simulations to estimate properly the covariance matrix.

\item Combining Wiener filtering for the field and its derivative bring the best constraints on $f_{_{\rm NL}}$ when using MFs. In particular, Wiener filtering does better than Gaussian smoothing. Note that most of the information on non Gaussianity is contained at the smallest possible scales, of the order of the beam size. 

\item Point sources foregrounds introduce a bias on the measurement of $f_{_{\rm NL}}$ that can be estimated and thus corrected for accurately. Note that with appropriate masking of the brightest sources followed by a simple in-painting procedure, this bias becomes negligible except at the 217\,GHz frequency. 

\item Galactic foregrounds introduce a complex bias on the measurement of $f_{_{\rm NL}}$ that depends on the fraction of sky covered, $f_{\rm sky}$, or in other words, depends on how much the most luminous part of the galaxy has been masked out. This bias can be again corrected for and is co\"incidentally negligible when the combination of Perimeter, Genus and number of clusters is used and $f_{\rm sky}=0.8$. With appropriate component separation the bias due to Galactic foregrounds should become negligible compared to the error bars on $f_{_{\rm NL}}$, at least for $f_{\rm sky} \la 0.8$.

\item With all the effects described above under control, we expect to be able to measure $f_{_{\rm NL}}$ using Minkowski Functionals with an error of the order
of $\Delta f_{_{\rm NL}}=10$.

\end{enumerate}
These are excellent news overall, since it suggests that Minkowski Functionals should be capable of putting interesting constraints on $f_{_{\rm NL}}$, even in view of the performance of fully optimal estimators for that purpose. It is reasonable to assume that they should also provide non-trivial constraints on other types of primordial non-Gaussianity even if there are no prediction yet at the time of the analysis which would allow building dedicated and somewhat more stringent indicators. Finally, although rather detailed, our analyses could be improved by lifting the following limitations before practical applications are considered:
\begin{enumerate}

\item Our model for testing component separation quality is rather na\"ive and two major issues remain to be addressed before drawing definite conclusions: (a) component separation methods do not remove galactic components the same way in each part of the sky nor each component: the description of Galactic residuals in terms of a contribution simply proportional to them has to be improved; (b) here, galactic components are ``added'' everywhere: component separation can subtract CMB signal too, an effect that we did not consider here.
In fact the analysis of  component separation method quality needs a specific study for each method at use \citep{2008A&A...491..597L}.

\item In this study we did not consider the use of foreground templates in the reference maps \citep{2002ApJ...566...19K, 2011ApJS..192...18K} to assess for the residuals. The use of templates, if they are reliable, is expected to correct for the biases introduced by the foregrounds. We can thus consider the results of our analyses as pessimistic in that respect.

\item We did not characterise biases induced by secondary anisotropies: they could be important, even dominant, as a bi-spectrum is created from the covariance between weak lensing and Sunyaev-Zeldovich effect or Integrated Sachs-Wolfe effect (ISW). Indeed, previous studies \citep{1999PhRvD..59j3002G,2008PhRvD..77j7305S,2009PhRvD..80h3004H} have warned about these spurious signals and a future work about their effects on Minkowski functionals is planned.

\item Our analyses neglected correlations in the noise. In a future work we shall refine them when reliable simulations of correlated noise are available. 

\end{enumerate}
In this paper, we have not used explicitly the analytic formulations of Minkowski functionals for our non Gaussianity studies, as what was done for example in \citet{2006ApJ...653...11H, 2008MNRAS.389.1439H}. Actually, using the ``Skewness parameters'' \citep{2003ApJ...584....1M} to study biases induced by secondaries, galactic foreground and point sources would allow us to use the numerous bi-spectra studies in an effective way. On the other hand, the advantage of our Monte-Carlo method is that it can be generalised to any statistics, e.g. for instance the skeleton length in the excursion \citep{2006MNRAS.366.1201N}, or any type of non Gaussianity, e.g. that induced by cosmic strings \citep[e.g.,][and references therein]{1988Natur.335..410B,2010AdAst2010E..66R}.

\section{Acknowledgements}
We thank G. Roudier for providing us the hit maps, J.-F. Cardoso for useful discussions about component separation and G. Castex for advices on point source masks.

\appendix

\section[Minkowski Functionals: definitions and theory]{Minkowski Functionals: definitions and theory}
\label{theory}

For a field $f(x)$ of zero average and variance $\sigma^{2}_0$ defined on the two-dimensional sphere $\mathbb{S}^{2}$, an overdense excursion set writes 
\begin{equation} 
\Sigma \equiv \lbrace x \in \mathbb{S}^{2} \vert \: f(x) > \nu\sigma_0  \rbrace.
\end{equation}
The boundary of the excursion is
\begin{equation}
\partial  \Sigma \equiv \lbrace x \in \mathbb{S}^{2} \vert \: f(x) = \nu\sigma_0  \rbrace. 
\end{equation}
Then the three Minkowski functionals on the sphere write
\begin{equation} {\rm Area:}\ V_{0}(\nu)=\dfrac{1}{4\pi}\int_{\Sigma}{\rm d}\Omega, \end{equation}
\begin{equation} {\rm Perimeter:}\ V_{1}(\nu)=\dfrac{1}{4\pi}\dfrac{1}{4}\int_{\partial \Sigma} {\rm d}l, \end{equation}
\begin{equation} {\rm Genus:}\ V_{2}(\nu)=\dfrac{1}{4\pi}\dfrac{1}{2\pi}\int_{\partial \Sigma} \kappa \, {\rm d}l, \end{equation}
where ${\rm d}\Omega$ and ${\rm d}l$ are respectively elements of solid angles (surface) and of angle (distance), $\kappa$ is the geodesic curvature. Note that the Genus can be also expressed as the number of components\footnote{A component is a connected subset of the excursion.} in the excursion minus the number of holes in the excursion. 

The fourth functional we use in this paper, $V_3(\nu)$, is defined, for $\nu > 0$,  as the number of components in the excursion. Symmetrically, for $\nu <0$, it is the number of underdense components (or the number of components in the excursion $\lbrace x \in \mathbb{S}^{2} \vert \: f(x) < \nu\sigma_0  \rbrace$).

In the Gaussian limit, the functionals can be expressed the following way \citep[see, e.g.][]{2010PhRvD..81h3505M,1983rafi.book.....V}:
\begin{equation}
V_{k}(\nu)=A_{k}v_{k}(\nu),
\end{equation} 
with
\begin{eqnarray}
v_{k}(\nu) &= &\exp(-\nu^{2}/2) H_{k-1}(\nu), \quad k \leq 2 \label{eq:nuk1} \\
v_{3}(\nu) &= & \dfrac{e^{-\nu^{2}}}{\mathrm{erfc}\left(\nu /\sqrt{2}\right)}, \label{eq:nuk2}
\end{eqnarray}
and
\begin{equation}
H_{n}(\nu)={\rm e}^{\nu^{2}/2}\left(-\dfrac{\rm d}{{\rm d}\nu}\right)^{n} {\rm e}^{-\nu^{2}/2}.
\end{equation}
The amplitude $A_k$ depends only on the shape of the power spectrum $C_{\ell}$:
\begin{eqnarray} 
A_{k} &= &\dfrac{1}{(2\pi)^{(k+1)/2}}\dfrac{\omega_{2}}{\omega_{2-k}\omega_{k}}\left( \dfrac{\sigma_{1}}{\sqrt{2}\sigma_{0}}\right)^{k},\quad k \leq 2 \\
A_{3} &= &\dfrac{2}{\pi}\left( \dfrac{\sigma_{1}}{\sqrt{2}\sigma_{0}}\right)^{2}
 \end{eqnarray}
where $\omega_{k}\equiv\pi^{k/2}/\Gamma(k/2+1)$, which gives $\omega_{0}=1$, $\omega_{1}=2$, $\omega_{2}=\pi$ and $\sigma_{0}$ and $\sigma_{1}$ are respectively the rms of the field and its first derivatives.

In the weakly non Gaussian regime, one can write, at leading order \citep[see, e.g.][]{2010PhRvD..81h3505M}
\begin{equation} 
v_{k} \simeq v_{k}^{(0)}+ v_{k}^{(1)}\sigma_{0},
\end{equation}
where $v_{k}^{(0)}$ is given by eqs.~(\ref{eq:nuk1}) and (\ref{eq:nuk2}), while the first order non Gaussian correction writes
\begin{equation}
 v_{k}^{(1)}(\nu) =\dfrac{S}{6}H_{k+2}(\nu) - \dfrac{kS_{\mathrm{I}}}{4} H_{k}(\nu)-\dfrac{k(k-1)S_{\mathrm{II}}}{4}H_{k-2}(\nu),
\end{equation}
with
\begin{equation}  
H_{-1}(\nu)\equiv {\rm e}^{\nu^{2}/2}\int_{\nu}^{\infty} {\rm e}^{-\nu^{2}/2} {\rm d}\nu={\rm e}^{\nu^{2}/2} \sqrt{\dfrac{\pi}{2}}\mathrm{erfc}\left(\dfrac{\nu}{\sqrt{2}}\right),
\end{equation}
and where
\begin{eqnarray}
S &= &\dfrac{\langle f^{3}\rangle}{\sigma_{0}^{4}}, \\
S_{\mathrm{I}} &= &\dfrac{\langle f^{2}\nabla^{2}f\rangle}{\sigma_{0}^{2}\sigma_{1}^{2}}, \\
S_{\mathrm{II}} &= &\dfrac{2\langle \vert\nabla f\vert^{2}\nabla^{2}f\rangle}{\sigma_{1}^{4}}
\end{eqnarray}
are skewness parameters \citep{2003ApJ...584....1M}. Each of them is a weighted integral of the bi-spectrum and is thus directly proportional to $f_{\rm NL}$.

\section[Algorithm used to compute Minkowski Functionals]{Algorithm used to compute Minkowski Functionals} \label{algorithm}
%
The code we developed is available by simple e-mail request. The numerical technique we use for measuring Minkowski functionals on HEALPix maps, consists, for each threshold value of the temperature, of two steps. 

In the {\it first step} of our algorithm, an on-grid cluster analysis is performed in order to define the connected ensembles of pixels that have temperature values above the threshold. The output of this step is a map of integers (flags) that assign a negative
``outside'' flag for pixels below the threshold and a positive flag that corresponds to the cluster number of each
connected region above the threshold.  Optionally, the boundary pixels (defined either within high excursion regions
or just outside them) can be marked.  The implementation follows closely the {\sc cnd\_reg2d} procedure developed
first for the Adhesion model \citep{AdhesionPreprint, 1990MNRAS.242..200K, 1992ApJ...393..437K}. 

The on-grid cluster analysis then proceeds as follows. At the onset, all pixels are considered as ``unseen''.
In the outer loop, the code scans the map, checking the temperature values
until the first pixel exceeding the threshold is encountered.  Along the way, the checked pixels that have been
found to be below the threshold are marked with the ``seen, below'' flag.  The new pixel above the threshold is 
assigned the flag that corresponds to the cluster order number (zero for the first found). Its neighbours are 
investigated. The ones below the threshold are marked as ``seen, below'', the ones above the threshold are given
the same cluster number flag and are put onto stack for further analysis.  Next the pixel from a stack is drawn,
its neighbours are checked in the same way, with the ones above the threshold further added to the stack.
This inner loop proceeds until the stack is exhausted which signifies that all connected pixels belonging
to the first cluster are found. The control is reverted to the outer scanner that proceeds checking the pixels,
skipping over those already analysed, until the next previously unseen pixel above the threshold is found. This
pixel  acquires the next cluster order number and then the inner, stack-based, loop finds
all the pixels connected to it, and so on.

This code is fast and linear in the number of pixels since the neighbours of each pixel are analysed only once.
Optionally, the boundary pixels can be marked differently from pixels inner to the regions. Different decisions
about what constitute connected pixels can be easily implemented. In the current implementation, we consider
all pixels that have at least a common vertex belonging to the same cluster (as we shall see below this has
to be taken into account when considering the Euler characteristic of the total excursion set).
The immediate by-products of the cluster analysis are the volume (area) of all connected regions above the threshold,
which constitutes the first Minkowski Functional, and their number $N_{\rm cluster}$.\footnote{To compute $N_{\rm cluster}$ for negative thresholds, one just multiplies the map by -1, and repeats the procedure.} 

The other two Minkowski functionals of the 2D excursion sets, the Euler characteristics and the perimeter
of the set are computed in the {\it second step} of our algorithm,  using the just obtained clustering information.

The Euler characteristic of each individual cluster, and, after summation over all the clusters, of the whole excursion
set above the threshold, can be obtained in one pass by analysing the pixel grid vertices on the cluster boundaries.
Gauss-Bonnet theorem links Euler characteristics of the region to the integration of  the curvature
of its boundary. However, since topological properties are invariant under any continuous transformation
of the boundary, the need to explicitly evaluate the boundary curvature is eliminated by considering
the curvature to be accumulated just in the outside vertices of the boundary pixels. Thus one only needs to
assign the appropriate curvature weights to the grid vertices and sum over them. 
Similar procedure on the Cartesian grid has been described in \cite{2012PhRvD..85b3011G}.
A 3D version of the code is also readily available.

Clearly, the vertex contribution
is determined solely by the temperature distribution of the pixels that form this vertex - which are below and
which are above the threshold. Necessary weights can be boot-strapped by considering the elementary situations.
Most of the grid vertices in HEALPix pixelisation are regular, being formed by four adjacent pixels.
\begin{enumerate}
\renewcommand{\theenumi}{(\alph{enumi})}
 \item  Consider a single pixel cluster above the threshold.  Its Euler characteristic is $\chi=1$. It has 4 boundary
vertices, all four equivalent, formed by one adjacent pixel above the threshold and three below. They should contribute
equally, thus we assign the weight $1/4$ to any vertex of this type.
\item Consider next a two pixels cluster, formed by pixels having a common side. The boundary has six vertices, four corner
ones of the type (a) and two new ones, which are formed by two side-by-side adjacent pixels above the threshold and two below it.
The cluster has $\chi=1$, thus new vertices must contribute a weight equal to $0$.
\item Consider a three pixels cluster that forms a corner. It has 5 vertices of type (a), which weights add up
to $5/4$, two vertices of type (b) and one new vertex that is formed by three pixels above the threshold and one below it.
For the total $\chi=1$ this vertex must contribute the weight $-1/4$.
\item The last possibility arises when we consider the boundary vertex formed by two pixels above the threshold
that are touching just at the vertex. Its weight depends whether we consider the clusters to be linked through the vertex
or disjoint. In the former case, which corresponds to our choice in cluster analysis, the weight is $1-6/4=-1/2$,
while in the latter case it is $2-6/4=+1/2$. However, for statistical analysis of the total Euler characteristic
of the excursion sets, both choices, while exact for the corresponding clustering decision, will lead to biased results.
Indeed, one may argue that in a situation when the discretised field has high and low pixels mixing at a vertex, it is 
equally probable that the regions below the threshold or above it connect through this vertex.
Assigning the weight zero in this case will reflect such a ``statistical'' consideration and this is the choice
that we make in our statistical analyses.
\item The vertex that is formed by four pixels above the threshold is not a boundary vertex and its weight is zero.
\end{enumerate}

The HEALPix grid on a sphere has eight special vertices that are formed only by three pixels. Their weights need to be defined
separately.  Consider elementary clusters that have one of such vertices.
\begin{enumerate}
\renewcommand{\theenumi}{(\alph{enumi})}
\setcounter{enumi}{5}
 \item A one pixel cluster with a special vertex has three regular vertices of type (a). Thus, a special vertex
 with one pixel above the threshold and two below has the weight $1/4$.
\item A two pixels cluster with a special vertex on a side shows that a special vertex 
formed by two pixels above the threshold and one below has the weight $0$.
\item A three pixels cluster with a special vertex inside has only 3 exterior regular vertices of
type (a). Thus, a special vertex formed by all three pixels above the threshold, albeit being an inner one,
has a weight $1/4$. These eight special vertices are  providing the Euler characteristic of the entire spherical
manifold,  $\chi=2$, when all pixel values lie above the threshold.
\end{enumerate}

Once the weights are defined, the computation of the Euler characteristic of individual clusters and the entire excursion
set is a simple one pass loop over vertices to determine their type from the temperature value at the adjacent pixels and
add the weights based on the index of pixels that are above the threshold. Our method does not involve any differentiation
nor integration of the field. Moreover, if only the total $\chi$ is needed, the cluster analysis step can be omitted.
The code works in the presence of masks of arbitrary complexity, treating the field as defined on
a manifold that itself has a non-trivial Euler characteristic. 

The measurement of the perimeter of the excursion set is somewhat more complex. Our technique is again based on 
a scan over the boundary grid vertices, for each of which we linearly interpolate the field based on the values in
the four (or three) adjacent pixels. Interpolation is least-squares, linear, performed in the plane tangent to the sphere
at the vertex. Interpolation requires first derivatives of the field which are precomputed by standard HEALPix routines
and stored. After finding the best-fit plane for the local field behaviour, we compute the intersection of this plane with
the polygon formed by the pixel centres around the vertex. The length of the intersection is then added to the perimeter of 
the cluster the vertex belongs to. We found that linear interpolation is sufficient given the level of accuracy offered by
the  pixelisation and there is no advantage to go for a higher order interpolation in comparison with increasing
the pixelisation level.  

The code we developed along these lines is available by simple e-mail request.

\section[Inversion of the covariance matrix: a convergence study of the $\chi^2$]{Inversion of the covariance matrix: a convergence study of the $\chi^2$} \label{app:numsim}

The inversion of matrix $C$ in eq.~(\ref{eq:monchi22}) is performed using Singular Value Decomposition (SVD):
$C=UDU^{T}$ or
  \begin{equation} C_{ij}=\sum_{k=1}^{n}U_{ik}d_{k}U_{jk},
 \end{equation}
hence
 \begin{equation} \chi^{2}=\sum_{i=1}^{n}  \sum_{j=1}^{n} \sum_{k=1}^{n}\dfrac{1}{d_{k}}U_{ik}U_{jk} \left[ \hat{y}_{i}- \bar{y}_{i}( f_{_{\rm NL}} )\right] \left[\hat{y}_{j}- \bar{y}_{j}( f_{_{\rm NL}} )\right],
 \end{equation}
  where  $U$ is an orthogonal matrix and $D$ a diagonal matrix such that its diagonal terms $d_k$, the singular values, are ranked in descending order.  Here, these latter are expected to be positive, since $C$ is definite positive by construction. When there is a large contrast between the singular values,  $r \equiv d_1/d_n \gg  1$,  the calculation of the inverse of $C$ becomes an ill-conditioned problem if $C$ is not computed with sufficient accuracy. In our calculations, the accuracy is mainly controlled by the number $m$ of simulations used to estimate $C$, but it also depends on other parameters such as the number of bins, the value of $\nu_{\rm max}$, the smoothing scale $\ell$, the presence of masks, etc. In particular, the ratio $r$ increases when the space between successive bins in the excursion is reduced, because they become more correlated. A sufficiently accurate calculation of $C$ can be costly. For instance, it takes about 3 mn at present time to mesure MFs on a HEALPix map with $N_{\rm side}=2048$, $\nu_{\rm max}=3.5$, $n_{\rm bins}=26$ and a Gaussian smoothing of $\theta_{_{\rm FWHM}}^{\rm S}=10'$, an operation that can become prohibitive if repeated many times, this is why it is worth investigating in details what should a reasonable value of $m$ to have a few percent error on the estimate of the $\chi^2$. 

In this appendix, instead of studying the accuracy in the determination of the inverse of $C$, we analyse directly the convergence of the $\chi^2$ as a function of $m$ for $f_{\rm NL}=0$. We focus on the perimeter functional, ${\hat y}={\hat \nu}_1$, but similar results would be obtained for other functionals. To perform the Gaussian simulations for computing the covariance matrix, we shall use the following typical configuration: the  HEALPix resolution is $N_{\rm side}=2048$ and the power-spectrum is calculated as detailed in Appendix \ref{planck}. The maps are convolved with  a Gaussian beam of size $\theta_{_{\rm FWHM}}^{\rm b}=7.1'$, then supplemented with a white noise of  $0.33\mu$K.deg, and finally smoothed with a Gaussian smoothing kernel with $\theta_{_{\rm FWHM}}^{\rm S}=10'$. Our concept of infinity corresponds to  $m=20\,000$ simulations for estimating the covariance matrix in the ``ideal'' case. 

Here we want to be able to estimate $m$ for various choices of the parameters which are the most influent in the convergence of the covariance matrix, namely the number of bins, $n$ and the excursion range, $\nu_{\rm max}$. We already computed a covariance matrix  $C_{20k}$ for a rather large value of $m=20\,000$. However, the optimal value of $m$ remains unknown and might be larger than 20\,000 in some cases. Now, to make a convergence study, we should realise a large number of trials, each of them with a subset of $m$ realisations, in order to estimate the error in the estimate of $C$ with $m$ simulations. To avoid that procedure to be prohibitive, instead of generating random maps and measuring the genus on each of them, we use the property that the random vectors ${\hat x}\equiv {\hat y}^{\rm G}-{\bar y}^{\rm G}$ are nearly Gaussianly distributed to simulate directly and rapidly a Gaussian distribution of $m$ vectors ${\hat x}$ of zero average and covariance $C_{20k}$. To do that, we use the following standard procedure. We draw
\begin{equation}
{\hat x}_{\rm N}=N(0,1)\ {\rm of}\ {\rm size}\ n,
\end{equation}
then the vector
\begin{equation}
{\hat x}=U (\sqrt{D}.{\hat x}_{\rm N})
\end{equation}
verifies, as required,
\begin{equation}
\left\langle \  {\hat x} {\hat x}^{T}\right\rangle =UDU^{T}=C_{20k}.
\end{equation}
The point is to see whether using $m$ realisations, ${\hat x}^{(i)}$, $i=1,\cdots,m$,  of vector ${\hat x}$ to estimate the covariance matrix is sufficient to compute the $\chi^2$ accurately enough. For one data realisation with $\fnl=0$, 
\begin{equation}
{\hat y} \equiv {\hat x}^{(0)}+\bar{y}^{\rm G}, 
\end{equation}
where ${\hat x}^{(0)}$ is generated randomly the same way as just above, the estimator of the $\chi^2$ writes
\begin{equation}
\chi^2_{\rm MC}(m)= \left[ {\hat y}-\bar{y}(f_{_{\rm NL}}) \right]^T C_{\rm MC}^{-1}(m) \, \left[{\hat y}-\bar{y}(f_{_{\rm NL}}) \right],
\end{equation}
with the covariance matrix estimated from $m$ realisations,
\begin{equation}
C_{\rm MC}(m)\equiv\frac{1}{m} \sum_{i=1}^m {\hat x^{(i)}} {\left[ \hat x^{(i)} \right]}^T,
\end{equation}
while the ``true'' $\chi^2$ should write
\begin{equation}
\chi^2_{\rm true}= \left[ \hat y -\bar{y}(f_{_{\rm NL}}) \right]^T C^{-1} \, \left[{\hat y} -\bar{y}(f_{_{\rm NL}}) \right],
\end{equation}
where we approximate here $C^{-1}$ by the inverse of $C_{20k}$. 
The relative difference between the exact $\chi^2$ and the one driven from $m$ simulations writes
\begin{equation}
\epsilon_m({\tilde x}) \equiv \frac{\chi^2_{\rm MC}(m)-\chi^2_{\rm true}}{\chi^2_{\rm true}},
\end{equation}
where ${\tilde x} \equiv (x^{(0)},x^{(1)},\cdots,x^{(m)})$. To assess the convergence, we need the second moment of $\epsilon_m$ over many realisations $M$ of ${\tilde x}$ [this corresponds to $M\times (m+1) \times n$ random numbers in total] to be small
\begin{equation}
\langle \left[ \epsilon_m({\tilde x}) \right]^2 \rangle \leq \epsilon^2.
\end{equation}
We performed this exercise for $\epsilon=2\%$,  with $M=1000$ and for the following set of values of $m=1000\, i , \: i\in\lbrace1, \ldots, 20\rbrace$. The results of our analyses are summarised in Table \ref{table:number_m1}. 

\section[Simulating Planck data]{Simulating Planck data} \label{planck}

\renewcommand{\arraystretch}{1.8}
\begin{table*}
 \begin{center}
 \begin{minipage}{120mm}
\begin{tabular}{ @{}  l  c  c  c @{} }
\hline
Channels & Beam size $\theta_{_{\rm FWHM}}^{\rm b}$ &  Noise (nominal mission) & Noise (extended mission) \\
\hline
100\,GHz  & 10' & 1.1$\,\mu$K.deg & 0.7$\,\mu$K.deg \\
143\,GHz  & 7.2' & 0.7$\,\mu$K.deg & 0.5$\,\mu$K.deg \\
217\,GHz & 5' & 1.1$\,\mu$K.deg & 0.7$\,\mu$K.deg \\
 100+143+217\,GHz & 7.2' \footnote{The beam for this combination of channels is not known but should be in-between those of the 143\,GHz and 217\,GHz channel, which are the most CMB constraining at high and very high angular resolution. We have conservatively used the resolution of the 143\,GHz for this study } &  0.5$\,\mu$K.deg & 0.33$\,\mu$K.deg  \\
\hline
\end{tabular}
  \caption{Planck characteristics, in terms of resolution and levels of noise, used in this study.}
  \label{table:noise_levels}
  \end{minipage}
 \end{center}
 \end{table*}

\subsection{Observations with Planck}

Planck is the third and latest generation of space observatory designed to observe the anisotropies of the cosmic microwave background (CMB) over the entire sky, at a high sensitivity and angular resolution. It observes the sky in nine frequency bands: the Low Frequency Instrument \citep[LFI;][]{2010A&A...520A...3M} covers the 30, 44	and 70\,GHz bands	with	amplifiers	cooled	to	20 K; the	High	Frequency	Instrument \citep[HFI;][]{2011A&A...536A...6P} covers the 100, 143, 217, 353, 545 and 857\,GHz	bands	with	bolometers	cooled	to	0.1 K.

In order to assess characteristics of Planck which are particularly relevant for the specific purposes of this paper, i.e. the noise level and the resolution of the CMB map analysed, we consider the expected performances of the three HFI channels at 100, 143 and 217\,GHz, and we adopt two durations for the mission: 15 months and 30 months which we shall refer to in the following as the nominal and extended mission.

The specifications of the three channels that are studied here are gathered in table~\ref{table:noise_levels}. The beams are assumed to be isotropic (``circular”) Gaussian. The beam size in table~\ref{table:noise_levels} is given in terms of ``FWHM'' scale as detailed in \S~\ref{app:simulations}. 

\subsection{Simulations} \label{app:simulations}

The simulation procedure used in this article can be summarised as follows
\begin{eqnarray}
{\rm map}& = & {\rm CMB}(a_{\ell,m},f_{\rm NL})*{\rm beam}[\theta_{\rm FWHM}^{\rm b}(f)] \nonumber \\
               & + & {\rm Galactic\ Foreground}*{\rm beam}[\theta_{\rm FWHM}^{\rm b}(f)]  \nonumber \\
               & + &{\rm Sources\ Foreground}*{\rm beam}[\theta_{\rm FWHM}^{\rm b}(f)]  \nonumber \\
               & + & {\rm noise}(f) + {\rm Point\ Sources\ Mask\ inpainted} (\beta, f) \nonumber \\
               & + &{\rm smoothing}(\theta_{\rm FWHM}^{\rm S}) {\, \rm or\ Wiener\ filtering} \nonumber \\
               & + &{\rm galactic\ Mask}(f_{\rm sky}) ,
\end{eqnarray}
where $f$ is the considered frequency.  We mainly describe here how the first line of this equation is dealt with, while other aspects are examined in various sections of the article. 

To generate the Gaussian part of the CMB, we use harmonic coefficients $a_{\ell,m}$ derived from standard Cold Dark Matter (CDM) cosmology, with the best parameters obtained from WMAP7+BAO+$H_{0}$ \citep{2011ApJS..192...18K}: $\Omega_{\Lambda}=0.728$, $\Omega_{c}h^{2}=0.1123$, $\Omega_{b}h^{2}=0.0226$, $H_{0}=70.4$ km/s/Mpc, $n_{s}=0.967$, $\tau=0.085$ and $\Delta_{\mathcal{R}}^{2}(k_{0})=2.42\times 10^{-9}$. 

The maps are generated using HEALPix pixelisation \citep{2005ApJ...622..759G}. Most of the calculations performed in that paper use $N_{\rm side}=1024$ and assume a truncation of the harmonic modes at $\ell=\ell_{\rm max}=2000$, but we also examined other
values of $N_{\rm side}$ and $\ell_{\rm max}$ as shown on Table~\ref{table:sensitivity_nside_lmax}. 

For the non Gaussian part of the CMB, we used updated versions of simulations of harmonic coefficients provided by \citet{2009ApJS..184..264E} which allow us to choose $N_{\rm side}$ and $\ell_{\rm max}$ up to $\ell=3500$.

Convolution, in particular with the Gaussian window, was performed in harmonic space using HEALPix package. In our conventions, the Gaussian kernel, writes, in harmonic space, 
\begin{equation}
W_{\ell}=\exp[-\ell(\ell+1)\dfrac{\theta_{\rm S}^{2}}{2}] \label{eq:beameq} 
\end{equation} 
where $\theta_{\rm S}$ is the Gaussian size. Here we will use instead the corresponding full width at half maximum (FWHM): $\theta_{_{\rm FWHM}}^{\rm S}=\theta_{\rm S} \times 2.35$.

\label{lastpage}

\end{document}